\documentclass[prd,twocolumn,nofootinbib,longbibliography]{revtex4-1}
\usepackage{verbatim}
\usepackage{amssymb}
\usepackage{amsmath}
\usepackage{amsfonts}
\usepackage{bm}
\usepackage{hyperref}
\usepackage{graphicx}
\usepackage{tensor}
\usepackage{mathrsfs}
\usepackage{color}
\usepackage{scrextend}

\usepackage{enumitem}

\newcommand{\nnm}{\nonumber}
\newcommand{\doe}{\partial}
\newcommand{\be}{\begin{equation}}
\newcommand{\ee}{\end{equation}}
\newcommand{\bea}{\begin{eqnarray}}
\newcommand{\eea}{\end{eqnarray}}
\newcommand{\bdm}{\begin{displaymath}}
\newcommand{\edm}{\end{displaymath}}
\newcommand{\bse}{\begin{subequations}}
\newcommand{\ese}{\end{subequations}}

\newcommand{\mr}{\mathrm}
\newcommand{\tr}{\textrm}
\newcommand{\mc}{\mathcal}

\newcommand{\py}{\phantom{yo}}

\newcommand{\bs}{\boldsymbol}

\newcommand{\cd}{\;\!\!\cdot\;\!\!}

\newcommand{\dw}{{\mc G}}
\newcommand{\st}{\;\!\!*\;\!\!}
\newcommand{\ms}{\mathsf}

\begin{document}

\title{Scattering of two spinning black holes in post-Minkowskian gravity, \\ to all orders in spin, and effective-one-body mappings}
\date{\today}
\author{Justin Vines}
%\email{justin.vines@aei.mpg.de}
\affiliation{Max Planck Institute for Gravitational Physics (Albert Einstein Institute),
\\ Am M{\"u}hlenberg 1, 14476 Potsdam-Golm, Germany}

\begin{abstract}
We demonstrate equivalences, under simple mappings, between the dynamics of three distinct systems---(i) an arbitrary-mass-ratio two-spinning-black-hole system, (ii) a spinning test black hole in a background Kerr spacetime, and (iii) geodesic motion in Kerr---when each is considered in the first post-Minkowskian (1PM) approximation to general relativity, i.e.\ to linear order $G$ but to all orders in $1/c$, and to all orders in the black holes' spins, with all orders in the multipole expansions of their linearized gravitational fields.  This is accomplished via computations of the net results of weak gravitational scattering encounters between two spinning black holes, namely the net $O(G)$ changes in the holes' momenta and spins as functions of the incoming state.  The results are given in remarkably simple closed forms, found by solving effective Mathisson-Papapetrou-Dixon-type equations of motion for a spinning black hole in conjunction with the linearized Einstein equation, with appropriate matching to the Kerr solution.
The scattering results fully encode the gauge-invariant content of a canonical Hamiltonian governing binary-black-hole dynamics at 1PM order, for generic (unbound and bound) orbits and spin orientations.  We deduce one such Hamiltonian,
which reproduces and resums the 1PM parts of all such previous post-Newtonian results, and which directly manifests the equivalences with the test-body limits via simple effective-one-body mappings.
\end{abstract}

\date{\today}

\maketitle

%\tableofcontents

\section{Introduction}\label{sec:intro}

The problem of determining the motion of two black holes (BHs) under their mutual gravitational interaction according to general relativity, the binary BH problem, persists as a key challenge of classical gravitational physics, being both of fundamental theoretical interest and pressingly relevant for astrophysical applications.  Its study is crucial to the burgeoning field of gravitational-wave (GW) astronomy, with the first four GW signals confidently detected by Earth-bound observatories each having originated from (what very much appear to be) the final inspirals and mergers of binary BHs \cite{Abbott:2016blz,Abbott:2016nmj,Abbott:2017vtc,Abbott:2017oio}.

%The LIGO observatories have now confidently detected three gravitational-wave signals passing Earth \cite{Abbott:2016blz,Abbott:2016nmj,Abbott:2017vtc}, each attributed to the final inspiral and merger of a binary system composed of two black holes (BHs).  While these events already provide unique probes of strong-field gravitational dynamics and BH physics, they represent only the beginning of the science potential of gravitational-wave astronomy.  This potential will grow significantly in the coming years, not only with more signals from networks of ground-based detectors, but also with increasingly sensitive pulsar timing arrays {\color{red}[??]} and with the launch of space-based gravitational-wave detectors such as eLISA \cite{Seoane:2013qna}.

%The analysis of gravitational waves from a binary coalescence requires accurate waveform templates, and this continues to drive efforts to find more accurate and more general solutions to the two-body problem in general relativity (and in alternative theories of gravity \cite{TheLIGOScientific:2016src,Yunes:2016jcc}).  

Alongside direct numerical integration of the Einstein field equations, relativistic binary dynamics is fruitfully studied within the following complementary analytic approximation schemes; 
for masses $m$, speeds $v$, and orbital distances $r$,
\begin{itemize}
\item the \emph{post-test-body} (or \emph{self-force}) approximation assumes an extreme mass ratio, $m_1\ll m_2$, and perturbs about the limit of test-body motion in a stationary background BH spacetime, but is valid in the strong-field, fast-motion regime \cite{Poisson:Pound:Vega:2011,Harte:review,Barack:2014,Pound:review};
\item the \emph{post-Newtonian} (PN) approximation assumes weak fields and low speeds, $Gm/rc^2\sim v^2/c^2\ll1$, and perturbs about the limit of Newtonian gravity (formally, an expansion in $1/c$) \cite{Futamase:2007zz,Schafer:2009dq,Blanchet:2013haa,poisson2014gravity,Rothstein:2014sra,Porto:2016pyg}; and
\item the \emph{post-Minkowskian} (PM) approximation assumes weak fields, $Gm/rc^2\ll 1$, but unrestricted speeds, $v^2/c^2\lesssim 1$, and perturbs about the limit of special relativity (formally, an expansion in $G$) \cite{Bertotti:1956,Bertotti:1960,Rosenblum:1978,Bel:1981,Damour:1981PhLA,Portilla:1979,Portilla:1980,Westpfahl1979,Westpfahl:1985,Wespfahl:1987,Ledvinka:2008,Foffa:2013gja,Damour:2016s}.
\end{itemize}
Of the latter two perturbation schemes, those directly applicable to comparable-mass systems, the PN expansion has received much more attention in recent years.  This is because the most promising sources for current GW detectors are \emph{bound} (in fact, nearly circular-orbit) systems of compact objects such as BHs or neutron stars, satisfying $v^2/c^2\sim Gm/rc^2$ (weak fields imply low speeds), thus being most naturally treated in the PN scheme.  The PM approximation, conversely, is most natural for treating \emph{unbound} systems, i.e.\ scattering situations, which can access the regime $v^2/c^2\sim 1\gg Gm/rc^2$.  Both schemes are nonetheless applicable to both bound and unbound systems, and they provide overlapping and complementary information on two-body dynamics.\footnote{Note that we are referring to applying both the PN and PM schemes to the binary's \emph{orbital dynamics, in the near zone}.  This is not to be confused with the use of the PM approximation to describe GW propagation in the far zone, which is necessary in either case, requiring a PN-near-zone--PM-far-zone matching procedure in the PN case; see e.g.\ \cite{Blanchet:2013haa}.  The 1PM (linear-in-$G$) dynamics considered here is purely conservative, with radiative effects entering at higher orders in $G$.
%At the 1PM level considered here, the dynamics is purely conservative, with dissipative effects entering at higher orders in $G$.
}

This paper considers two-BH scattering according to the 1PM approximation (to linear order in $G$, to all orders in $1/c$), but the results have direct relevance to bound binary BHs as well.  As discussed e.g.\ in \cite{Damour:2009sm,Bini:2012ji,Damour:2014afa,Damour:2016s,Bini:2017wfr} and below, the net results of scattering encounters can be seen to encode \emph{gauge-invariant} information characterizing the dynamics of both unbound \emph{and bound} systems.

While 1PM scattering of \emph{point-masses} (structureless, spherically symmetric, nonspinning bodies) has been treated by several authors \cite{Portilla:1979,Portilla:1980,Westpfahl1979,Westpfahl:1985,Ledvinka:2008,Damour:2016s}, one of our novel contributions in this paper is to consider 1PM scattering with the effects of the bodies' spins and \emph{all} higher multipole moments, \emph{to all orders in spin}, for the specific case of bodies with spin-induced multipoles matching those of spinning BHs.

A spinning body, such as a BH, generally has axisymmetric mass multipole moments $\mc I_\ell$ and momentum/current multipole moments $\mc J_\ell$, scaling with increasing powers of the body's spin.  The \mbox{$2^\ell$-pole} moments of a BH of mass $m$ and spin $S$ are given by the simple relation 
\be\label{mcMell}
\mc M_\ell=\mc I_\ell+\frac{i}{c}\mc J_\ell= m(ia)^\ell,
\ee 
where $a=S/mc$ \cite{Hansen:1974zz}.  Spin and spin-induced multipole effects are typically (appropriately) treated perturbatively in PN calculations, truncating the multipole series or spin expansion at a finite order.
%\footnote{Orders of the spin/multipole expansion are are appropriately grouped with particular orders of the PN expansion.  In using $1/c$ as the formal PN expansion parameter, with an ``$n$PN'' contribution (e.g.\ to an action or Hamiltonian) scaling as $1/c^{2n}$, the order-counting appropriate for rapidly rotating BHs is manifested by expressing spins $S$ in terms of dimensionless spin parameters $\chi\in(0,1]$, with $S=mca=Gm^2\chi/c$.  Using the $\chi$ parametrization makes each factor of spin come with an extra factor of $G$, which, in using $G$ as the formal expansion parameter for the PM expansion, would make higher powers of spin appear at higher PM orders.  This is not the counting we use for the PM expansion here, choosing instead to express spins in terms of $S$ or $a$, introducing no further factors of $G$ with spin terms, and maintaining $G$ as our formal expansion parameter.  Quite apart from that matter of bookkeeping, }  
In the PM approximation, however, it is arguably natural to treat all multipoles, or all powers of spin, on an equal footing.\footnote{In using $1/c^2$ as a formal PN expansion parameter, counting orders of $v^2/c^2\sim Gm/rc^2$, the appropriate PN scaling of spin/multipole effects for rapidly rotating BHs is made manifest by expressing spins in terms of the dimensionless parameters $\chi\in[0,1)$ with $S=mca=Gm^2\chi/c$.

In using $G$ as a formal PM expansion parameter, one would assign different PM orders to spin effects depending on whether they were expressed in terms of $S$'s or $a$'s or $\chi$'s.  Our ``convention'' here is to use $S$'s or $a$'s, not $\chi$'s, in which case all powers of spin will contribute ``at linear order in $G$'' or ``at 1PM order.''

Physically, $Gm/rc^2$ and $a/r$ represent two \emph{a priori} distinct dimensionless small parameters, and though these are comparable for high-spin BHs $(\chi\lesssim 1)$, we will work to linear order in $Gm/rc^2$ but to all orders in $a/r$.  Our basic motivation for doing so is that this is the (maximal) level of approximation accessible by use of the linearized Einstein equation on a flat background---this is indeed the more appropriate definition of the 1PM approximation.}

This paper's central result is a computation to 1PM order of the net changes in the momenta and spins of two spinning BHs (with arbitrary masses) due to a weak gravitational scattering encounter, working to all orders in the BHs' multipole expansions, to all orders in their spins.  The results, in (\ref{D1Z}) and (\ref{DpDa2}) below, are given as strikingly compact, specially covariant expressions, with the spin-multipole series resummed in closed form.

A significant outcome of these calculations is that the results reveal simple relationships between the dynamics of arbitrary-mass-ratio two-BH systems and the dynamics of test bodies in stationary background spacetimes, at 1PM order.  

Before turning to the situation with spins, we should discuss the work of \cite{Damour:2016s}, which revisited classic results \cite{Portilla:1979,Portilla:1980,Westpfahl1979,Westpfahl:1985} for 1PM scattering of point-masses, or nonspinning BHs (and which motivated the investigations of this paper).  It was shown in \cite{Damour:2016s} that the scattering dynamics of an arbitrary-mass-ratio two-point-mass system is fully equivalent at 1PM order to unbound geodesic (test-point-mass) motion in a background Schwarzschild spacetime, under a simple mapping.  The only nontrivial aspect of the mapping is an \emph{energy map}, relating the energy of the test body relative to the background and the total center-of-mass-frame energy of the two-body system, which can be seen to arise from simple special-relativistic kinematics applied to scattering states at infinity.  This turned out to be the same as (and gave new insight into) the energy map at the core of effective-one-body (EOB) models for relativistic binary dynamics \cite{Buonanno99,Buonanno00,Damour1SEOB,DJS,Barausse:2009xi,Damour:2012mv,pan2014,2016LNP...905..273D,Balmelli:Damour:NLOSS}.  Such EOB models, premised on modeling arbitrary-mass-ratio two-body motion after test-body motion in a fixed background while also incorporating information from PN and self-force calculations and numerical relativity simulations, have played an important role in the detection and analysis of (including tests of general relativity with) the first gravitational wave signals \cite{Abbott:2016blz,Abbott:2016nmj,Abbott:2017vtc,Abbott:2017oio,TheLIGOScientific:2016wfe,Abbott:2016izl,TheLIGOScientific:2016src}.

We find results somewhat analogous to those of \cite{Damour:2016s}, with the point-masses there replaced by spinning BHs here.  It is pertinent in the spinning BH case to consider the dynamics of not two but three distinct systems,
\begin{enumerate}
\item[(i)] an arbitrary-mass-ratio two-spinning-BH system,
\item[(ii)] a spinning test BH (a test body with the spin-induced multipole structure of a BH) in a background Kerr spacetime, and
\item[(iii)] a test point-mass following a geodesic in a background Kerr spacetime,
\end{enumerate}
the latter two being limits of the first.  From (i) to (ii) we take the \emph{test-body limit}, formally taking the mass of the test BH to zero---more precisely, taking the mass ratio $m/M$ to zero, in which case $m$ scales out---, while keeping its rescaled spin $a=S/mc$ finite to retain nonzero spin/multipole effects, as in the Mathisson-Papapetrou-Dixon (MPD) model of extended-test-body motion \cite{Mathisson:1937,Mathisson:2010,Papapetrou:1951pa,Dixon:1979,Dixon:2015vxa}.  From (ii) to (iii), we take the spin of the test BH to zero.

Using our scattering results for system (i) and limits thereof, we are ultimately able to demonstrate complete equivalences between the dynamics of systems (i), (ii), and (iii) at 1PM order, under simple \emph{EOB mappings} which involve the same EOB energy map of \cite{Buonanno99,Damour:2016s} along with new (energy-dependent) mappings of the BHs spins.
%The aims of the present paper are, firstly, to derive new results for the 1PM scattering of \emph{two spinning BHs}, working to all orders in the BHs' spins (or to all orders in the multipole expansions of their linearized gravitational fields), and secondly, to show that the results reveal two distinct EOB equivalences which generalize the findings of \cite{Damour:2016s} for point-masses to the case of spinning BHs:
%\begin{enumerate}
%\item [(A)] the 1PM dynamics of an arbitrary-mass-ratio two-spinning-BH system is equivalent to the 1PM dynamics of a spinning test BH in a stationary Kerr field,
%\end{enumerate}
%and more surprisingly,
%\begin{enumerate}
%\item [(B)] each of those is equivalent at 1PM order to geodesic (test-point-mass) motion in a Kerr spacetime.
%\end{enumerate}
%With each BH having an infinite series of multipole moments proportional to increasing powers of the BH's spin \cite{Hansen:1974zz,Vines:2016qwa}, 
The equivalence of (i) or (ii) with (iii) involves the remarkable property that, for both systems (i) and (ii), at 1PM order,
\vspace{0.2cm}
\begin{addmargin}[1em]{0em}
the double sum over BH-1-multipole--BH-2-multipole couplings \mbox{$(\mc M_{1k}\otimes\mc M_{2\ell})$} factorizes into a single sum over the multipoles of one effective spinning BH ($\mc M_{0\ell}$) coupled to an effective point-mass; schematically,
\begin{alignat}{3}
\sum_{k=0}^\infty \sum_{\ell=0}^\infty \bigg[\mc M_{1k}\otimes\mc M_{2\ell}&=m_1(ia_1)^k\otimes m_2(ia_2)^\ell\bigg]
\nnm\\
\to\quad m_1\otimes\sum_{\ell=0}^\infty \bigg[{\mc M}_{0\ell} &=m_2\big(i(a_1+a_2)\big){}^\ell\bigg],
\end{alignat}
\end{addmargin}
cf.\ (\ref{mcMell}).
Hints of this property were first noted in 
\cite{Damour1SEOB}, which demonstrated the corresponding factorization of the leading-PN-order spin-squared (quadrupole-monopole and dipole-dipole) couplings in the canonical Hamiltonian for a binary BH; such Hamiltonian factorizations were later argued in \cite{Vines:2016qwa} to occur at all even orders in spin at the leading PN orders.  We find here that, in an appropriate sense, more evident in terms of covariant scattering results than in terms of canonical Hamiltonians, seen most directly in (\ref{Lintfactor}) below, the factorization occurs at all (even and odd) orders in spin and at all orders in $1/c$, to linear order in $G$.  %(Furthermore, the multipole series can be summed in closed form.)

Our treatment of spinning BH dynamics at 1PM order is 
based on a classical \emph{worldline-skeleton} effective action treatment of spinning extended (test) bodies, yielding a form of the MPD dynamics.   Following the seminal works \cite{Hanson:Regge:1974} in flat spacetime and \cite{Bailey1975} in curved spacetime, the worldline action approach has been developed, applied, and reviewed by many authors, e.g.\ in \cite{Porto:2005ac,Blanchet:2013haa,Steinhoff:2014,Marsat:2014xea,Levi:Steinhoff:2015:1,Porto:2016pyg}.   The latter works have focused mostly on explicit results in the PN approximation, but many of their intermediate results are readily applicable to the PM approximation.  In particular, an effective action for a spinning BH valid to all orders in the multipole expansion at linear order in curvature (and thus complete at 1PM order, yielding at 1PM order an equivalent form of the action we employ here), has been given in \cite{Levi:Steinhoff:2015:1}; its form at that order is fully determined by general covariance and other appropriate symmetries along with matching to the linearized Kerr solution.  

A related approach to computing spin and multipole contributions in two-body dynamics considers the classical limits of quantum fields coupled to gravity.  Minimally coupled massive spin-$(n/2)$ fields are seen to lead to effective particles with a spin-multipole structure matching that of a BH up to the $2^n$-pole (up to order spin$^n$), as has been explored in the PN context in \cite{Holstein:2008sx,Vaidya:2014kza}.  Recent work in \cite{Guevara:2017csg} has treated arbitrary-spin fields, with full BH results obtained in the limit of an infinite-spin field, and has given results for the classical part of the quantum two-body scattering amplitude, without expanding in $1/c$, at both the tree and one-loop levels, corresponding respectively to 1PM order and partial information at 2PM order; we leave to future work a detailed comparison between those 1PM results and ours, but initial assessments indicate agreement.  Inherent in \cite{Guevara:2017csg}'s derivation of the scattering amplitudes is the ``double copy'' relation between gauge theory and gravity \cite{Kawai:1985xq,Bern:2008qj}, which has significant potential for advancing perturbative gravity calculations as discussed e.g.\ in \cite{Monteiro:2014cda,Monteiro:2015bna,Luna:2016hge,Goldberger:2016iau,Goldberger:2017frp}; see also \cite{Cheung:2016say} for an intriguing new perspective on double-copy-type properties of gravity amplitudes.  We look forward to further developments along these lines, with improved understanding of the relationships between these and other differing approaches to gravitational dynamics and related topics \cite{Poisson:Pound:Vega:2011,Harte:review,Barack:2014,Pound:review,FNSV,Harte:2016vwo,Harte:2017gpd}, and to explorations of the possibility of higher-order analogs of the effective-one-body equivalences noted in this paper.

Near the completion of this work, a complementary treatment of scattering of spinning bodies in the 1PM approximation, working to linear order in the bodies' spins, was given in \cite{Bini:2017xzy}.  The linear-in-spin parts of our all-orders-in-spin BH results
are in complete agreement with the findings of \cite{Bini:2017xzy}.  We reproduce their effective gyrogravitomagnetic ratios in (\ref{gyros}) below, as special cases of analogous coefficients (\ref{allgyros}) at all orders in spin.  (Note that, to linear order in spin, at the pole-dipole level, the dynamics is universal, independent of the structure and composition of the bodies; any body is equivalent to a BH at the pole-dipole level.)

A brief summary of the body of this paper and some further details of our results are as follows:

Section \ref{sec:scat} presents our derivation of the net $O(G)$ changes in the linear momenta and spins of two arbitrary-mass spinning BHs [system (i)] which weakly gravitationally scatter, given as functions of quantities describing the incoming state at past infinity---namely, momenta, spins, and a vectorial impact parameter.

In Section \ref{sec:aligned}, we briefly consider the special case of \emph{aligned spins}, where the BHs' spins are aligned with the system's orbital angular momentum.  In this case, the motion is confined to a plane, and the full information of the net scattering process is encapsulated in a small $O(G)$ scattering angle---for system (i), the angle by which both BHs are deflected in the system's center-of-mass frame, and for systems (ii) and (iii), the angle by which the test body is deflected in the background frame.  As a direct generalization of the point-mass results in \cite{Damour:2016s}, we demonstrate the equivalences amongst all three systems via mappings directly between the aligned-spin scattering angles (as functions of energies and angular momenta in the respective frames).

For the generic case of misaligned spins, we do not directly discuss mappings directly between scattering results (which are no longer encapsulated by a single scattering angle).  Rather, in Section \ref{sec:canhams}, we consider the original setting for EOB mappings \cite{Buonanno99}, namely, canonical Hamiltonians for (unbound or bound) binary dynamics.  The scattering results uniquely encode the gauge-invariant information of a canonical Hamiltonian for a generic binary BH at 1PM order, allowing us to deduce one such Hamiltonian (in a certain gauge), along with its test-body limits.  We find equivalences amongst systems (i)--(iii), for generic spin orientations, via simple EOB mappings between their Hamiltonians.  These involve the EOB energy map and new energy-dependent spin maps, as well as shifts of the relative position vectors related to the frame-dependent definitions of the bodies' center-of-mass worldlines.
Our final 1PM spinning-binary-BH Hamiltonian, given by (\ref{HBBH}) below, when expanded in $1/c$ and in powers of the spins, is equivalent via canonical transformations to the 1PM parts of all previous PN results for binary-BH Hamiltonians (or equivalent descriptions of the dynamics)---e.g.\ from \cite{Damour:2014jta, Damour:2015isa, Bernard:2015njp, Damour:2016abl,Hartung:2011te,Marchand:2017pir,Hartung:Steinhoff:Schafer:2012,Marsat:2012fn,Bohe:2012mr,Levi:Steinhoff:2015:2,Levi:Steinhoff:2015:3,Hergt:2007ha, Hergt:2008jn, Levi:Steinhoff:2014:2,Vaidya:2014kza, Marsat:2014xea,Vines:2016qwa,Blanchet:2013haa,Porto:2016pyg}, as detailed below following (\ref{aKLO})---and fully agrees with the recent 1PM linear-in-spin results of \cite{Bini:2017xzy}.

We conclude in Section \ref{sec:conclude}.

Appendix \ref{app:Kerr} reviews relevant properties of the Schwarzschild and Kerr spacetimes and their harmonic-gauge linearizations.
Appendix \ref{app:gencov} discusses a generally covariant formulation of the dynamics of spinning extended test bodies.
Appendix \ref{app:Sred} discusses properties of the reduced action for the two-BH system at 1PM order.

\section{First-post-Minkowskian scattering of spinning black holes}\label{sec:scat}

\subsection{On scattering at 1PM order}\label{sec:on1PMscat}

The equivalence at 1PM order between arbitrary-mass-ratio point-mass scattering and Schwarzschild geodesics, established in \cite{Damour:2016s}, could be considered somewhat unsurprising---and the same can be said of the equivalence between systems (i) and (ii) from Sec.~\ref{sec:intro}, with spins---given the nature of the 1PM scattering approximation, which can be effectively summarized as follows \cite{Portilla:1979,Portilla:1980,Westpfahl1979,Westpfahl:1985,Ledvinka:2008,Damour:2016s,Bini:2017xzy}.  To compute the 1PM deflection of a first point-mass, body 1, deviating only slightly from inertial (straight-line) motion in a background Minkowski spacetime, due to its gravitational interaction with a second point-mass, body 2,
\begin{enumerate}
\item [(a)] body 1 can be taken to follow a geodesic in the linearized field sourced by body 2---because corrections to this, from the influence of body 1's own field, are ${O}(G^2)$---, and
\item [(b)] the field sourced by body 2 can be computed using its zeroth-order (inertial) motion---as the corrections to its field due to body 2's own deflection are also ${O}(G^2)$.
\end{enumerate}
Thus, to linear order in $G$ (assuming $Gm/rc^2\ll 1$ even at closest approach), body 1 responds precisely as would a test particle in the stationary linearized (Schwarzschild) field sourced by an undeflected body 2.  The same logic applies with $1\leftrightarrow 2$.  (While this may make it seem unsurprising, the 1PM point-mass EOB equivalence established in \cite{Damour:2016s} is separate from and goes beyond these simple observations, as discussed in Sec.~\ref{sec:aligned} below.)

Here we generalize the analysis of 1PM scattering to the case where the point-masses are replaced by spinning BHs---or, insofar as can be discerned within the context of 1PM scattering, by bodies whose linearized gravitational fields match the linearized Kerr solution, to all orders in the multipole expansion.
The reasoning of (a) and (b) above extends almost verbatim to this case, with body 2's field now being the linearized Kerr field, and with the ``geodesic'' in (a) replaced by a ``spinning-test-BH trajectory.''  By this we mean a solution of the equations governing the dynamics of a spinning extended test body in curved spacetime, often framed as the MPD equations
\cite{Mathisson:1937,Mathisson:2010,Papapetrou:1951pa,Dixon:1979,Dixon:2015vxa}, with the multipole moments entering the equations being those appropriate for a spinning BH.

We proceed with a discussion of how such test-BH equations of motion are fully determined at 1PM order by the structure of linearized general relativity, enforcing effective stress-energy conservation, and matching to the Kerr solution.  We set $c=1$ henceforth.

\subsection{Post-Minkowskian action and field equations}

On a background Minkowski spacetime with metric $\eta_{\mu\nu}$, signature $(-,+,+,+)$, and flat connection $\doe_\mu$, we expand the perturbed metric as
\be
g_{\mu\nu}=\eta_{\mu\nu}+h_{\mu\nu}+O(G^2),
\ee
where $h\sim O(G)$ is the linear metric perturbation.  We treat $g_{\mu\nu}(x)$, $h_{\mu\nu}(x)$, etc.\ as tensors fields on the flat background; all index-raising, -lowering, squaring of vectors, etc.\ is done with $\eta$ unless otherwise noted.  The perturbation $h$ is subject to the linear gauge freedom
\be
h_{\mu\nu}\to h_{\mu\nu}+2\doe_{(\mu}\xi_{\nu)},
\ee
for any vector field $\xi^\mu(x)$.  It can be convenient (and it is always possible) to use this freedom to make $h$ satisfy the harmonic/Lorenz gauge condition,
\be\label{mcPgauge}
\mc P^{\mu\nu\alpha\beta}\doe_\nu h_{\alpha\beta}=0,
\ee
$\mc P$ being the trace-reverser,
\begin{alignat}{3}
\mc P_{\mu\nu}{}^{\alpha\beta}=\delta_{(\mu}{}^{(\alpha}\delta_{\nu)}{}^{\beta)}-\frac{1}{2}\eta_{\mu\nu}\eta^{\alpha\beta}.
\end{alignat}

A total effective action for the gravitational field coupled to some other fields $\Psi$ (which could describe one more localized bodies) can be expanded at 1PM order as
\bse\label{mcS}
\begin{alignat}{3}\label{mcStot}
\mc S_\mr{tot}[\Psi,h]=\mc S_\mr{grav}[h]+\mc S_\mr{int}[\Psi,h]+\mc S_\mr{kin}[\Psi]+O(G^2).
\end{alignat}
The action for the linearized gravitational field, the graviton kinetic term, can be taken in its harmonic gauge-fixed form,
\be\label{mcSgrav}
\mc S_\mr{grav}[h]=\frac{-1}{64\pi G}\int d^4x\;\doe_\rho h_{\mu\nu}\;\mc P^{\mu\nu\alpha\beta}\;\doe^\rho h_{\alpha\beta},
\ee
and the linear interaction term is generically given by
\be\label{mcSint}
\mc S_\mr{int}[\Psi,h]=\frac{1}{2}\int d^4x\;h_{\mu\nu}\;T^{\mu\nu}[\Psi],
\ee
where 
\be
T^{\mu\nu}[\Psi]=2\frac{\delta\mc S_\mr{tot}[\Psi,h]}{\delta h_{\mu\nu}}(h=0)
\ee
 is the zeroth-order stress-energy tensor of the other fields $\Psi$.  Both $T^{\mu\nu}[\Psi]$ and the kinematic term
\be
\mc S_\mr{kin}[\Psi]=\mc S_\mr{tot}[\Psi,h=0]
\ee
\ese
depend here only on the other fields $\Psi$, not on $h$; they should take forms appropriate for flat spacetime.

Varying the action (\ref{mcS}) with respect to $h$ yields
the 
harmonic-gauge linearized field equation,
\be\label{mcPFE}
\Box h_{\mu\nu}=-16\pi G\,\mc P_{\mu\nu\alpha\beta}\,T^{\alpha\beta},
\ee
with $\Box=\doe_\mu\doe^\mu$, with $h$ satisfying (\ref{mcPgauge}) being equivalent to stress-energy conservation, $\doe_\mu T^{\mu\nu}=0$, on the Minkowski background.

\subsection{Localized bodies; effective worldline-skeleton sources; linear and angular momenta}

To describe a body of finite extent surrounded by vacuum, in the limit where its size is small compared to the curvature length scale, it is natural to take the effective degrees of freedom $\Psi$ describing the body to be an arbitrarily parametrized worldline $x=z(\tau)$, with tangent $\dot z^\mu={dz^\mu}/{d\tau}$, along with some other fields $\psi(\tau)$ defined along the worldline.  It is natural also to posit an effective \emph{skeleton} stress-energy tensor $T^{\mu\nu}$ with distributional support only on the worldline.
With no dependence on $h$ at zeroth order, a general form for such a $T^{\mu\nu}$ is
\bse
\be\label{Tpsi}
T^{\mu\nu}(x)=\int d\tau\; \hat{\mc T}^{\mu\nu}(\psi,\doe)\;\delta^4(x-z)+O(G),
\ee
where $\hat{\mc T}^{\mu\nu}(\psi,\doe)$ is a differential operator of the form
\be\label{Thatform}
{\mc T}^{\mu\nu}(\psi)+{\mc T}^{\mu\nu\alpha}(\psi)\,\doe_\alpha+{\mc T}^{\mu\nu\alpha\beta}(\psi)\,\doe_\alpha\doe_\beta+\ldots,
\ee
\ese
with $\mc T^{\mu\nu\cdots}(\psi)$, the moments of $T^{\mu\nu}$, depending only on the worldline fields $\psi(\tau)$.

The first two moments, the monopole $\mc T^{\mu\nu}$ and dipole $\mc T^{\mu\nu\alpha}$, are constrained by zeroth-order stress-energy conservation, $\doe_{\mu}T^{\mu\nu}=O(G)$, such that \cite{Tulczyjew:1959,Dixon:1979}
\begin{alignat}{6}
\hat{\mc T}^{\mu\nu}&=\;\,{\mc T}^{\mu\nu}&&+\;\;{\mc T}^{\mu\nu\alpha}\,\doe_\alpha&&+{\mc T}^{\mu\nu\alpha\beta}\,\doe_\alpha\doe_\beta+\ldots
\\\nnm
&=\dot z^{(\mu} p^{\nu)}&&+\dot z^{(\mu}S^{\nu)\alpha}\,\doe_\alpha&&+{\mc T}^{\mu\nu\alpha\beta}\,\doe_\alpha\doe_\beta+\ldots,
\end{alignat}
where the two worldline fields $p^{\mu}(\tau)$ and $S^{\mu\nu}(\tau)$---repesctively, the body's linear momentum vector and its intrinsic angular momentum tensor (ang.\ mom.\ about the worldline) or spin tensor---must satisfy
\begin{alignat}{3}\label{MP0}
\frac{dp^\mu}{d\tau}=O(G),\qquad \frac{dS^{\mu\nu}}{d\tau}=2p^{[\mu}\dot z^{\nu]}+O(G),
\end{alignat}
which are the 0PM versions of the MPD equations \cite{Mathisson:1937,Mathisson:2010,Papapetrou:1951pa,Dixon:1979,Dixon:2015vxa}.
Stress-energy conservation alone is insufficient to constrain the evolution of both the momenta $(p,S)$ and a worldline, and the evolution equations (\ref{MP0}) can indeed be considered along any worldline $z(\tau)$; (\ref{MP0}) are in fact the restriction to a worldline of the field equations
\begin{alignat}{3}
\doe_\nu p^\mu&=O(G),
\\\nnm
\doe_\alpha S^{\mu\nu}&=2 p^{[\mu}\delta^{\nu]}{}_\alpha+O(G),
\end{alignat}
for fields $p^\mu(x)$ and $S^{\mu\nu}(x)$ on spacetime.  
In flat spacetime, with $G\to0$, these imply the constancy of $p^\mu$ and the transformation law for the special-relativistic angular momentum $S^{\mu\nu}(x)$ of an isolated system about some point $x$,
\be\label{Stransf}
S^{\mu\nu}(x)=S^{\mu\nu}(x')+2p^{[\mu}(x-x')^{\nu]}
\ee
[$+O(G)$].

For the effective description of a localized body, evolution equations for the worldline $z$ together with the momenta $(p,S)$ along $z$ can be determined by a further algebraic constraint of the form
\be\label{SSCv}
S^{\mu\nu}v_\nu=0,
\ee
known as a ``spin supplementary condition''  (SSC) \cite{Barker:1975ae,Barker1979,Kyrian:2007zz}, where $v^\mu$ is some timelike vector field depending e.g.\ on $p$, $\dot z$, other worldline fields, or background fields evaluated along the worldline.  The SSC requires the vanishing of the body's mass dipole vector, $\propto S^{\mu\nu}v_\nu$, in the frame of $v^\mu$.  This is generally sufficient to determine evolution equations for $p$, $S$ and $z$---but which may still depend on other degrees of freedom $\psi$ along the worldline, via the dependence of the higher-order stress-energy moments $\mc T^{\mu\nu\alpha\beta\cdots}$ on such $\psi$.

\subsection{No-hair (black-hole-like) spinning bodies; covariant dynamics}

Our interest here is in the case where the body's stress-energy depends on no further degrees of freedom beyond its worldline $z(\tau)$ and its Poincar\'e charges $p_\mu(\tau)$ and $S_{\mu\nu}(\tau)$, and all the higher moments are given as functions $\mc T^{\mu\nu\alpha\beta\ldots}(\dot z,p,S)$.  We could call these \emph{no-hair} bodies, or BH-like bodies, at least at 1PM order, as any deviations from these properties for BHs, due e.g.\ to tidal effects or quasinormal mode excitation, are present only at $O(G^2)$ or beyond if at all; see e.g.\ \cite{ Taylor:2008xy,Damour:2009vw,Kol:2011vg,Pani:2015hfa}.
The dynamics of such a body is fully determined by effective stress energy conservation and the constitutive relations giving the higher moments $\mc T^{\mu\nu\alpha\beta\ldots}(\dot z,p,S)$.  

In general, one can and should formulate an effective description for such a body which is at least defined covariantly (and possibly valid to some order) in a general, possibly strong-curvature spacetime.  We discuss this, following e.g.\ \cite{Porto:2005ac,Steinhoff:2014,Marsat:2014xea,Levi:Steinhoff:2015:1,Porto:2016pyg}, in Appendix \ref{app:gencov}, seeing an effective action formulation in which the evolution equations for $(p,S)$, of the form of the MPD equations, obtained by varying the total covariant action with respect to the worldline fields, are equivalent to the curved-spacetime covariant conservation of the corresponding effective stress-energy tensor, obtained by varying with respect to the metric.  %The symmetry-allowed couplings relevant at linear order in curvature are restricted to ones whose coefficients can be unambiguously identified by matching to the Kerr solution, as in \cite{Levi:Steinhoff:2015:1}, or as in Sec.~\ref{sec:LinKerr} below for the sub-case of 1PM order; this does not seem to be the case at 2PM order, or more generally with terms nonlinear in the curvature.

For the bottom line at 1PM order, we can return to the perturbative reasoning of the 1PM effective action (\ref{mcS}).  There, the evolution equations at $O(G)$ for the ``other'' fields $\Psi=(z,p,S)$ can be determined by varying the total action (\ref{mcStot}) with respect to those fields (and auxiliary fields $\Lambda$; see Appendix \ref{app:gencov})---where we can take $\mc S_\mr{kin}[z,p,S,\Lambda]$ and $T^{\mu\nu}[z,p,S]$ to have their flat-spacetime forms, independent of the metric perturbation.

One natural way to specify the kinematics of a no-hair body is to do so in terms of the worldline fields defined by the \emph{covariant SSC},
\be\label{CovSSC}
S^{\mu\nu}p_\nu=0,
\ee
also known as the Tulczjew-Dixon SSC \cite{Tulczyjew:1959,Dixon:1979}.  Given (\ref{CovSSC}), the components of $S^{\mu\nu}$ are determined by $p^\mu$ and the (mass-rescaled) covariant spin vector $a^\mu$,
\be\label{acovS}
a^\mu=\frac{1}{2p^2}\epsilon^\mu{}_{\nu\alpha\beta}p^\nu S^{\alpha\beta}\quad\Leftrightarrow
\quad
S^{\mu\nu}=\epsilon^{\mu\nu}{}_{\alpha\beta} p^\alpha a^\beta,
\ee
with $a\cdot p=a^\mu p_\mu=0$, where $\epsilon_{\mu\nu\alpha\beta}$ is the volume form (with $\epsilon_{0123}=1$ on a Minkowski basis).  We define also the mass $m$ and timelike unit vector $u^\mu$ from
\be
p^\mu=mu^\mu, \qquad p^2=-m^2,\qquad u^2=-1.
\ee
The zeroth-order stress-energy (\ref{Tpsi}) can be taken to depend only on $p^\mu$ and $a^\mu$ along $z(\tau)$,
\begin{alignat}{3}\label{Tpad}
T^{\mu\nu}(x)&=\int d\tau\;\hat{\mc T}^{\mu\nu}(p,a,\doe)\;\delta^4(x-z),
\end{alignat}
for some differential operator $\hat{\mc T}^{\mu\nu}$ of the form (\ref{Thatform}) with $\psi=(p,a)$.  The interaction term (\ref{mcSint}) then reads
\bse\label{SLint}
\begin{alignat}{3}\label{SintL}
\mc S_\mr{int}&=\frac{1}{2}\int d^4x\;h_{\mu\nu}\;T^{\mu\nu}=\int d\tau\; \mc L_\mr{int}(z,p,a)[h],
\\\label{Lint}
\mc L_\mr{int}&=\frac{1}{2}\,\hat{\mc T}^{\mu\nu}(p,a,-\doe)\;h_{\mu\nu}\,(x=z),
\end{alignat}
\ese
with the minus sign coming from integration by parts after inserting (\ref{Tpad}) into (\ref{SintL}).  As discussed in Appendix \ref{app:gencov}, given (\ref{CovSSC})--(\ref{Lint}), the appropriate kinematic term $\mc S_\mr{kin}[z,p,a,\Lambda]$ is such that $\delta(\mc S_\mr{int}+\mc S_\mr{kin})/\delta(z,p,a,\Lambda)=0$ implies the 1PM (MPD) evolution equations
\begin{alignat}{3}\label{MPD1PMcov}
\frac{dp_\mu}{d\tau}&=\frac{\doe}{\doe z^\mu}\mc L_\mr{int},
\\\nnm
m\frac{da^\mu}{d\tau}&=\left(-\epsilon^{\mu\nu}{}_{\alpha\beta}u^\alpha a^\beta\frac{\doe}{\doe a^\nu}+u^\mu a^\nu\frac{\doe}{\doe z^\nu}\right)\mc L_\mr{int},
\end{alignat}
$+O(G^2)$.  Differentiating the covariant SSC (\ref{CovSSC}) and using (\ref{MPD1PMcov}) with (\ref{acovS}) allows one to solve for the worldline tangent $\dot z$ in terms of the momentum $p$. It will be sufficient for our purposes here to note that one finds
\be\label{puzd}
\frac{p^\mu}{m}=u^\mu=\frac{\dot z^\mu}{\sqrt{-\dot z^2}}+O(G).
\ee
It will be convenient in the following to fix the parameter $\tau$ to be the (Minkowski) proper time along $z(\tau)$, with $\dot z^2=-1$, and thus $\dot z^\mu=u^\mu+O(G)$.

\subsection{Linearized fields of no-hair bodies; \\the linearized Kerr field}\label{sec:LinKerr}

The dynamics (\ref{MPD1PMcov}) from the Lagrangian (\ref{Lint}) is ultimately specified by the differential operator $\hat{\mc T}^{\mu\nu}(p,a,\doe)$.  In (\ref{Lint}), it acts on the metric perturbation (due to other sources) to determine the body's motion.  The same operator $\hat{\mc T}^{\mu\nu}$ also determines the field perturbation sourced by the body.

 A Green's function solution to the linearized field equation (\ref{mcPFE}) is given by
\be\label{hT}
h_{\mu\nu}(x)=4G\,\mc P_{\mu\nu\alpha\beta}\int d^4x'\; \mc G(x,x')\;T^{\alpha\beta}(x'),
\ee
where the scalar Green's function $\mc G(x,x')$ satisfies the sourced wave equation
\bse
\be
\Box\mc G(x,x')=-4\pi\delta^4(x-x'),
\ee
and has the symmetry property
\be\label{mcGex}
\doe_\mu\mc G(x,x')=-\doe'_{\mu}\mc G(x,x'),
\ee
\ese
where $\doe_\mu={\doe}/{\doe x^\mu}$ and $\doe'_{\mu}={\doe}/{\doe x'{}^\mu}$; Poincar\'e invariance in fact implies $\mc G(x,x')=\mc G(x-x')$.  The physical solution is found from the retarded Green's function,
\bse
\begin{alignat}{3}
\mc G_\mr{ret}(x,x')&=\theta_\mr{ret}(x,x')\,\mc G_\mr{sym}(x,x'),
\\
\theta_\mr{ret}(x,x')&=\left\{\begin{array}{l} 1,\quad\tr{$x$ in the future of $x'$,}\\0,\quad\tr{otherwise,}\end{array}\right.
\\
\mc G_\mr{sym}(x,x')&=\delta\left(\frac{(x-x')^2}{2}\right),
\end{alignat}
\ese
where the symmetric Green's function $\mc G_\mr{sym}$ is a delta function on the Minkowski lightcone $(x-x')^2=0$.

Substituting (\ref{Tpad}) into (\ref{hT}), using integration by parts and the symmetry (\ref{mcGex}), yields the metric perturbation sourced by a body with worldline $z(\tau)$ and momentum and spin $(p,a)$ along $z$, in the form
\begin{alignat}{3}\label{hmunuhatT}
h_{\mu\nu}(x)
%&=4G\int d\tau\int d^4x'\;\mc G(x,x')\;\hat{\mc T}^{\mu\nu}(\psi,\doe')\;\delta^4(x'-z)\nnm\\&=4G\int d\tau\int d^4x'\;\delta^4(x'-z)\;\hat{\mc T}^{\mu\nu}(\psi,-\doe')\;\mc G(x,x')\nnm\\
&=4G\,\mc P_{\mu\nu\alpha\beta}\int d\tau\;\hat{\mc T}^{\alpha\beta}(p,a,\doe)\;\mc G_\mr{ret}(x,z).
\end{alignat}

With the body alone in a flat background, satisfying $\doe_\mu T^{\mu\nu}=0$, the worldline $x=z(\tau)$ is a Minkowski geodesic with unit tangent $u^\mu$, the fields $p^\mu=mu^\mu$ and $a^\mu$ are covariantly constant, and (\ref{hmunuhatT}) simplifies to
\be\label{hfromr}
h_{\mu\nu}=4G\,\mc P_{\mu\nu\alpha\beta}\,\hat{\mc T}^{\alpha\beta}(p,a,\doe)\,\frac{1}{r},
\ee
where the scalar field
\begin{alignat}{3}\label{rzx}
\frac{1}{r}&=\frac{1}{r_{[z]}}(x)=\int d\tau\;\mc G\big(x,z(\tau)\big)
\\\nnm
&=\Big[(x-z)^2+\big(u\cd(x-z)\big)^2\Big]^{-1/2}
\\\nnm
&=\Big[\big({\perp}_u(x-z)\big)^2\Big]^{-1/2},\qquad ({\perp}_ux)^\mu=(\delta^\mu_\nu+u^\mu u_\nu)x^\nu,
\end{alignat}
---where $\mc G$ could be $\mc G_\mr{ret}$ or $\mc G_\mr{sym}$, and $z$ in the last two lines can be any point on the geodesic---is the inverse of the distance of the field point $x$ from the Minkowski geodesic $z(\tau)$ in its rest frame, satisfying
\be\label{rprops}
u^\mu\doe_\mu\frac{1}{r}=0,
\qquad
\Box\frac{1}{r}=\int d\tau\;\delta^4(x-z).
\ee

The crucial observation to be made here is that we can find a differential operator $\hat{\mc T}^{\mu\nu}$ appropriate for the effective description of a spinning BH at 1PM order by finding a harmonic-gauge linearized field $h_{\mu\nu}$ of the form (\ref{hfromr}) corresponding to the Kerr spacetime.

The appropriate form of the field is discussed in Appendix \ref{app:Kerr}, drawing from \cite{Harte:2016vwo}, where it was shown that the \emph{exact} Kerr metric $g^\mr{Kerr}_{\mu\nu}$ takes the form
\be
g^\mr{Kerr}_{\mu\nu}=\eta_{\mu\nu}+h^\mr{Kerr}_{\mu\nu}+2\doe_{(\mu}\xi_{\nu)},
\ee
where the vector $\xi^\nu$, effecting what is nothing more than a gauge transformation at linear order, is given by (\ref{xiK}), and where $h^\mr{Kerr}_{\mu\nu}$ is an exact solution of the linearized field equation (\ref{mcPFE}) and the harmonic gauge condition (\ref{mcPgauge}) whose trace-reversal $\bar h^\mr{Kerr}_{\mu\nu}=\mc P_{\mu\nu}{}^{\alpha\beta}h^\mr{Kerr}_{\alpha\beta}$ is given by
(\ref{hharmK}) or (\ref{hKfin}) as
\begin{alignat}{3}
&\bar h^\mr{Kerr}_{\mu\nu}=u_\mu u_\nu\left(1-\frac{1}{2!}(a\cdot\doe)^2+\frac{1}{4!}(a\cdot\doe)^4-\ldots\right)\frac{4Gm}{r}
\nnm\\\label{hbarKerr}
&+u_{(\mu}\epsilon_{\nu)\rho\alpha\beta} u^\rho a^\alpha\doe^\beta\left(1-\frac{1}{3!}(a\cdot\doe)^2+\ldots\right)\frac{4Gm}{r}.
\end{alignat}
The BH has momentum $p^\mu=m u^\mu$ (along the direction of the spacetime's timelike Killing vector) and spin tensor $S^{\mu\nu}=\epsilon^{\mu\nu}{}_{\alpha\beta}p^\alpha a^\beta$ about its worldline $r=0$. 
 We see in the first line the contributions from the BH's mass multipole moments $\mc I_\ell\sim ma^{\ell\tr{ even}}$ (monopole${}\sim a^0$, quadrupole${}\sim a^2$, hexadecapole${}\sim a^4$ shown), with even powers of the spin vector $a^\mu$ contracted into derivatives of $1/r$, and in the second line contributions from the momentum/current multipole moments $\mc J_\ell\sim ma^{\ell\tr{ odd}}$ (dipole${}\sim a^1$, octupole${}\sim a^3$ shown), with odd powers of spin and odd derivatives.

Another useful way to express (\ref{hbarKerr}) is as
\begin{alignat}{3}\label{hexp}
h^\mr{Kerr}_{\mu\nu}&\,=\,4Gm\; \mc P_{\mu\nu\alpha\beta}\;\exp(a\st\doe)^\alpha{}_\gamma \;\frac{u^\gamma u^\beta}{r},
\end{alignat}
giving via (\ref{hfromr}) the operator
\bse
\be\label{hatTKerr}
\hat{\mc T}^{\mu\nu}(p,a,\doe)=m\, \exp(a\st\doe)^{(\mu}{}_\rho\;u^{\nu)} u^\rho,
\ee
where
\begin{alignat}{3}
\exp(a\st\doe)^\mu{}_\nu&=\delta^\mu{}_\nu+(a\st\doe)^\mu{}_\nu
\\\nnm
&\quad\;\;\;+\frac{1}{2!}(a\st\doe)^\mu{}_\rho(a\st\doe)^\rho{}_\nu+\ldots
\end{alignat}
is the exponential of the generator
\begin{alignat}{3}\label{astdoe}
(a\st\doe)^\mu{}_\nu&\equiv(*(a\wedge\doe))^\mu{}_\nu\phantom{\Big|}
\\\nnm
&=\epsilon^\mu{}_{\nu\alpha\beta}a^\alpha\doe^\beta,
\end{alignat}
satisfying
\be
(a\st\doe)^\mu{}_\nu (a\st\doe)^\nu{}_\alpha \frac{u^\alpha}{r}=-(a\cd\doe)^2\frac{u^\mu}{r}
\ee
\ese
by virtue of (\ref{rprops}) and $\epsilon_{\mu\nu\rho\sigma}\epsilon^{\alpha\beta\gamma\delta}=-4!\,\delta_{[\mu}{}^\alpha\delta_\nu{}^\beta\delta_\rho{}^\gamma\delta_{\sigma]}{}^\delta$.

%\begin{alignat}{3}h_{\mu\nu}&=\Bigg[\mc P_{\mu\nu\alpha\beta}\dot z^\beta \left(1-\frac{1}{2!}(a\cd\doe)^2+\frac{1}{4!}(a\cd\doe)^4+\ldots\right)\nnm\\&\quad+\dot z_{(\mu}\epsilon_{\nu)\rho\sigma\alpha} a^\rho\doe^\sigma\left(1-\frac{1}{3!}(a\cd\doe)^2+\ldots\right)\Bigg]\frac{4G p^\alpha}{r}\end{alignat}

\subsection{Linear interaction of two spinning black holes}\label{sec:linint}
Having in hand the form (\ref{hatTKerr}) of the $\hat{\mc T}^{\mu\nu}$ operator for a spinning BH, we can return to (\ref{Lint}) and consider the interaction Lagrangian $\mc L_\mr{int}$ for a first BH ``1'' with worldline $z_1(\tau_1)$, momentum $p_1^\mu=m_1u_1^\mu$, and (mass-rescaled) spin vector $a_1^\mu$ in the field of a second BH ``2'' with worldline $z_2(\tau_2)$, momentum $p_2^\mu=m_2u_2^\mu$, and spin $a_2^\mu$---taking BH 2 to be in its zeroth-order state---%evaluated with both BHs in their zeroth-order states,
\begin{alignat}{3}\label{manL}
&\mc L_\mr{int}=\frac{1}{2}\,\hat{\mc T}^{\mu\nu}(p_1,a_1,-\doe)\,h_{2\mu\nu}(x)\,\Big|_{x=z_1}
\\\nnm
&=2G\,\hat{\mc T}^{\mu\nu}(p_1,a_1,-\doe)\,\mc P_{\mu\nu\alpha\beta}\,\hat{\mc T}^{\alpha\beta}(p_2,a_2,\doe)\,\frac{1}{r_2}\,\bigg|_{x=z_1}
\\\nnm
&=u_{1\mu}\, \exp(a_1\st\doe)^\mu{}_\nu\, \mc Q^{\nu}{}_{\alpha} \,\exp(a_2\st\doe)^\alpha{}_\beta\, u_2^\beta \,\frac{Gm_1m_2}{r_2}\,\bigg|_{x=z_1}
\\\nnm
&=\big\langle u_1\big|\exp(a_1\st\doe)\;\mc Q\;\exp(a_2\st\doe)\,\big|u_2\big\rangle\,\frac{Gm_1m_2}{r_2}\,\bigg|_{x=z_1},
\end{alignat}
where $r_2(x)$ is as in (\ref{rzx}) with the geodesic $z\to z_2$, and
%\begin{widetext}\begin{alignat}{3}\mc S_\mr{int}&=\frac{1}{2}\int d^4x\;h_{2\mu\nu}(x) T_1^{\mu\nu}(x)\\&=2G\int d\tau_1\int d\tau_2\int d^4x\;\Big[\hat{\mc T}^{\mu\nu}(u_1,a_1\st\doe)\;\delta^4(x-z_1)\Big]\,\mc P_{\mu\nu\alpha\beta}\,\Big[\hat{\mc T}^{\alpha\beta}(u_2,a_2\st\doe)\;\mc G(x,z_2)\Big]\\&=2G\int d\tau_1\int d\tau_2\;\Big[\hat{\mc T}^{\mu\nu}(u_1,-a_1\st\doe)\;\mc P_{\mu\nu\alpha\beta}\;\hat{\mc T}^{\alpha\beta}(u_2,a_2\st\doe)\;\mc G(x,z_2)\Big](x=z_1)\\&=2G\int d\tau_1\left[\hat{\mc T}^{\mu\nu}(u_1,-a_1\st\doe)\;\mc P_{\mu\nu\alpha\beta}\;\hat{\mc T}^{\alpha\beta}(u_2,a_2\st\doe)\;\frac{1}{r_2(x)}\right](x=z_1)\\&=2Gm_1m_2\int d\tau_1\left[\exp(-a_1\st\doe)^\mu{}_\rho\; u_1^\nu u_1^\rho\;\mc P_{\mu\nu\alpha\beta}\;\exp(a_2\st\doe)^\alpha{}_\gamma\; u_2^\beta u_2^\gamma\;\frac{1}{r_2(x)}\right](x=z_1)\\&=Gm_1m_2\int d\tau_1\left[u_{1\mu}\; \exp(a_1\st\doe)^\mu{}_\nu\; \mc Q^{\nu}{}_{\alpha} \;\exp(a_2\st\doe)^\alpha{}_\beta\; u_2^\beta \;\frac{1}{r_2(x)}\right](x=z_1)\end{alignat}\end{widetext}\be\exp(a_\mathsf{a}\st\doe)^\alpha{}_\beta \exp(a_\mathsf{b}\st\doe)^\beta{}_\gamma\ee$\mathsf{a}\to\mr{t},\; \mathsf{b}\to\mr{b}$
\be\label{mcQ}
\mc Q^{\mu}{}_{\alpha}=2\mc P^\mu{}_{\nu\alpha\beta} u_{1}^{\nu} u_{2}^{\beta}.
\ee
We have used here \mbox{$\exp(a\st\doe)^{\mu}{}_\nu=\exp(-a\st\doe)_\nu{}^\mu$}, and in the last line introduced an index-free ``bra-ket'' notation.  The bra-ket notation indicates only the contractions of the spacetime indices, with all derivatives $\doe_\mu=\doe/\doe x^\mu$ acting to the right on $1/r_2(x)$.

Let us define the relative Lorentz factor
\be\label{gammaw}
\gamma=-u_1\cdot u_2,
\ee
and the normalized bivector $w=(u_1\wedge u_2)/\sqrt{\gamma^2-1}$ spanned by $u_1$ and $u_2$,
\be\label{defw}
w^{\mu\nu}=\frac{2u^{[\mu}_1u^{\nu]}_2}{\sqrt{\gamma^2-1}},\qquad w_{\mu\nu}w^{\mu\nu}=-2,
\ee
so that (\ref{mcQ}) becomes
\be
\mc Q^{\mu\alpha}=-\gamma\, \eta^{\mu\alpha}-\sqrt{\gamma^2-1}\,w^{\mu\alpha}.
\ee
Defining also the vector
\begin{alignat}{3}\label{wsta}
(w\st a)^\mu&\equiv((*w)\cdot a)^\mu\phantom{\Big|}
\\\nnm
&=\frac{1}{2}\epsilon^{\mu\nu\alpha\beta}w_{\alpha\beta}a_\nu,
\\\nnm
&=\epsilon^{\mu}{}_{\nu\alpha\beta}\frac{u_1^\alpha u_2^\beta}{\sqrt{\gamma^2-1}}a^\nu,
\end{alignat}
for any $a^\mu$, we have [cf.\ (\ref{astdoe})]
\begin{alignat}{3}\label{wstacddoe}
w\st a\cd\doe&\;\equiv\; (w\st a)^\mu \doe_\mu \;=\;-\frac{1}{2}w_{\mu\nu}(a\st\doe)^{\mu\nu}\phantom{\Big|}
\\\nnm
&\;=\;\frac{1}{2}\epsilon^{\mu\nu\alpha\beta}w_{\alpha\beta}a_\nu\doe_\mu
\;=\;\frac{\epsilon_{\mu\nu\alpha\beta}u_2^\mu u_1^\nu a^\alpha\doe^\beta}{\sqrt{\gamma^2-1}}.
\end{alignat}

Importantly, in manipulating the Lagrangian (\ref{manL}), we can make use of the fact---following from $\dot z_\ms{a}=u_\ms{a}+O(G)$, $\dot p_\ms{a}=O(G)$, $\dot a_\ms{a}=O(G)$ as in (\ref{MPD1PMcov})--(\ref{puzd})---that
\be
u_1\cd\doe\, f(x,p_{\ms a},a_{\ms a})\Big|_{x=z_1}=\frac{d}{d\tau_1}f(z_1,p_{\ms a},a_{\ms a})+O(G)f,
\ee
(with $\ms{a}=1,2$) to take any terms in $\mc L_\mr{int}$ with a factor of $u_1\cdot\doe$ and drop the resultant total-derivative terms and $O(G^2)$ terms from the action $\mc S_\mr{int}=\int d\tau_1\,\mc L_\mr{int}$, at the expense of field redefinitions (which will be irrelevant for scattering).  We can also drop any terms with factors of $u_2\cdot\doe$ or $\Box$ acting on $1/r_2$ thanks to (\ref{rprops}).  Thus, in manipulating the Lagrangian (\ref{manL}), we can take 
\bse\label{opprops}
\be\label{opids}
u_1\cd\doe\;\dot{=}\; 0,\qquad u_2\cd\doe\;\dot{=}\; 0,\qquad \Box\;\dot{=}\;0,
\ee
as operator identities in the sense
\begin{alignat}{3}\label{doteq}
&A\;\dot{=}\;B
\\\nnm
&\quad\Leftrightarrow\quad \int d\tau_1(...A...)\frac{1}{r_2}\bigg|_{x=z_1}=\int d\tau_1(...B...)\frac{1}{r_2}\bigg|_{x=z_1},
\end{alignat}
\ese
neglecting $O(G^2)$ and total derivative terms.

In expanding the Lagrangian (\ref{manL}), one finds the basic contractions
\bse
\be\label{uQu}
\big\langle u_{1}\big| \mc Q\,\big| u_2\big\rangle\,=\,u_{1\mu}\mc Q^\mu{}_\alpha u_2^\alpha\,=\,2\gamma^2-1,
\ee
and
\begin{alignat}{3}
&\big\langle u_{1}\big| \mc Q\,(a_\ms{a}\st\doe)\,\big| u_2\big\rangle
\\\nnm
=\,&\big\langle u_{1}\big|(a_\ms{a}\st\doe)\, \mc Q\,\big| u_2\big\rangle&\;=\;2\gamma\sqrt{\gamma^2-1}\;w\st a_\ms{a}\cd\doe,
\end{alignat}
for $\ms{a}=1,2$.  Given (\ref{opids}) and $\epsilon$-identities, we find
\begin{alignat}{3}\label{uaQau}
&\big\langle u_{1}\big| \mc Q\,(a_\ms{a}\st\doe)\,(a_\ms{b}\st\doe)\,\big| u_2\big\rangle
\\\nnm
\dot{=}\;&\big\langle u_{1}\big|(a_\ms{a}\st\doe)\, \mc Q\,(a_\ms{b}\st\doe)\,\big| u_2\big\rangle
\\\nnm
\dot{=}\;&\big\langle u_{1}\big|(a_\ms{a}\st\doe)\,(a_\ms{b}\st\doe)\, \mc Q\,\big| u_2\big\rangle
\\\nnm
\dot{=}\;&(2\gamma^2-1)(w\st a_\ms{a}\cd\doe)(w\st a_\ms{b}\cd\doe)\phantom{\Big|}
\end{alignat}
with $\ms{a}=1,2$ and independently $\ms{b}=1,2$, noting also
\be
(w\st a_\ms{a}\cd\doe)(w\st a_\ms{b}\cd\doe)\;
\dot{=}\;-(a_\ms{a}\cd\doe)(a_\ms{b}\cd\doe),
\ee
\ese
and that the scalar operators \mbox{$a\cd\doe$} and \mbox{$w\st a\cd\doe$} all commute.

It is straightforward from (\ref{uQu})--(\ref{uaQau}) to verify \emph{the \mbox{remarkable} factorization} of (\ref{manL}),
\begin{alignat}{3}
\mc L_\mr{int}&=\big\langle u_1\big|\exp(a_1\st\doe)\;\mc Q\;\exp(a_2\st\doe)\,\big|u_2\big\rangle\,\frac{Gm_1m_2}{r_2}\bigg|_{x=z_1},
\nnm\\\label{Lintfactor}
&\;\dot{=}\; \big\langle u_1\big|\mc Q\;\exp\big((a_1+a_2)\st\doe\big)\,\big|u_2\big\rangle\,\frac{Gm_1m_2}{r_2}\bigg|_{x=z_1}.
\end{alignat}
The first line gives the interaction Lagrangian for a two-spinning-BH system with mass-rescaled spins $a_1^\mu$ and $a_2^\mu$, while the second line would be the interaction term for a nonspinning point-mass coupled to a BH with mass-rescaled spin $a_1^\mu+a_2^\nu$.  Here, similar to (\ref{doteq}), we use 
\be
A\;\dot{=}\;B\quad\Leftrightarrow\quad\int d\tau_1\, A=\int d\tau_1 \,B,
\ee
neglecting $O(G^2)$ and total derivative terms.

Continuing to simplify (\ref{Lintfactor}) with (\ref{gammaw})--(\ref{uaQau}), defining
\be\label{a0}
a_0^\mu=a_1^\mu+a_2^\mu,
\ee
with \mbox{$(w\st a_0)^\beta=\epsilon_{\mu\nu\alpha}{}^\beta u_2^\mu u_1^\nu a_0^\alpha/{\sqrt{\gamma^2-1}}$} as in (\ref{wsta}), we find
\begin{widetext}
\begin{alignat}{3}\label{Lint12}
\mc L_\mr{int}&\;\dot{=}\; Gm_1m_2\bigg[(2\gamma^2-1)\cosh(w\st a_0\cd\doe)
+2\gamma\sqrt{\gamma^2-1}\,\sinh(w\st a_0\cd\doe) \bigg]\frac{1}{r_2}\bigg|_{x=z_1}
=\frac{Gm_1m_2}{2}\sum_{s=\pm 1}
\frac{e^{2s\beta}}{r_2(z_1+sw\st a_0)}
\nnm\\\nnm
&\;\dot{=}\; Gm_1m_2\bigg[(2\gamma^2-1)\cos(a_0\cd\doe)
+2\gamma\epsilon_{\mu\nu\alpha\beta}u_2^\mu u_1^\nu a_0^\alpha \doe^\beta\frac{\sin(a_0\cd\doe)}{a_0\cd\doe}\bigg]\frac{1}{r_2}\bigg|_{x=z_1}
\\
&\;\dot{=}\; Gm_1m_2\;\Re\,\bigg[(2\gamma^2-1)
-i
\,2\gamma\epsilon_{\mu\nu\alpha\beta}u_2^\mu u_1^\nu \frac{a_0^\alpha \doe^\beta}{a_0\cd\doe}\bigg]\exp(ia_0\cd\doe)\frac{1}{r_2}\bigg|_{x=z_1},
\end{alignat}
\end{widetext}
if the reader will allow the notation in the last two lines, noting that there is no actual need to invert \mbox{$a_0\cd\doe$}.
In the second equality of the first line, the infinite series of differential operators has been resummed by recognizing translation operators,
$\exp(c\cd\doe)f(x)=f(x+c)$,
for $x$-independent vectors $c^\mu$, and we have used the definition of the ``rapidity'' $\beta$,\begin{alignat}{3}\cosh\beta&=\gamma=-u_1\cd u_2,&\sinh\beta&=\sqrt{\gamma^2-1},\nnm\\\cosh{2\beta}&=2\gamma^2-1,&\sinh{2\beta}&=2\gamma\sqrt{\gamma^2-1},\nnm\\e^{\pm2\beta}&=\big(\gamma\pm\sqrt{\gamma^2-1}\,\big)^2.\quad\end{alignat}

As discussed in Sec.~\ref{sec:on1PMscat}, we can find the 1PM-accurate evolution of BH 1 by using its 1PM test-BH MPD equations in the stationary field sourced by BH 2 in its zeroth order state.  These are given by (\ref{MPD1PMcov})--(\ref{puzd}) with $(z,p,a)(\tau)\to(z_1,p_1,a_1)(\tau_1)$ and with $\mc L_\mr{int}$ precisely as in (\ref{Lint12}).  One can verify that, at linear order in $G$, these are the same equations of motion that would be obtained from varying the total reduced action for the two-BH system (after having ``integrated out'' the field $h$) with respect to $(z_1,p_1,a_1)$.  This is a well-known circumstance for reduced actions at linear order---see e.g.\ discussion in \cite{Bini:2012gu}---with cancelling contributions from the field action $\mc S_\mr{grav}[h]$ and symmetric pairs of interaction terms $\mc S_\mr{int}[\Psi,h]$; this is discussed further in Appendix \ref{app:Sred}.

\subsection{Net scattering deflections}\label{sec:pmscat}

To determine the net $O(G)$ changes in the momentum $p_1^\mu$ and spin $a_1^\mu$ of BH 1, from past infinity to future infinity, due to its scattering with BH 2, we can take the equations of motion as discussed in the last paragraph of the previous section and integrate them over the entire history of BH 1's zeroth-order state, in which $z_1(\tau_1)$ is a Minkowski geodesic and $p_1$ and $a_1$ are constant---noting that corrections to the $O(G)$ ``force'' and ``torque'' of (\ref{MPD1PMcov}) due to the $O(G)$ variation of BH 1 from its zeroth-order state are $O(G^2)$.  We obtain
\begin{alignat}{3}\label{Holo1}
\Delta p_{1\mu} &=\int d\tau_1\,\frac{\doe\mc L_\mr{int}}{\doe z_1^\mu},
\\\nnm
m_1\Delta a_1^\mu&=\int d\tau_1\left(-\epsilon^{\mu\nu}{}_{\alpha\beta}u_1^\alpha a_1^\beta\frac{\doe\mc L_\mr{int}}{\doe a_1^\nu}+u_1^\mu a_1^\nu\frac{\doe\mc L_\mr{int}}{\doe z_1^\nu}\right),
\end{alignat}
with $\mc L_\mr{int}$ given by (\ref{Lint12}), integrating over the entire worldline $z_1$.  These are finally functionals only of two zeroth-order BH states $(z_\ms{a},p_\ms{a},a_{\ms{a}})_{\ms{a}=1,2}$, which can be identified with the incoming states for the scattering process, with the worldlines $z_\ms{a}$ being Minkowski geodesics and the momenta and spins $p_\ms{a}$ and $a_\ms{a}$ being constant vectors.

We can parametrize the two geodesics as
\begin{subequations}\label{imp}
\begin{align}\label{z10}
x^\mu&=z_1^\mu(\tau_1)=z_{10}^\mu+u_1^\mu \tau_1,
\\\label{z20}
x^\mu&=z_2^\mu(\tau_2)=z_{20}^\mu+u_2^\mu \tau_2,
\end{align}
while enforcing
\be\label{bperp}
b\cdot u_1=b\cdot u_2=0,
\ee
where
\be\label{bmu0}
b^\mu=z_{10}^\mu-z_{20}^\mu,
\ee
\end{subequations}
which uniquely define $z_{10}$ and $z_{20}$ as the points of mutual closest approach of the two worldlines, with the vectorial ``impact parameter'' $b^\mu$,
the spacelike separation vector at closest approach, being orthogonal to both worldlines.  See Fig.~\ref{fig1} below.

Upon inserting (\ref{Lint12}) into (\ref{Holo1}), we have integrals of the following form, with constant vectors $c^\mu$, which are straightforwardly evaluated using (\ref{rzx}) and (\ref{imp}),
\begin{alignat}{3}\label{theint}
&\int d\tau_1\,\doe_\mu\frac{1}{r_2(x+c)}\bigg|_{x=z_1}
\\\nnm
&=\int d\tau_1\,\doe_\mu\Big[\big({\perp}_{u_2}(x-z_2+c)\big)^2\Big]^{-1/2}\bigg|_{x=z_1}
\\\nnm
&=\frac{-2}{\sqrt{\gamma^2-1}}\frac{b_\mu+\Pi_{\mu\nu} c^\nu}{(b+\Pi c)^2},
\end{alignat}
where
\be
\Pi^{\mu}{}_{\nu}=\epsilon^{\mu\rho}{}_{\alpha\beta}\epsilon_{\nu\rho\gamma\delta}\frac{u_1^\alpha u_2^\beta u_1^\gamma u_2^\delta}{\gamma^2-1}
\ee
is the projector into the plane orthogonal to both $u_1^\mu$ and $u_2^\mu$.  From (\ref{bperp}), the impact parameter $b^\mu$ satisfies $b=\Pi b$.  Also, $\Pi(w\st a_0)=w\st a_0$, while $\Pi a_0\ne a_0$ in general, but note that
\be
(w\st a_0)^2=(\Pi a_0)^2, \qquad (\Pi a_0)^\mu=\Pi^\mu{}_\nu a_0^\nu.
\ee
A final useful identity follows from the fact that an antisymmetrization over five or more indices in four dimensions vanishes,
\begin{align}\label{A5}
0&=5\doe_{[\mu}\epsilon_{\alpha\beta\gamma\delta]}u_1^\alpha u_2^\beta a_\ms{a}^\gamma \doe^\delta
\\\nnm
&\;\dot{=}\;\doe_{\mu}\epsilon_{\alpha\beta\gamma\delta}u_1^\alpha u_2^\beta a_\ms{a}^\gamma \doe^\delta-\epsilon_{\mu\alpha\beta\gamma}u_1^\alpha u_2^\beta \doe^\gamma\, a_\ms{a}\cd \doe,
\end{align}
for $\ms{a}=0,1,2$, where $\dot{=}$ holds in the sense of (\ref{opprops}).

Combining (\ref{Lint12}), (\ref{Holo1}) and (\ref{theint})--(\ref{A5}), one finds after some manipulation that the results of (\ref{Holo1}) can be expressed as
\begin{subequations}\label{D1Z}
\begin{alignat}{3}\label{DpDa1}
\Delta p_{1}^{\mu}&=Gm_1m_2\,\Re Z^\mu,\phantom{\bigg|}
\\\nnm
\Delta a_1^\mu&=Gm_2\Big(
\epsilon^{\mu\nu}{}_{\alpha\beta}u_1^\alpha a_1^\beta\,\Im Z_\nu+u_1^\mu a_1^\nu\,\Re Z_\nu \Big),
\end{alignat}
given in terms of the real and imaginary parts of the complex vector
\begin{alignat}{3}\label{Zintro}
Z_\mu&=\sum_{s=\pm1}\frac{-e^{2s\beta}}{\sqrt{\gamma^2-1}}\Big[\eta_{\mu\nu}+is(*w)_{\mu\nu}\Big]\frac{(b+sw\st a_0)^\nu}{(b+sw\st a_0)^2}
\nnm\\
&=\frac{-2\Big[(2\gamma^2-1)\eta_{\mu\nu} +2i\gamma\epsilon_{\mu\nu}{}_{\alpha\beta} u_1^\alpha u_2^\beta\Big](b+i\Pi a_0)^\nu}{\sqrt{\gamma^2-1}\,(b+i\Pi a_0)^2},
\end{alignat}
\end{subequations}
which depends only on the two velocities $u_1$ and $u_2$, the impact parameter $b$, and the projection $\Pi a_0$ orthogonal to both $u_1$ and $u_2$ of the sum $a_0=a_1+a_2$ of the mass-rescaled covariant spin vectors.

With the spins set to zero, (\ref{D1Z}) becomes
\be\label{Dp1}
\Delta p_{1\mu}=-2Gm_1m_2\,\frac{2\gamma^2-1}{\sqrt{\gamma^2-1}}\,\frac{b_\mu}{b^2},
\ee
matching the 1PM point-mass scattering deflections found in \cite{Portilla:1979,Portilla:1980,Westpfahl1979,Westpfahl:1985,Ledvinka:2008,Damour:2016s}.

The results (\ref{DpDa1}) can also be expressed as $p_1^\mu$ and $a_1^\mu$ undergoing the same linearized Lorentz transformation $\delta^\mu{}_\nu+\theta_1^\mu{}_\nu$,
\be
\Delta p_1^\mu=\theta_1^\mu{}_\nu p_1^\nu,
\qquad
\Delta a_1^\mu=\theta_1^\mu{}_\nu a_1^\nu,
\ee
with
\be\label{theta}
\theta^{\mu\nu}_1=Gm_2\,\Re\,\Big(2u_1^{[\mu}Z_{\phantom{1}}^{\nu]}+i\epsilon_{\phantom{1}}^{\mu\nu}{}_{\alpha\beta}u_1^\alpha Z^\beta\Big).
\ee

The results for $\Delta p_2^\mu$ and $\Delta a_2^\mu$ due to the same scattering encounter are found by a simple exchange of the BHs' identities, $1\leftrightarrow 2$, entailing $b^\mu\leftrightarrow-b^\mu$ and $Z^\mu\leftrightarrow-(Z^\mu)^*$, yielding
\begin{alignat}{3}\label{DpDa2}
\Delta p_{2}^{\mu}&=-Gm_1m_2\,\Re Z^\mu,\phantom{\bigg|}
\\\nnm
\Delta a_2^\mu&=Gm_1\Big(
\epsilon^{\mu\nu}{}_{\alpha\beta}u_2^\alpha a_2^\beta\,\Im Z_\nu-u_2^\mu a_2^\nu\,\Re Z_\nu \Big).
\end{alignat}
The net conservation [to $O(G)$] of the two-body system's total linear momentum $p_1^\mu+p_2^\mu$ is manifest in the results (\ref{DpDa1}) and (\ref{DpDa2}).  The conservation of angular momentum is less manifest, as the changes in spins are balanced by the change in an appropriate (covariant-SSC, center-of-mass-frame) orbital angular momentum vector $L^\mu$, as defined in (\ref{Lvec}) below.  While the calculation has not directly provided $\Delta L^\mu$, it can be deduced from the conservation of the total angular momentum vector $J^\mu$ defined in (\ref{Jvec}).

One can verify directly from (\ref{DpDa1})--(\ref{Zintro}), noting 
\be\label{Zu0}
Z\cdot u_1=Z\cdot u_2=0,
\ee
that the rest mass $m_1^2=-p_1^2$ is conserved because $p_1\cdot\Delta p_1=0$, the spin magnitude $a_1^2$ is conserved because $a_1\cdot\Delta a_1=0$, and the orthogonality condition $p_1\cdot a_1=0$ is preserved because $\Delta (p_1\cdot a_1)=p_1\cdot\Delta a_1+\Delta p_1\cdot a_1=0$, and similarly for body 2.  This all holds to linear order in $G$, using $p^\mu=m u^\mu$ for the zeroth-order states, and noting $\Delta p^\mu=m\Delta u^\mu$ since $\Delta m=0$, for each body.

\subsection{Two-body and test-body cases; kinematics and reference frames}\label{sec:frames}

The 1PM scattering results of the previous section are applicable both to arbitrary-mass-ratio two-BH scattering and to the scattering of a test BH in a stationary BH background; in the latter case, we simply ignore $\Delta p^\mu$ and $\Delta a^\mu$ for one of the bodies.  While the results (\ref{D1Z}) and (\ref{DpDa2}) provide fully specially covariant descriptions of the bodies' asymptotic ingoing and outgoing states for both situations, it will be useful to specialize our descriptions of the processes to appropriate frames of reference---the conserved center-of-mass (cm) frame for the two-body case, and the rest frame of the background body for the test-body case---and to express the results in terms quantities defined with respect to those frames.  Here we define and collect relationships between energies and linear and angular momenta in the respective frames, for the zeroth-order scattering states, for which we can reason as in special relativity.

\subsubsection{Two-body case in the center-of-mass frame}\label{sec:twobody}

The two-body system's total momentum $p_\mr{tot}^\mu$, the velocity $u_\mr{cm}^\mu$ of the cm frame with $u_\mr{cm}^2=-1$, and the system's total energy ${E}$ in the cm frame are defined by
\be\label{ptot}
p_\mr{tot}^\mu=p_1^\mu+p_2^\mu={E}u_\mr{cm}^\mu,  \qquad {E}^2=-p_\mr{tot}^2.
\ee
The individual energies in the cm frame are then given by 
$E_{1,2}=-u_\mr{cm}\cdot p_{1,2}$, and the individual momenta can be split into parts along and orthogonal to $u_\mr{cm}^\mu$ according to
\begin{align}\label{p1split}
p_1^\mu=m_1 u_1^\mu&=E_1 u_\mr{cm}^\mu+p_{\perp}^\mu,
\\\nnm
p_2^\mu=m_2u_2^\mu&=E_2 u_\mr{cm}^\mu -p_{\perp}^\mu,\qquad p_{\perp}\cdot u_\mr{cm}=0,
\end{align}
with the \emph{relative momentum} $p_{\perp}^\mu$ orthogonal to $u_\mr{cm}^\mu$.  Useful relations following from (\ref{ptot})--(\ref{p1split}) and (\ref{gammaw}) are
\begin{subequations}\label{2bEp}
\begin{alignat}{3}\label{Eeq}
E\,=\,E_1+E_2\,&=\,\sqrt{m_1^2+m_2^2+2m_1m_2\gamma},
\\\label{pperpsq}
p_{\perp}^2\,=\,E_1^2-m_1^2\,&=\,E_2^2-m_2^2\,=\,\frac{m_1^2m_2^2}{E^2}(\gamma^2-1),
\\\label{u1wedgeu2}
u_1^{[\mu}u_2^{\nu]}&=-\sqrt{\gamma^2-1}\,u_\mr{cm}^{[\mu}p_{\perp}^{\nu]}/p_{\perp}.
\end{alignat}
\end{subequations}

\begin{figure}
\begin{center}
\includegraphics[scale=.5]{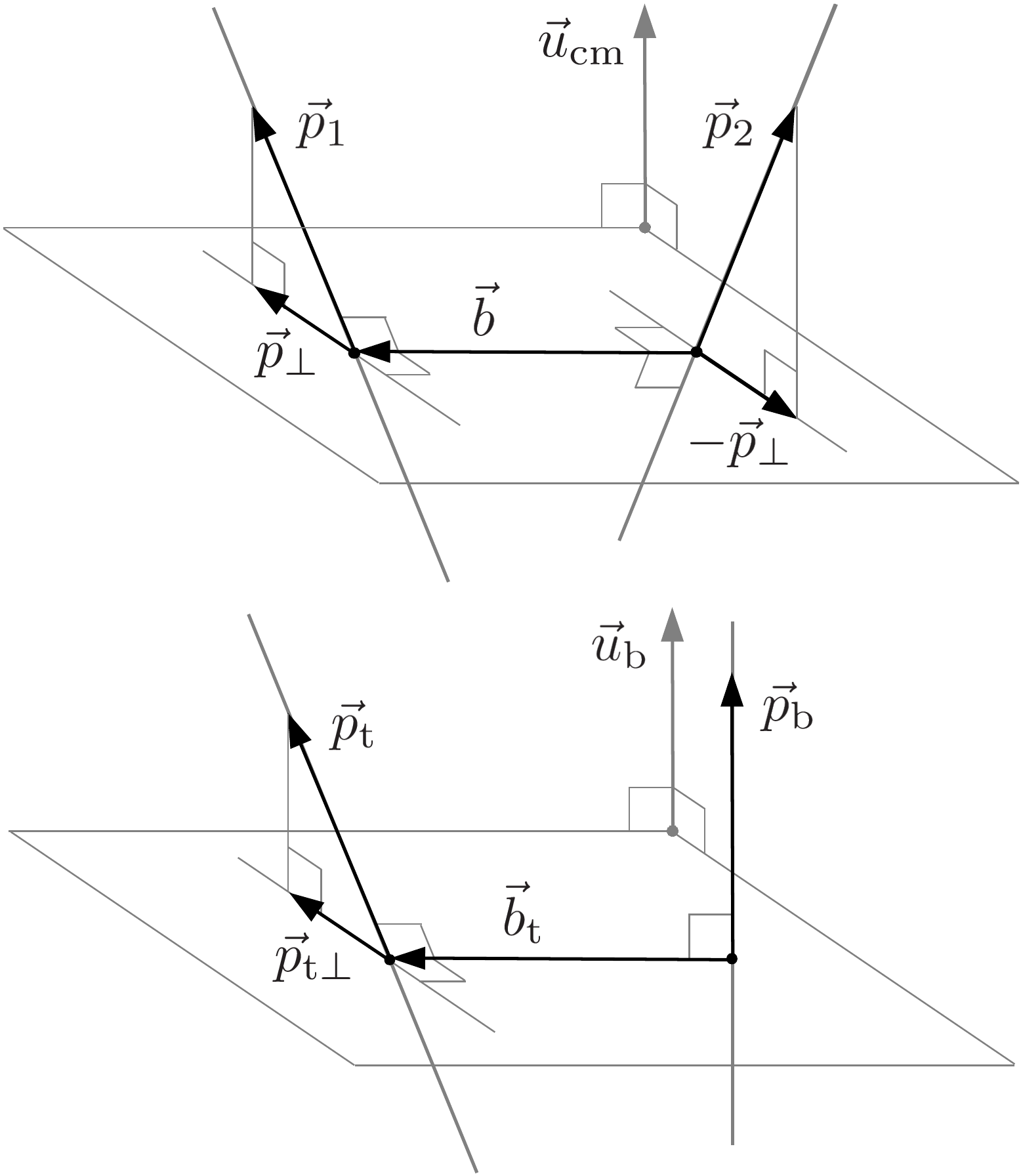}
\caption{Spacetime diagrams, with the future being upward, of the zeroth-order scattering states.  \emph{Top}: The two body case, with the bodies' momenta $\vec p_1$ and $\vec p_2$ pointing along their worldlines, and with the impact parameter $\vec b$ and relative momenta $\pm\vec p_{\perp}$ contained in the plane orthogonal to the cm-frame velocity $\vec u_\mr{cm}$.   \emph{Bottom}:  The test-body case, with the test and background bodies' momenta $\vec p_\mr{t}$ and $\vec p_\mr{b}$ pointing along their worldlines, and with $\vec b_\mr{t}$ and $\vec p_{\mr{t}{\perp}}$ contained in the plane orthogonal to the background velocity $\vec u_\mr{b}$.}\label{fig1}
\end{center}
\end{figure}

For the net 1PM BH-BH scattering described by (\ref{DpDa1})--(\ref{DpDa2}), we see from (\ref{ptot}) that
\bse
\be\label{Deltaucm}
\Delta p_\mr{tot}^\mu=0 \qquad\Rightarrow\qquad \Delta E=0,\qquad \Delta u_\mr{cm}^\mu=0.
\ee
Since $u_{1,2,\mr{cm}}\cdot\Delta p_{1,2}=0$, from (\ref{Zu0})--(\ref{ptot}), we also have
\be
\Delta E_1=\Delta E_2=\Delta \gamma=0,
\ee
and then from (\ref{p1split}),
\be\label{Deltapp}
\Delta p_{\perp}^\mu=\Delta p_1^\mu=-\Delta p_2^\mu.
\ee
\ese

The system's conserved total angular momentum tensor $J^{\mu\nu}$ about any spacetime point $x$ is given by
\be\label{Jtot}
J^{\mu\nu}(x)=2p_1^{[\mu}(x-z_1)^{\nu]}+S_1^{\mu\nu}+(1\to2),
\ee
where $z_1$ and $z_2$ can be any points on the bodies' worldlines.  From this we can define the following conserved cm-frame total angular momentum vector $J^\mu$, 
which is orthogonal to $u_\mr{cm}^\mu$ and independent of $x$; it follows from (\ref{z10})--(\ref{imp}), (\ref{p1split}), and (\ref{Jtot}) that, for any $x$,
\begin{alignat}{3}\label{Jvec}
J^{\mu}&=\frac{1}{2} u_\mr{cm}^\nu  \epsilon_\nu{}^\mu{}_{\alpha\beta} J^{\alpha\beta}(x)
\\\nnm
&=L^\mu+\frac{1}{2}u_\mr{cm}^\nu  \epsilon_\nu{}^\mu{}_{\alpha\beta}(S_1^{\alpha\beta}+S_2^{\alpha\beta}),
\end{alignat}
where
\be\label{Lvec}
L^\mu=u_\mr{cm}^\nu  \epsilon_\nu{}^\mu{}_{\alpha\beta} b^\alpha p_{\perp}^\beta
\ee
is the orbital angular momentum vector in the cm frame, the contribution to $J^\mu$ that survives when the bodies' spin tensors are set to zero.  In view of the fact that $u_\mr{cm}^\mu$, $p_{\perp}^\mu$ and $b^\mu$ are all mutually orthogonal [cf.\ (\ref{imp}) and (\ref{p1split})], its magnitude is given by
\be\label{Lmag}
L=bp_{\perp}.
\ee

Note that everything in the previous paragraph can be applied with the worldlines (and thus impact parameters) and spin tensors defined by arbitrary SSCs (\ref{SSCv}), not just the covariant SSC (\ref{CovSSC}).  Differing SSC choices will lead to differing values for $L^\mu$ and $S_{1,2}^{\mu\nu}$ while leaving $J^\mu$ invariant.  We apply the above relations with covariant SSCs (\ref{CovSSC}) in Sec.~\ref{sec:aligned}, and return to discuss transformations to other SSCs in Sec.~\ref{sec:canhams}.

\subsubsection{Test-body case in the background frame}\label{sec:frames:test}

Now consider the case of a test body of mass $m_\mr{t}$ being described in the rest frame of a background body with momentum $p^\mu_\mr{b}=m_\mr{b}u^\mu_\mr{b}$.  We split the test body's momentum $p^\mu_\mr{t}$ into parts along and orthogonal to $u^\mu_\mr{b}$,
\be\label{splitpt}
p_\mr{t}^\mu=m_\mr{t} u_\mr{t}^\mu= E_\mr{t}  u_\mr{b}^\mu+ p_{\mr{t}{\perp}}^\mu,\qquad p_{\mr{t}{\perp}}\cdot u_\mr{b}=0,
\ee
where $E_\mr{t}$ is the test body's energy in the background frame,
\be\label{Et}
E_\mr{t}= m_\mr{t} \gamma_\mr{t},\qquad \gamma_\mr{t}=- u_\mr{b}\cdot u_\mr{t},
\ee
with $\gamma_\mr{t}$ being the Lorentz factor of the test body relative to the background, and where $p_{\mr{t}{\perp}}^\mu$ is the relative momentum orthogonal to $u_\mr{b}^\mu$.  Note that
\begin{alignat}{3}\label{tbp2}
p_{\mr{t}{\perp}}^2&= m_\mr{t}^2 (\gamma_\mr{t}^2-1),
\\\label{utwedgeub}
u_\mr{t}^{[\mu}u_\mr{b}^{\nu]}&=-\sqrt{\gamma_\mr{t}^2-1}\,u_\mr{b}^{[\mu}p_{\mr{t}{\perp}}^{\nu]}/p_{\mr{t}{\perp}}.
\end{alignat}

The total angular momentum of (only) the test body about the background body (which will not be conserved in the scattering of the test body) is
\be\label{Jttot}
 J_\mr{t}^{\mu\nu}=-2p_\mr{t}^{[\mu} z_\mr{t}^{\nu]}+S^{\mu\nu}_\mr{t},
\ee
where $z_\mr{t}$ can be any point on the test body's worldline, with the background body assumed at rest at the origin for simplicity, and $S_\mr{t}^{\mu\nu}$ is the test body's spin tensor.  Let us parametrize the test body's worldline as
\be\label{xtz}
z_\mr{t}^\mu(\tau_\mr{t})=b_\mr{t}^\mu+u_\mr{t}^\mu\tau_\mr{t},\qquad  b_\mr{t}\cdot u_\mr{b}=b_\mr{t}\cdot u_\mr{t}=0,
\ee
so that $b_\mr{t}^\mu$ is the vectorial impact parameter.  Then the test body's angular momentum vector with respect to the background frame is
\begin{alignat}{3}\label{Jtvec}
J_\mr{t}^\mu&=\frac{1}{2}u_\mr{b}^\nu\epsilon_\nu{}^\mu{}_{\alpha\beta} J_\mr{t}^{\alpha\beta}
\\\nnm
&=L_\mr{t}^\mu+\frac{1}{2}u_\mr{b}^\nu\epsilon_\nu{}^\mu{}_{\alpha\beta} S_\mr{t}^{\alpha\beta},
\end{alignat}
where its orbital angular momentum is
\begin{alignat}{3}\label{Ltvec}
L^\mu_\mr{t}&=u_\mr{b}^\nu\epsilon_\nu{}^\mu{}_{\alpha\beta} b_\mr{t}^\alpha p_{\mr{t}{\perp}}^\beta
\end{alignat}
with magnitude
\be\label{Ltmag}
L_\mr{t}=b_\mr{t} p_{\mr{t}{\perp}}.
\ee

While all the above considerations in this or the previous section are applicable for worldlines (and thus impact parameters) and spin tensors defined in terms of arbitrary SSCs (\ref{SSCv}), let us now specialize again to the covariant SSC (\ref{CovSSC}) for all bodies, and define as in (\ref{acovS}) mass-rescaled covariant spin vectors $a_\mr{t}^\mu$ and $a_\mr{b}^\mu$ for the test body and the background body.

Note that, apart from the definitions of the energies, relative momenta, and angular momenta (which are all frame-dependent quantities), the setup for the test-body--background system, in particular concerning the geometry of the worldlines, is identical to the setup for the two-body system from above, under the identifications
\begin{alignat}{3}\label{twototest}
&u_1^\mu \to u_\mr{t}^\mu,\quad u_2^\mu \to u_\mr{b}^\mu,
\quad
\gamma\to\gamma_\mr{t},\quad b^\mu\to b_\mr{t}^\mu,
\nnm\\
&a_1^\mu\to a_\mr{t}^\mu,\quad a_2^\mu\to a_\mr{b}^\mu,
\end{alignat}
noting that $b^\mu$ is invariant under the boost which relates the cm frame for the two-body system to the rest frame of body 2, since $u_1\cd b=u_2\cd b=0$.  The specially covariant results for the scattering of the test BH off the background BH are thus obtained from (\ref{D1Z}) with $\Delta p_1^\mu\to \Delta p_\mr{t}^\mu$ and $\Delta a_1^\mu\to\Delta a_\mr{t}^\mu$ and the replacements (\ref{twototest}), while $\Delta p_\mr{b}^\mu=0=\Delta a_\mr{b}^\mu$ for the background BH.  It follows, with (\ref{splitpt})--(\ref{Et}), that
\be\label{DeltaEt}
\Delta E_\mr{t}=\Delta\gamma_t=0,\qquad \Delta p_{\mr{t}{\perp}}^\mu=\Delta p_\mr{t}^\mu.
\ee

\section{Scattering angles in the nonspinning and aligned-spin cases and effective-one-body mappings}\label{sec:aligned}

Let us re-emphasize that the 1PM BH-BH scattering results from Sec.~\ref{sec:pmscat} can be applied and specialized to both of the following situations as described in Sec.~\ref{sec:frames}:
%While the MPD equations or equivalent formulations of spinning extended-body dynamics (applicable to self-gravitating bodies as well as test bodies) have been applied extensively to the problem of two-body dynamics in the PN expansion, working perturbatively in $1/c$ [???],  and while the 1PM scattering deflection for point-masses [???] is a now classic result, this paper presents the first analysis of 1PM scattering of bodies with spin or higher multipoles of which we are aware.  
\begin{itemize}
\item \emph{the two-body case}, in which each body is deflected (and torqued) by the other's field, and which is best described in the system's cm frame, and
\item \emph{the test-body case}, in which only one (test) body is dynamical, being scattered by the second (background) body which is stationary, and which is best described in the rest frame of the background body.
\end{itemize}
In terms of the physical response of a given body (irrespective of the frame in which it is described), the first case is simply two interrelated copies of the second, to linear order in $G$.

The previous statement constitutes one sense in which the two-body results can be deduced from the test-body results, but it does not constitute a ``proper EOB mapping.''  The latter would produce the full information of the two-body results (both ``copies'') from some mapping of a single copy of the test-body results, and would be expected to depend crucially on the frames in which the systems are described.  

Such a mapping directly relating two- and test-body 1PM scattering results was found for nonspinning point-masses in \cite{Damour:2016s}, and we find here that this straightforwardly generalizes to two spinning BHs in the case where the spins are aligned with the orbital angular momentum.
  With zero or aligned spins, the spin vectors are constant, the motion is confined to a plane, and a complete description of the net scattering process is provided by the angle (in the respective frame) by which the BHs are deflected.  
  
  The EOB relations map the two- and test-body scattering angles into another under maps of the systems' masses, energies, and angular momenta.
In the nonspinning case, as in \cite{Damour:2016s}, one arrives at the EOB energy map.  In our generalization including aligned spins (to all orders in the BHs' spin-multipole expansions) we arrive at additional energy-dependent mappings of the BHs' spins.

We do not discuss here a proper EOB mapping at the level of the scattering results for the case of misaligned spins; we return to this fully generic case in the context of canonical Hamiltonians in Sec.~\ref{sec:canhams}.

\subsection{Covariant results with aligned spins}

Let us first express the scattering results for the case of aligned spins in a specially covariant form which can be applied to both the two- and test-body cases, working with bodies ``1'' and ``2,'' but which can later become the test body and the background body under (\ref{twototest}).  Here, again, we can use flat-spacetime relations to manipulate the zeroth-order quantities on which the $O(G)$ scattering results depend.

To describe the aligned-spin configuration, let us use an orthonormal tetrad $e_{\hat a}{}^\mu$ with $\hat a=\hat 0,\hat 1,\hat 2,\hat 3$, satisfying $\eta_{\mu\nu}e_{\hat a}{}^\mu e_{\hat b}{}^\nu=\eta_{\hat a\hat b}$ and $\epsilon_{\mu\nu\alpha\beta}e_{\hat 0}{}^\mu e_{\hat 1}{}^\nu e_{\hat 2}{}^\alpha e_{\hat 3}{}^\beta=1$.  We take the velocities to be in the plane spanned by $e_{\hat 0}$ and $e_{\hat 2}$, the impact parameter to be along $e_{\hat 1}$, and the spins to be along $\pm e_{\hat 3}$,
\begin{alignat}{3}\label{aligntetrad}
u_1^{[\mu} u_2^{\nu]}=-\sqrt{\gamma^2-1}\,e_{\hat 0}{}^{[\mu} e_{\hat 2}{}^{\nu]},\qquad && a_1^\mu&=a_1\,e_{\hat 3}{}^\mu,
\nnm\\
b^\mu=b\,e_{\hat 1}{}^\mu,\qquad && a_2^\mu&=a_2\,e_{\hat 3}{}^\mu,
\end{alignat}
where $b$ is the magnitude of the (covariant-SSC) impact parameter $b^\mu$, while $a_1$ and $a_2$ are the magnitudes of the rescaled spins $a_1^\mu$ and $a_2^\mu$ times $\pm 1$, allowing for the spins to be aligned or anti-aligned with $e_{\hat 3}$.  We fix here only the (oriented) plane spanned by $u_1$ and $u_2$ so that we can align $e_{\hat 0}$ to $u_\mr{cm}$ for the two-body case and to $u_\mr{b}$ for the test-body case; cf.\ (\ref{u1wedgeu2}), (\ref{utwedgeub}), and the paragraph containing (\ref{twototest}).

With the aligned-spin configuration (\ref{aligntetrad}), we find that (\ref{D1Z}) simplify to
\begin{alignat}{3}\label{Deltap1aligned}
\Delta p_1^\mu&=-\frac{2Gm_1m_2}{b^2-a_0^2}\left(\frac{2\gamma^2-1}{\sqrt{\gamma^2-1}}b-2\gamma a_0\right)e_{\hat 1}{}^\mu,
\nnm\\
\Delta a_1^\mu&=0,
\end{alignat}
where $a_0=a_1+a_2$.  From (\ref{DpDa2}), we also have $\Delta p_2^\mu=-\Delta p_1^\mu$ and $\Delta a_2^\mu=0$.  The aligned-spin results (\ref{Deltap1aligned}) apply both for the two-body system and for the test-body with $\Delta p_1^\mu\to\Delta p_\mr{t}^\mu$ and $\Delta a_1^\mu\to\Delta a_\mr{t}^\mu$ under the identifications in (\ref{twototest}).

\subsection{Aligned-spin scattering angles}
\subsubsection{Two-body case}
We see from (\ref{p1split}), (\ref{Deltapp}) and (\ref{Deltap1aligned}) that both bodies 1 and 2 are scattered by the same angle $\chi$ (in the $e_{\hat 1}$-$e_{\hat 2}$ plane) in the cm frame (with $e_{\hat 0}=u_\mr{cm}$), which is given in the small-angle approximation by the magnitude of $\Delta p_1^\mu=\Delta p_{\perp}^\mu=-(\Delta p_{\perp})e_{\hat 1}{}^\mu$ divided by the magnitude of $p_{\perp}^\mu=p_{\perp}e_{\hat 2}{}^\mu$,
\be
\chi=\frac{\Delta p_{\perp}}{p_{\perp}}.
\ee
From (\ref{Deltap1aligned}), using (\ref{pperpsq}) and (\ref{Lmag}) to express the result in terms of only the rest masses, $\gamma$ [and $E(m_1,m_2,\gamma)$ from (\ref{Eeq})], the (signed) spin values $a_1$ and $a_2$, and the magnitude $L$ of the (covariant-SSC) orbital angular momentum (\ref{Lvec}), we find
\be\label{chitwo}
\chi=\frac{2Gm_1m_2}{L\sqrt{\gamma^2-1}}\,\frac{{2\gamma^2-1}-2\gamma({\gamma^2-1})\dfrac{m_1m_2}{EL}(a_1+a_2)}{1-(\gamma^2-1)\dfrac{m_1^2m_2^2}{E^2L^2}(a_1+a_2)^2}.
\ee
\subsubsection{Test-body case}
Given (\ref{DeltaEt}), the angle $\chi_\mr{t}$ (in the $e_{\hat 1}$-$e_{\hat 2}$ plane) in the background frame (with $e_{\hat 0}=u_\mr{b}$) by which the test body is scattered is given by the magnitude of $\Delta p_{\mr{t}}^\mu=\Delta p_{\mr{t}{\perp}}^\mu=-(\Delta p_{\mr{t}{\perp}})e_{\hat 1}{}^\mu$, from (\ref{Deltap1aligned}) with (\ref{twototest}), divided by the magnitude of $p_{\mr{t}{\perp}}^\mu=p_{\mr{t}{\perp}}e_{\hat 2}{}^\mu$,
\be
\chi_\mr{t}=\frac{\Delta p_{\mr{t}{\perp}}}{p_{\mr{t}{\perp}}}.
\ee
Using (\ref{tbp2}) and (\ref{Ltmag}) to express the result in terms of only the rest masses, $\gamma_\mr{t}$, the signed spin magnitudes $a_\mr{t}$ and $a_\mr{b}$, and the magnitude $L_\mr{t}$ of the test-body's (covariant-SSC) orbital angular momentum (\ref{Ltvec}), we find
\be\label{chitest}
\chi_\mr{t}=\frac{2Gm_\mr{t}m_\mr{b}}{L_\mr{t}\sqrt{\gamma_\mr{t}^2-1}}\,\frac{{2\gamma_\mr{t}^2-1}-2\gamma_\mr{t}(\gamma_\mr{t}^2-1)\dfrac{m_\mr{t}}{L_\mr{t}}(a_\mr{t}+a_\mr{b})}{1-(\gamma_\mr{t}^2-1)\dfrac{m_\mr{t}^2 }{L_\mr{t}^2}(a_\mr{t}+a_\mr{b})^2}.
\ee

\subsection{Effective-one-body mappings of scattering angles}

\subsubsection{Nonspinning case; the energy map}
First consider the two scattering angles, $\chi$ from (\ref{chitwo}) for the two-body system, and $\chi_\mr{t}$ from (\ref{chitest}) for the test body, with all spins set to zero,
\begin{alignat}{3}\label{chinospin}
\chi_\tr{no spin}&=\frac{2G\mu M}{L}\frac{2\gamma^2-1}{\sqrt{\gamma^2-1}},
\\\nnm
\chi_\tr{t,no spin}&=\frac{2Gm_\mr{t}m_\mr{b}}{L_\mr{t}}\frac{2\gamma_\mr{t}^2-1}{\sqrt{\gamma_\mr{t}^2-1}}.
\end{alignat}
If we map the rest masses according to the usual ``Newtonian EOB mapping,'' with the background mass $m_\mr{b}$ being the total mass $M$, and the test mass $m_\mr{t}$ being the reduced mass $\mu$,
\begin{alignat}{3}\label{restmassmap}
m_\mr{b}&=M,\qquad& m_\mr{t}&=\mu,
\\\label{Mmu}
M&=m_1+m_2,\qquad&\mu&=\frac{m_1m_2}{M},
\end{alignat}
then we see that one simple relationship between the two- and test-body cases is given by
\be\label{nospinmapping}
\gamma_\mr{t}=\gamma,\quad L_\mr{t}=L\quad\Rightarrow\quad \chi_\tr{t,no spin}=\chi_\tr{no spin}.
\ee
Given (\ref{restmassmap})--(\ref{Mmu}), the relation $\gamma_\mr{t}=\gamma$ implies that the two-body system's cm-frame total energy $E$ as in (\ref{Eeq}) is related to the test body's background-frame energy $E_\mr{t}$ as in (\ref{Et}) by
\be\label{EOBEm}
\gamma_\mr{t}=\gamma\qquad\Leftrightarrow\qquad E_\mr{t}=\mu+\frac{E^2-M^2}{2M}.
\ee
The relation on the right is the \emph{EOB energy map} proposed (and derived in a PN context) in \cite{Buonanno99} to relate the Hamiltonian of a test body (replacing $E_\mr{t}$) to the Hamiltonian of a finite-mass-ratio two-body system (replacing $E$); the same relation was used earlier in \cite{Brezin:1970zr} to relate the energy levels of two- and test-body bound states in quantum electrodynamics.   It is seen to arise here from simple special relativistic kinematics [(\ref{Eeq}) and (\ref{Et})] applied to the scattering states at infinity, while identifying the frame-invariant relative Lorentz factors between the two pairs of bodies [and using (\ref{restmassmap})--(\ref{Mmu})].  The observations of (\ref{chinospin})--(\ref{EOBEm}), demonstrating a complete EOB equivalence inherent in the 1PM scattering of two point-masses, were first made in \cite{Damour:2016s}.

\subsubsection{Spinning case; new spin maps}

Returning to the scattering angles $\chi$ (\ref{chitwo}) and $\chi_\mr{t}$ (\ref{chitest}) with aligned spins, we easily find a generalization of the EOB mapping (\ref{nospinmapping}) from \cite{Damour:2016s} to the aligned-spin case; given the rest-mass maps (\ref{restmassmap}), we see that
\begin{alignat}{3}
&\gamma_\mr{t}=\gamma,\qquad L_\mr{t}=L,
\nnm\\\label{chimapspin}
& a_\mr{b}+a_\mr{t}=\frac{M}{E}(a_1+a_2)
\\\nnm
&&&\quad\Rightarrow\quad \chi_\mr{t}=\chi,
\end{alignat}
which establishes the equivalence at 1PM order between the arbitrary-mass-ratio two-spinning-BH system and the spinning test BH in a Kerr background, at the level of the aligned-spin scattering angles.  The mapping uses the same identifications of (covariant-SSC) orbital angular momenta and Lorentz factors, leading to the same energy map (\ref{EOBEm}), as well as a new energy-dependent mapping of the spins in (\ref{chimapspin}).\footnote{We see from (\ref{pperpsq}), (\ref{Lmag}), (\ref{tbp2}), and (\ref{Ltmag}), with (\ref{restmassmap}), that the mappings (\ref{nospinmapping}) and (\ref{chimapspin}) entail
\be\label{bpmap}
\Big(\gamma=\gamma_\mr{t}\;\;\Rightarrow\;\; p_{\perp}=\frac{M}{E}p_{\mr{t}{\perp}}\Big),\;\; L=L_\mr{t}\;\;\Rightarrow\;\; b=\frac{E}{M}b_\mr{t}.
\ee

We note that an alternative form of the mapping (\ref{nospinmapping}) between the nonspinning scattering angles, identifying the impact parameters instead of the angular momenta, could be given as
\be\label{nospinmappingalt}
\gamma=\gamma_\mr{t},\quad b=b_\mr{t}\quad\Rightarrow\quad \chi_\tr{no spin}=\frac{E}{M}\chi_\tr{t,no spin}.
\ee
A corresponding alternative to the aligned-spin mapping (\ref{chimapspin}) is
\be\label{spinmappingalt}
\gamma=\gamma_\mr{t},\;\;\; b=b_\mr{t},\;\;\; a_\mr{b}+a_\mr{t}=a_1+a_2\;\;\;\Rightarrow\;\;\; \chi=\frac{E}{M}\chi_\mr{t}.
\ee
Both entail $L=({M}/{E})L_\mr{t}$.  
For all of the mappings, (\ref{nospinmapping}) and (\ref{chimapspin}), and (\ref{nospinmappingalt})--(\ref{spinmappingalt}), it holds that $L\chi=L_\mr{t}\chi_\mr{t}$.
}

Furthermore, it is clear that both $\chi$ (\ref{chitwo}) and $\chi_\mr{t}$ (\ref{chitest}) can be obtained from the 1PM scattering angle $\chi_\mr{geod}$ for a \emph{geodesic} with energy $m_\mr{t}\gamma_\mr{t}$ and angular momentum $L_\mr{t}$ in the equatorial plane of a Kerr spacetime with mass $m_\mr{b}$ and spin $m_\mr{b}a$ [obtained from (\ref{chitest}) with $a_\mr{t}\to0$, $a_\mr{b}\to a$],
\be
\chi_\mr{geod}=\frac{2Gm_\mr{t}m_\mr{b}}{L_\mr{t}\sqrt{\gamma_\mr{t}^2-1}}\,\frac{2\gamma_\mr{t}^2-1-2\gamma_\mr{t}({\gamma_\mr{t}^2-1})m_\mr{t} a/L_\mr{t}}{1-(\gamma_\mr{t}^2-1)(m_\mr{t} a/L_\mr{t})^2},
\ee
via 
\begin{alignat}{3}
\gamma_\mr{t}=\gamma,\qquad L_\mr{t}=L,\qquad m_\mr{t}&=\mu,\qquad m_\mr{b}=M,
\nnm\\\label{chimapspina}
 a=a_\mr{b}+a_\mr{t}=\frac{M}{E}(a_1+a_2)
\\\nnm
&\;\;\Rightarrow\quad \chi_\mr{geod}=\chi_\mr{t}=\chi.
\end{alignat}
This establishes the equivalences amongst all three systems at 1PM order at the level of the aligned-spin scattering angles, under simple and arguably natural mappings.
We will discuss the relationship between these new \emph{EOB spin maps} and those used in previous EOB models of spinning BH binaries at the end of Sec.~\ref{sec:HamEOB}.

It is important to note that the mapping (\ref{chimapspin}) identifies the magnitudes of the \emph{orbital} angular momenta, $L^\mu$ from (\ref{Lvec}) and $L_\mr{t}^\mu$ from (\ref{Ltvec}), defined in terms of the worldlines/impact parameters defined by the \emph{covariant SSCs} (\ref{CovSSC}) for all bodies.  These are related to the total angular momenta $J^\mu$ and $J_\mr{t}^\mu$ and the covariant-SSC spin tensors by (\ref{Jvec}) and (\ref{Jtvec}).  With aligned spins (\ref{aligntetrad}), all the $L$- and $J$-vectors are collinear with all the covariant spin vectors $s^\mu=ma^\mu$, and one finds upon combining several relations from Sec.~\ref{sec:frames} that their magnitudes are related by
\bse
\begin{alignat}{3}
J&=L+E_1a_1+E_2a_2,&\qquad(L&=L_\mr{cov})
\\
J_\mr{t}&=L_\mr{t}+E_\mr{t}a_\mr{t},&\qquad(L_\mr{t}&=L_\mr{t,cov})
\end{alignat}
\ese
where we have indicated that these (like all $L$'s in Sec.~\ref{sec:aligned}) are the covariant-SSC orbital angular momenta.  While the values of the $L$'s and spin tensors $S^{\mu\nu}$ and their relations to the $J$'s, defined for arbitrary SSCs by (\ref{Jvec})--(\ref{Lvec}) and (\ref{Jtvec})--(\ref{Ltvec}), will differ for different SSC choices, the values of the total angular momentum vectors $J^\mu$ and $J_\mr{t}^\mu$ will be invariant under these choices.

\section{Canonical Hamiltonians for generic binary black holes and effective-one-body mappings}\label{sec:canhams}

A convenient and widely used way to encode the conservative dynamics of a binary of spinning objects is with a reduced canonical Hamiltonian function $H(\bs R,\bs P,\bs S_1,\bs S_2)$, where the canonical variables are a relative position vector $\bs R(t)$, its conjugate momentum $\bs P(t)$, and two canonical spin vectors $\bs S_1(t)$ and $\bs S_2(t)$, all being 3-vectors in a (background) Euclidean space.  The equations of motion follow from
\bse\label{HamEOM}
\be\label{dqdt}
\frac{dq}{dt}=\{q,H\}
\ee
for any function $q(\bs R,\bs P,\bs S_1,\bs S_2)$, with the canonical Poisson brackets
\be\label{PBs}
\{R^i,P_j\}=\delta^i_j,\qquad\{S_1^i,S_1^j\}=\epsilon^{ij}{}_kS_{1}^k,
\ee
\ese
and similarly with $S_1\to S_2$, with all other Poisson brackets vanishing.  
A given physical system does not have a unique such Hamiltonian function; rather, it has a class of Hamiltonian functions related by canonical transformations---diffeomorphisms of the phase space which preserve the canonical bracket algebra (\ref{PBs}).

One finds that the information in the Hamiltonian which is gauge-invariant (or invariant under canonical transformations) is uniquely determined by the net scattering results as discussed in Sec.~\ref{sec:pmscat}, and one can deduce a valid Hamiltonian from the scattering results---provided that one can translate between the covariant 4D description of Sec.~\ref{sec:pmscat} and the canonical 3D description as in (\ref{HamEOM}).

The 3D canonical phase space structure (\ref{PBs}) is known to be associated with a 4D MPD-type description of spinning body with momentum $p^\mu=m u^\mu$ which uses a ``canonical SSC'' \cite{Barausse:Racine:Buonanno:2009,VKSH},
\bse\label{SSCS}
\be\label{CanSSC}
S_\mr{can}^{\mu\nu}(U_\nu+u_\mu)=0,
\ee
or Pryce-Newton-Wigner SSC \cite{Pryce:1935, Pryce:1948, Newton:Wigner:1949}, where $U^\mu$ is some fixed (background) timelike vector field---as opposed to the covariant SSC (\ref{CovSSC}),
\be\label{COVSSC}
S_\mr{cov}^{\mu\nu}p_\mu=0,
\ee
\ese
as used above, where $S_\mr{cov}^{\mu\nu}$ was written simply as $S^{\mu\nu}$.  One key to relating our covariant scattering results to canonical Hamiltonians is thus the translation between the worldlines and spin tensors defined by these two SSCs, for both test-body and two-body situations, considered for zeroth-order scattering states at infinity and thus reasoning as in flat spacetime, which is the subject of Sec.~\ref{sec:NW}.  

In the final Sec.~\ref{sec:HamEOB}, we present results for Hamiltonians which reproduce (the canonical-SSC translations of) the covariant scattering deflections/holonomies, for each of the three cases---(iii) geodesics in a Kerr background, (ii) a spinning test BH in a Kerr background, and (i) an arbitrary-mass-ratio two-spinning-BH system---and we see that they exhibit simple and natural EOB mappings between them.

\subsection{Canonical worldlines and spins}\label{sec:NW}

\subsubsection{A general body in a general frame}\label{sec:NWgen}

For a body with momentum $p^\mu=m u^\mu$ (which is invariant under changes of SSC), the two spin tensors and two (parallel) worldlines defined by the canonical SSC (\ref{CanSSC}) with a background frame $U^\mu$ and by the covariant SSC (\ref{COVSSC}) are related by
\bse\label{SSCsolve}
\begin{alignat}{3}\label{Scandz}
S_\mr{can}^{\mu\nu}&=S_\mr{cov}^{\mu\nu}+2p^{[\mu}\delta z^{\nu]},\phantom{\Big|}
\\\label{dzcancov}
\delta z^\mu&=(z_\mr{can}-z_\mr{cov})^\mu,\qquad u\cd \delta z=0,
\end{alignat}
\ese
at zeroth order, which is simply an application of the transformation law (\ref{Stransf}) for angular momentum in flat spacetime.

Equations (\ref{SSCS})--(\ref{SSCsolve}) can be used to solve for the canonical quantities in terms of the covariant quantities or vice versa.  Let us split the momentum into parts along and orthogonal to $U^\mu$ according to
\be\label{pppp}
p^\mu=mu^\mu=\mc EU^\mu+p_{\perp}^\mu,
\ee
with $U\cdot p_{\perp}=0$, where $\mc E=-U\cdot p$ is the body's energy in the frame of $U^\mu$.  One then finds that the \emph{canonical spin vector},
\be\label{canSvec}
S^\mu=-\frac{1}{2}\epsilon^\mu{}_{\nu\alpha\beta}U^\nu S_\mr{can}^{\alpha\beta},
\ee
is related to the mass-rescaled covariant spin vector $a^\mu=-\epsilon^{\mu}{}_{\nu\alpha\beta}u^\nu S_\mr{cov}^{\alpha\beta}/{2m^2}$ of (\ref{acovS}) by
\begin{alignat}{3}\label{stoS}
S^\mu&=ma^\mu-\frac{m}{{\mc E}}(p_{\perp}\cd a)\left(U^\mu+\frac{p_{\perp}^\mu}{{\mc E}+m}\right)
\nnm\\
&=B(u\to U)^\mu{}_\nu \, ma^\nu,
\end{alignat}
where $B$ is the standard boost, e.g.\ as in (15)--(17) of \cite{Bini:2017xzy}, and that
\be\label{dzScanV}
\delta z^\mu=-\frac{\epsilon^\mu{}_{\nu\alpha\beta}U^\nu p_{\perp}^\alpha a^\beta}{({\mc E}+m)}=-\frac{\epsilon^\mu{}_{\nu\alpha\beta}U^\nu p_{\perp}^\alpha S^\beta}{m({\mc E}+m)}.
\ee

Noting that $p_{\perp}^\mu$, $S^\mu$, and $\delta z^\mu$, and are all spacelike vectors orthogonal to $U^\mu$, we can employ a boldface spatial 3-vector notation in the frame of $U^\mu$, with $\bs p=p_{\perp}^\mu$, to write
\be\label{deltabsz}
\delta\bs z=\bs z_\mr{can}-\bs z_\mr{cov}=\frac{\bs p\times\bs S}{m({\mc E}+m)},
\ee
giving the constant displacement between the two parallel zeroth-order worldlines $\bs z_\mr{can}$ and $\bs z_\mr{cov}$ (orthogonal to their common velocity $\propto\bs p$) in the $U^\mu$ frame.

%If $\Pi$ denotes the projector into the plane orthogonal to both $U^\mu$ and $u^\mu$ (or $p_{\perp}^\mu$), then the projection $\Pi\bs a=\Pi^\mu{}_\nu a^\nu$ of the covariant spin vector $a^\mu$ is equal to ($1/m$ times) that of the canonical spin vector $S^\mu$,\be\label{mPia}m\,\Pi \bs a=\Pi\bs S,\ee as follows from (\ref{stoS}).

\subsubsection{Test-body case in the background frame}

Now take the body of Sec.~\ref{sec:NWgen} to be the test body of Sec.~\ref{sec:frames:test} with momentum $p_\mr{t}^\mu=m_\mr{t}u_\mr{t}^\mu$, with the canonical-SSC frame (and the frame used for a spatial 3-vector notation) $U^\mu$ taken to be the frame of the background body, $U^\mu\to u_\mr{b}$.

%The result (\ref{mPia}) translates into \be\label{mtPiat} m_\mr{t}\,\Pi \bs a_\mr{t}=\Pi\bs S_\mr{t}, \ee where $\Pi$ projects orthogonal to $u_\mr{b}^\mu$ and $u_\mr{t}^\mu$ (and $\bs p_\mr{t}=p_{\mr{t}{\perp}}^\mu$).

With one copy of (\ref{xtz}) for each SSC, the result (\ref{deltabsz}) translates into the following relationship between the impact parameters defined by the canonical- and covariant-SSC worldlines,
\bse
\be\label{bbt}
\bs b_\mr{t,can}=\bs b_\mr{t,cov}+\frac{\bs p_\mr{t}\times\bs S_\mr{t}}{m_\mr{t}^2(\gamma_\mr{t}+1)},
\ee
with $\bs p_\mr{t}=p_{\mr{t}{\perp}}^\mu$.  With the canonical-SSC background-frame orbital angular momentum given by (\ref{Ltvec}) as
\be\label{Ltcan}
\bs L_\mr{t,can}=\bs b_\mr{t,can}\times\bs p_\mr{t},
\ee
the SSC-independent background-frame total angular momentum (of the test body, about the background body) is given by (\ref{Jtvec}) as
\be\label{JLSSt}
\bs J_\mr{t}=\bs L_\mr{t,can}+\bs S_\mr{t}.
\ee
\ese

\subsubsection{Two-body case in the center-of-mass frame}

Now take the body of Sec.~\ref{sec:NWgen} to be each of the two bodies 1 and 2 from Sec.~\ref{sec:twobody}, with the canonical-SSC frame (and the frame used for a spatial 3-vector notation) $U^\mu$ taken to be the cm frame, $U^\mu\to u_\mr{cm}^\mu$, as in (\ref{ptot}).  

%The result (\ref{mPia}) translates into \be\label{m1Pia1} m_1\,\Pi \bs a_1=\Pi\bs S_1,\qquad m_2\,\Pi \bs a_2=\Pi\bs S_2,\ee where $\Pi$ projects orthogonal to $u_1^\mu$ and $u_2^\mu$ (and $u_\mr{cm}^\mu$ and $\bs p=p_{\perp}^\mu$).

The result (\ref{deltabsz}), being careful with $\pm p_{\perp}$ for 1 vs.\ 2 as in (\ref{p1split}), translates into
\bse
\begin{alignat}{3}\label{bmb}
\bs b_\mr{can}-\bs b_\mr{cov}&=\bs z_\mr{1,can}-\bs z_\mr{1,cov}-\bs z_\mr{2,can}+\bs z_\mr{2,cov}
\nnm\\
&=\bs p\times\bs \Xi,
\end{alignat}
where
\be\label{Xi}
\bs\Xi=\frac{\bs S_1}{m_1(E_1+m_1)}+\frac{\bs S_2}{m_2(E_2+m_2)}.
\ee
With the canonical-SSC cm-frame orbital angular momentum given by (\ref{Lvec}) as
\be\label{Lcan}
\bs L_\mr{can}=\bs b_\mr{can}\times\bs p,
\ee
the SSC-independent cm-frame total angular momentum is given by (\ref{Jtvec}) as
\be\label{JLSS}
\bs J=\bs L_\mr{can}+\bs S_1+\bs S_2.
\ee
\ese

\subsection{Canonical Hamiltonians and EOB mappings}\label{sec:HamEOB}

Net 1PM scattering results in terms of canonical variables, giving $\Delta\bs p_\ms{a}$ and $\Delta \bs S_\ms{a}$ as functions of the incoming state, can be derived from a canonical Hamiltonian by taking the equations of motion resulting from (\ref{HamEOM}) and integrating them along entire zeroth-order histories, just as was done with the covariant equations of motion in (\ref{Holo1}), and as is considered also in \cite{Bini:2017xzy}.  Also as emphasized in \cite{Bini:2017xzy}, the naturally invariant quantities to be matched between differing descriptions of the dynamics are the net Lorentz transformations $\Lambda^\mu{}_\nu$,
\bse
\begin{alignat}{3}
p^\mu+\Delta p^\mu&=\Lambda^\mu{}_\nu p^\nu,
\\
a^\mu+\Delta a^\mu&=\Lambda^\mu{}_\nu a^\nu,
\end{alignat}
\ese
as in (\ref{theta}), which map to 3D spatial rotations in terms of canonical variables $\bs p$ and $\bs S$, giving the ``scattering holonomy'' and/or ``spin holonomy'' \cite{Bini:2017xzy}, expressed in terms of invariantly defined total energies and angular momenta (in respective frames).

By enforcing such matching with the covariant scattering results from Secs.~\ref{sec:pmscat} and \ref{sec:frames}, translated into canonical variables via Sec.~\ref{sec:NW}, for both two- and test-body cases in the respective frames, one arrives (modulo canonical transformations) at the three 1PM Hamiltonians presented below.

We define the second and third Hamiltonians here via the EOB mappings which are found to relate them to their predecessors: 
\begin{itemize}
\item starting from the Kerr-geodesic Hamiltonian $H_\mr{geod}$, 
\item the test-BH-in-Kerr Hamiltonian $H_\mr{t}$ is obtained from $H_\mr{geod}$ via a shift of the relative position and a simple (energy-independent) spin mapping, and then 
\item the arbitrary-mass-ratio two-spinning-BH Hamiltonian $H$ is obtained from $H_\mr{t}$ via the EOB energy map [as in (\ref{EOBEm})] and energy-dependent spin mappings.
\end{itemize}

Each of the Hamiltonians is given in terms of canonical relative position and momentum vectors $\bs R$ and $\bs P$, which are related to the physical worldlines and 3-momenta (at infinity) in each case as follows,
\bse\label{RP}
\begin{alignat}{5}
&\tr{for $H_\mr{geod}$,}&  \bs R&=\bs z_\mr{t,can}=\bs z_\mr{t,cov},
\\\nnm
&\tr{for $H_\mr{t}$,}& \bs R&=\bs z_\mr{t,can}=\bs z_\mr{t,cov}+\frac{\bs p_\mr{t}\times\bs S_\mr{t}}{m_\mr{t}^2(\gamma_\mr{t}+1)},
\\\nnm
&\tr{for $H$,}&  \frac{E}{M}\bs R&=(\bs z_1-\bs z_2)_\mr{can}=(\bs z_1-\bs z_2)_\mr{cov}+\bs p\times\bs\Xi,
\end{alignat}
as in the impact parameter relations (\ref{bbt}) and (\ref{bmb}), with
\be
\bs P=\bs p_\mr{t}=\frac{E}{M}\bs p,\qquad \gamma_\mr{t}=\gamma,
\ee
\ese
as in (\ref{bpmap}).  

We define the mass-rescaled canonical spin 3-vectors
\be
\bs a_\ms{a}=\frac{\bs S_\ms{a}}{m_\ms{a}},
\ee
for $\ms a=1,2,\mr{t},\mr{b}$, etc., and while these do not in general agree with the projections orthogonal to $u_\mr{cm}^\mu$ or $u_\mr{b}^\mu$ of the covariant spin vectors $a_\ms{a}^\mu$, the spins enter the Hamiltonians only through their projections orthogonal to both $u_\mr{cm}^\mu$ and $p_{\perp}^\mu=\bs p$ or $u_\mr{b}^\mu$ and $p_{\mr{t}{\perp}}^\mu=\bs p_\mr{t}$, and these projections are the same for canonical or covariant spin vectors, as shown by (\ref{stoS}).

The mappings are ultimately fully specified by (\ref{RP}) along with the key relation $\bs a_\mr{b}+\bs a_\mr{t}=(M/E)(\bs a_1+\bs a_2)$ as in (\ref{chimapspina}) and
the usual Newtonian EOB mapping of the rest masses (defining the symmetric mass ratio $\nu$),
\begin{alignat}{3}
m_\mr{b}&=M=m_1+m_2,\phantom{\Big|}
\\\nnm
m_\mr{t}&=\mu=\frac{m_1m_2}{M}=\nu M,
\end{alignat}
which we apply preemptively below.

\subsubsection{Kerr geodesics}

A 1PM Hamiltonian for geodesics in the Kerr spacetime can be found from a Legendre transformation of the covariant Lagrangian (\ref{Lint12}) specialized to a nonspinning test BH and reparametrized in terms of the background-frame proper time $\tau_\mr{b}=t$.  For a test point-mass $\mu $ in a Kerr background with mass $M$ and spin $M\bs a$, one finds
\bse\label{Hg}
\begin{alignat}{3}\label{Hgeod}
&\frac{1}{\mu}H_\mr{geod}(\mu ,\bs R,\bs P;M,\bs a)
\\\nnm
&=\gamma-\frac{GM}{2\gamma}\sum_{s=\pm1}\frac{\big(\gamma+s\sqrt{\gamma^2-1}\,\big)^2}{\bigg|\bs R+\dfrac{s\bs P\times\bs a}{\mu \sqrt{\gamma^2-1}}\bigg|},
\end{alignat}
with
\be\label{gammaP}
\gamma=\sqrt{1+\frac{\bs P^2}{\mu ^2}},
\ee
\ese
or $|\bs P|=\mu \sqrt{\gamma^2-1}$.   The Hamiltonian (\ref{Hg}) is $H_\mr{geod}=\mu \gamma-\mc L_\mr{int}/\gamma$ with $\mc L_\mr{int}$ given by the appropriate specialization of (\ref{Lint12}).

\subsubsection{Test black hole in Kerr}

A 1PM test-BH-in-Kerr Hamiltonian $H_\mr{t}$, for a test BH of mass $\mu $ and spin $\mu  \bs a_\mr{t}$ in a Kerr background with mass $M$ and spin $M\bs a_\mr{b}$ is obtained from (\ref{Hg}) via
\bse
\begin{alignat}{3}\label{TBHgeod}
&H_\mr{t}(\mu ,\bs R,\bs P,\bs a_\mr{t};M,\bs a_\mr{b})
\\\nnm
&=H_\mr{geod}\Big(\mu ,\bs R\,-\frac{\bs P\times\bs a_\mr{t}}{\mu (\gamma+1)},\bs P;M,\bs a_\mr{b}+\bs a_\mr{t}\Big),
\end{alignat}
which involves a shift of the position variable arising directly from (\ref{bbt}), as in (\ref{RP}), and the map 
\be
\bs a=\bs a_\mr{b}+\bs a_\mr{t},
\ee
 relating the (mass-rescaled) background-BH spin $\bs a$ for $H_\mr{geod}$ to the background-BH spin $\bs a_\mr{b}$ and test-BH spin $\bs a_\mr{t}$ for $H_\mr{t}$.
Explicitly,
\begin{alignat}{3}\label{Htex}
\frac{H_\mr{t}}{\mu}=\gamma
-\frac{GM}{2\gamma}\sum_{s=\pm1}\frac{\big(\gamma+s\sqrt{\gamma^2-1}\,\big)^2}{\bigg|\bs R+\dfrac{\bs P}{\mu }\times\bigg(s\dfrac{\bs a_\mr{b}+\bs a_\mr{t}}{\sqrt{\gamma^2-1}}-\dfrac{\bs a_\mr{t}}{\gamma+1}\bigg)\bigg|}.
\end{alignat}
\ese

\subsubsection{Generic two-spinning-black-hole system}

One obtains the arbitrary-mass-ratio binary-spinning-BH Hamiltonian
\bse\label{HBBH}
\be
H(m_1,m_2,\bs R,\bs P,\bs a_1,\bs a_2)
\ee
from the EOB energy map applied to $H_\mr{t}$ (\ref{TBHgeod}),
\be\label{EOBEM}
H=\sqrt{M^2+2M(H_\mr{t}-\mu)}+O(G^2),
\ee
with $\bs R$ and $\bs P$ in $H_\mr{t}$ unchanged, and with the spins in $H_\mr{t}$ given by
\begin{alignat}{3}\label{atmap}
\bs a_\mr{b}&=\frac{2}{\Gamma(\Gamma+1)}\bs\sigma,
\\\nnm
\bs a_\mr{t}&=\frac{1}{\Gamma}\bs\sigma_*+\frac{\Gamma-1}{\Gamma(\Gamma+1)}\bs\sigma,
\end{alignat}
where
\be\label{sigmastar}
\bs\sigma=\frac{m_1\bs a_1+m_2\bs a_2}{M},\qquad\bs\sigma_*=\frac{m_2\bs a_1+m_1\bs a_2}{M},
\ee
and
\be
 \Gamma\equiv\frac{E}{M}=\sqrt{1+2\nu(\gamma-1)},
\ee
\ese
with $\gamma$ as in (\ref{gammaP}), and where $E=H(G\to0)$ is as in (\ref{Eeq}).\footnote{
Note that (\ref{atmap}) are equivalent to (\ref{aKgeod}) along with 
\bse
\be
\frac{\bs a_\mr{t}}{\mu(\gamma+1)}=\frac{\bs\Xi}{\Gamma^2}=\frac{1}{\Gamma^2}\left(\frac{\bs a_1}{E_1+m_1}+\frac{\bs a_2}{E_2+m_2}\right),
\ee
which follows from (\ref{RP}) with $\bs R$ for $H_\mr{t}$ and $\bs R$ for $H$ identified, using the relation
\be
\frac{M}{E_1+m_1}=\frac{\Gamma}{\mu(\gamma+1)}\left(m_1+\frac{\Gamma-1}{\Gamma+1}m_2\right),
\ee
\ese
sim.\ $1\leftrightarrow2$, following (via nontrivial algebra) from (\ref{2bEp}).
} 
When (\ref{atmap})--(\ref{sigmastar}) are inserted into (\ref{TBHgeod}), we see that the background Kerr spin $\bs a$ for $H_\mr{geod}$ maps to
\be\label{aKgeod}
\bs a=\bs a_\mr{b}+\bs a_\mr{t}=\frac{\bs a_1+\bs a_2}{\Gamma},
\ee
just as in (\ref{chimapspina}).

The effective Hamiltonian given by $H_\mr{t}$ (\ref{Htex}) under the spin map (\ref{atmap})---which yields the generic binary BH Hamiltonian $H$ via the energy map (\ref{EOBEM})---can be expanded as
\bse\begin{alignat}{3}
\frac{H_\mr{t}}{\mu}&=\gamma-\sum_{\ell=0}^\infty\sum_{k=0}^\ell g_{\ell k}\frac{(\bs P\times\bs\sigma\cd\bs\nabla)^\ell(\bs P\times\bs\sigma_*\cd\bs\nabla)^{\ell-k}}{k!(\ell-k)!\mu^\ell}\,\frac{GM}{|\bs R|},
\end{alignat}
with the coefficients of the spin$^\ell$ terms with $k$ factors of $\bs\sigma$ and $\ell-k$ factors of $\bs\sigma_*$ given by
\begin{alignat}{3}\label{allgyros}
g_{\ell k}=\sum_{s=\pm1}\frac{(\gamma+s\sqrt{\gamma^2-1}\,)^2}{2\gamma(\gamma^2-1)^{\ell/2}\Gamma^\ell}
&\left(s-\frac{\Gamma-1}{\Gamma+1}\sqrt{\frac{\gamma-1}{\gamma+1}}\right)^k
\nnm\\
\times&\left(s-\sqrt{\frac{\gamma-1}{\gamma+1}}\right)^{\ell-k}.
\end{alignat}
At linear order in spin, we find the effective gyrogravitomagnetic ratios
\begin{alignat}{3}\label{gyros}
g_{11}&=\frac{(2\gamma+1)(2\gamma+\Gamma)-1}{\gamma(\gamma+1)\Gamma(\Gamma+1)}&&=g^\mr{1PM}_S,
\\\nnm
g_{10}&=\frac{2\gamma+1}{\gamma(\gamma+1)\Gamma}&&=g^\mr{1PM}_{S_*},
\end{alignat}
\ese
which match the results for $g^\mr{1PM}_{S}$ and $g^\mr{1PM}_{S_*}$ given recently in \cite{Bini:2017xzy}.  In the PN (post-\emph{Newtwonian}, non-relativistic) limit, with $\bs P^2\to0$ $\Rightarrow$ $\gamma\to 1$, $\Gamma\to1$,
these reduce to the leading-PN-order values $g_{S}\to 2$ and $g_{S_*}\to3/2$.

The new relativistic spin map (\ref{atmap})---which, along with the energy map, turns $H_\mr{t}$ into $H$---becomes in the PN limit
\be\label{abatLO} 
\bs a_\mr{b}\to\bs\sigma, \qquad \bs a_\mr{t}\to\bs\sigma_*,
\ee
which is the leading-PN-order spin map applied to the (linear-in-the-test-spin) spinning-test-body-in-Kerr Hamiltonian in the Barausse-Buonanno strain of spinning EOB models originating in \cite{Barausse:2009xi}.

The other new relativistic spin map given by the extremes of (\ref{aKgeod})---which, along with the energy map and the shift of $\bs R$ in (\ref{TBHgeod}) [involving $\bs a_\mr{t}$ from (\ref{atmap})], turns $H_\mr{geod}$ into $H$---has the PN limit
\be\label{aKLO}
\bs a\to\bs a_1+\bs a_2\;\; (=\bs a_0),
\ee
which is the spin map applied to the Kerr-geodesic Hamiltonian in the Damour-Jaranowski-Sch\"afer strain of spinning EOB models originating in \cite{DJS}.

%For a test point-mass, the test BH with zero spin, the covariant equations of motion (\ref{MPD1PMcov})--(\ref{puzd}) for its worldline $z(\tau)$ and momentum $p^\mu=mu^\mu$ follow from the action \begin{alignat}{3} \mc S_\mr{geod}&=\int d\tau\Big[p_{\mu}\dot z^\mu+\mc L_\mr{int}(z,p)\Big], \end{alignat} where $\mc L_\mr{int}=(m/2)u^\mu u^\nu h_{\mu\nu}$.  This defines a geodesic in the metric $g=\eta+h+O(G^2)$. When $h$ is the linearized Kerr field of a background BH with momentum $p_\mr{b}=m_\mr{b}he Lagrangian (\ref{Lint12}) [under (\ref{twototest})]

Our final 1PM generic binary-spinning-BH Hamiltonian $H$, given by (\ref{Htex})--(\ref{HBBH}), when expanded in powers of the spins and in PN (post-\emph{Newtonian}) orders (effectively, in powers of $\bs P$), reproduces (modulo canonical transformations) the linear-in-$G$ parts of all previous results for binary BH Hamiltonians obtained from PN calculations.  This includes the linear-in-$G$ parts of the 4PN [${O}(1/c^8)$] nonspinning \cite{Damour:2014jta, Damour:2015isa, Bernard:2015njp, Damour:2016abl,Marchand:2017pir}, 3.5PN [${O}(1/c^7)$] linear-in-spin \cite{Hartung:2011te,Hartung:Steinhoff:Schafer:2012,Marsat:2012fn,Bohe:2012mr,Levi:Steinhoff:2015:2}, and 4PN quadratic-in-spin \cite{Levi:Steinhoff:2015:3} results, along with the complete 3.5PN cubic- \cite{Hergt:2007ha, Hergt:2008jn, Levi:Steinhoff:2014:2,Vaidya:2014kza, Marsat:2014xea} and 4PN quartic-in-spin \cite{  Hergt:2007ha, Hergt:2008jn, Levi:Steinhoff:2014:2,Vaidya:2014kza} results (which are linear in $G$), and more generally the complete leading-PN-order results at each order in spin (which are linear in $G$) \cite{Vines:2016qwa}, as well as the linear-in-$G$ parts of all lower-order results reviewed e.g.\ in \cite{Blanchet:2013haa,Porto:2016pyg}.  Our Hamiltonian resums all of these 1PM parts of PN results for binary-BH dynamics in a compact form, and provides new information at all higher orders in $1/c$, to all orders in the spins, to linear order in $G$.

\section{Conclusion}\label{sec:conclude}

We have considered the dynamics of two spinning BHs in the 1PM approximation to general relativity---to linear order in the deviation of the metric from that of flat spacetime, or to linear order in $G$---to all orders in $v/c$, and to all orders in the BHs' spin-multipole expansions.  

We have seen that a complete explicit specification of the 1PM dynamics for the fully generic case---with an arbitrary mass ratio, for generic (bound or unbound) orbits and generic (aligned or misaligned) spin orientations---can be built up from the 1PM \emph{scattering holonomy}---by which we mean Lorentz-transformation generator $\theta^\mu{}_\nu$ of (\ref{theta}) giving the net $O(G)$ changes $\Delta p^\mu=\theta^\mu{}_\nu p^\nu$ and $\Delta a^\mu=\theta^\mu{}_\nu a^\nu$ in the momentum and spin---\emph{for a spinning test BH in a Kerr background}.  

That two-spinning-BH scattering holonomy is simply related to the 1PM covariant interaction Lagrangian (\ref{Lint12}), which exhibits the remarkable property (\ref{Lintfactor}) of being equivalent to the 1PM Lagrangian for \emph{geodesic motion in a Kerr spacetime}, with the latter effective Kerr spacetime having a mass-rescaled spin $a_0^\mu=a_1^\mu+a_2^\mu$ being the sum of the two BHs' mass-rescaled spins.

That property, combined with properties of the EOB energy map (\ref{EOBEm}) and a careful treatment of the SSCs (\ref{SSCS}) defining the BHs' frame-dependent center-of-mass worldlines, has allowed us to produce a 1PM generic two-spinning-BH Hamiltonian (\ref{HBBH}) from the Hamiltonian (\ref{Hgeod}) for Kerr geodesics, via a simple EOB mapping.  \emph{The EOB mapping is ultimately fully determined by special relativistic kinematics applied to scattering states at infinity.}  This provides significant new structural input for the construction of EOB models of gravitational wave signals from binary BHs, particularly for BHs with large spins.

These surprising equivalences---between two-body and test-body motion, and between spinning-BH motion and geodesic motion---have been demonstrated here only to linear order in $G$, but they are suggestive of further surprises at higher orders.  %Hints of (at least partial) EOB equivalences at $O(G^2)$, to be discussed in future work, are present in the relationships between (i) the (partial) 2PM scattering results given in \cite{Westpfahl:1985}, and (ii) Schwarzschild geodesic scattering, and the same is true for the relationships between (i) available PN results for spinning binary BHs, (ii) spinning test-BH scattering in Kerr \cite{Bini:2017pee}, and (iii) the recent results of \cite{Guevara:2017csg} for classical scattering amplitudes for arbitrary-spin massive particles at the one-loop level.

\acknowledgements

We are grateful to Alessandra Buonanno, Abraham Harte, and Jan Steinhoff for helpful discussions and for comments on earlier versions of this draft.

%\section*{Appendices}

\appendix

%\addtocontents{toc}{\protect\setcounter{tocdepth}{1}}
%\makeatletter
%\addtocontents{toc}{%
%  \begingroup
%  \let\protect\l@chapter\protect\l@section
%  \let\protect\l@section\protect\l@subsection
%}
%\makeatother

\section{The Schwarzschild and Kerr metrics and their harmonic-gauge linearizations}\label{app:Kerr}

In a sense, the full structure of the exact Kerr metric, in the Kerr-Schild form \cite{Kerr2009}, is encoded in a certain congruence of null rays in Minkowski spacetime, corresponding to a certain \emph{twisted} (shear-free, geodesic) congruence of straight lines in Euclidean 3-space.

\subsection{Spherical coordinates and the radial congruence}

The same is true for the Schwarzschild spacetime, with a non-twisted congruence.  

In flat spacetime with inertial coordinates $x^\mu=(t,x^i)$, we define the following standard coordinates $x^i$ on the flat 3-spaces,
\begin{itemize}
\item Cartesian $(x,y,z)$,
\end{itemize}
\be\label{Cartesian}
x=\rho\cos\phi,\qquad y=\rho\sin\phi,\qquad z=z,
\ee
\begin{itemize}
\item cylindrical $(\rho,z,\phi)$,
\end{itemize}
\be\label{spherical}
\rho=r\sin\theta,\qquad z=r\cos\theta,\qquad\phi=\phi,
\ee
\begin{itemize}
\item spherical $(r,\theta,\phi)$,
\end{itemize}
and we define the vector field $\bs r$ giving the displacement from the origin, and its direction, the unit vector field $\bs n$,
\begin{alignat}{3}\label{r3vec}
\bs r&=r^i\doe_i&&=r\bs n=rn^i\doe_i
\nnm\\
&=x\doe_x+y\doe_y+z\doe_z
\nnm\\
&=r\doe_r,
\end{alignat}
with $n_i=\doe_ir$.  The radial unit vector $\bs n$ is tangent to a shear-free geodesic congruence, the set of straight half-lines from the origin; $\bs n$ is the unit tangent, $\bs n^2=1$, and $r$ is the corresponding affine parameter (particularly, the proper distance) along the geodesics.  

In the Minkowski space, consider the null vector field
\begin{alignat}{3}\label{ellup}
\ell^\mu\doe_\mu&=\doe_t-n^i\doe_i
\\\nnm
&=\doe_t-\doe_r,
\end{alignat}
which is tangent to the shear-free geodesic congruence given by the set of future-pointing half null rays with future endpoints on the worldline of the origin $r=0$, the rays of the worldline's incoming lightcones, along which $r$ is an affine parameter.  The corresponding 1-form is given by
\begin{alignat}{3}\label{ell1form}
\ell_\mu dx^\mu&=-dt-n_idx^i
\\\nnm
&=-dt-dr.
\end{alignat}
Defining the timelike unit vector field $u^\mu$ and spatial radial unit vector field $n^\mu$,
\be
u^\mu\doe_\mu=\doe_t,\qquad n^\mu\doe_\mu=n^i\doe_i=\doe_r,
\ee
the null vector field $\ell^\mu=u^\mu-n^\mu$ satisfies
\be\label{doeell}
\doe_\mu\ell_\nu=\doe_\mu n_\nu=\frac{1}{r}(\eta_{\mu\nu}+u_\mu u_\nu-n_\mu n_\nu),
\ee
since $\doe_i n_j=(\delta_{ij}-n_in_j)/r$ and $\doe_\mu u_\nu=0$ (which demonstrate that $\ell$ and $n$ are geodesic are shear-free), away from $r=0$.
The scalar field given by divergence of $\ell$,
\be\label{divSchw}
\doe_\mu\ell^\mu=\bs \nabla\cd \bs n=\frac{2}{r},
\ee
is harmonic, $\Box(1/r)=\bs\nabla^2(1/r)=0$, away from $r=0$. 

 Note that we continue to raise and lower indices with the Minkowski metric $\eta$, and we use $\doe_\mu$ for the covariant derivative on the Minkowski spacetime, and $\doe_i$ or $\bs\nabla$ for the covariant derivative on flat 3-space.

\subsection{The Schwarzschild metric in spherical-Kerr-Schild coordinates and in Schwarzschild coordinates}

The null vector field $\ell^\mu$ together with the scalar field \mbox{$2/r=\doe\cd\ell$} generate the (ingoing) Kerr-Schild form of the Schwarzschild metric with mass $m$, as a second metric on the manifold of the (background) Minkowski spacetime, according to
\be\label{gSchw}
g^\mr{Schw}_{\mu\nu}=\eta_{\mu\nu}+\frac{2m}{r}\ell_\mu\ell_\nu,
\ee
setting $G=1$ in this section.  The line element $ds^2_\mr{Schw}=g^\mr{Schw}_{\mu\nu}dx^\mu dx^\nu$ is  given by (\ref{ell1form}) and (\ref{gSchw}) as
\begin{alignat}{3}
ds^2_\mr{Schw}=-dt^2+dr^2+r^2d\Omega^2+\frac{2m}{r}(dt+dr)^2,
\end{alignat}
in the spherical-Kerr-Schild coordinates $(t,r,\theta,\phi)$ which are the standard spherical coordinates on the Minkowski spacetime, with $d\Omega^2=d\theta^2+\sin^2\theta\, d\phi^2$.  Introducing a new time coordinate $t_\mr{Schw}$ via
\be
dt=dt_\mr{Schw}+\frac{2m}{r-2m}dr,
\ee
yields the standard form
\be
ds^2_\mr{Schw}=-\left(1-\frac{2m}{r}\right)dt_\mr{Schw}^2+\frac{dr^2}{1-2m/r}+r^2d\Omega^2,
\ee
in Schwarzschild coordinates $(t_\mr{Schw},r,\theta,\phi)$.

\subsection{Harmonic-gauge linearization of Schwarzschild}

It follows from (\ref{doeell}) that the Schwarzschild metric (\ref{gSchw}) can be decomposed according to
\begin{alignat}{3}\label{KSSchw}
g^\mr{Schw}_{\mu\nu}&=\eta_{\mu\nu}+\frac{2m}{r}\ell_\mu\ell_\nu
\\
&=\eta_{\mu\nu}+ h_{\mu\nu}^\mr{Schw}+2\doe_{(\mu}\xi^\mr{Schw}_{\nu)},\phantom{\Big|}
\end{alignat}
where $h_{\mu\nu}^\mr{Schw}=\mc P_{\mu\nu\alpha\beta}\bar h^{\alpha\beta}_\mr{Schw}$ with
\be
\bar h^{\mu\nu}_\mr{Schw}=\frac{4m}{r}u^\mu u^\nu,
\ee
and
\be
\xi_\mr{Schw}^\mu=-2m\ln r\,u^\mu-mn^\mu.
\ee
The metric perturbation $h^\mr{Schw}_{\mu\nu}$ exactly solves the harmonic gauge condition (\ref{mcPgauge}) and the harmonic-gauge linearized vacuum field equation (\ref{mcPFE}) with the source
\be\label{TSchw}
T^{\mu\nu}_\mr{Schw}(x)=m\int dt\, u^\mu u^\nu \delta(x-z),
\ee
where $z^\mu(t)=u^\mu t$ is the worldline of the origin.
Considering the gauge-invariant form of the linearized field equation for a metric perturbation $h_{\mu\nu}=\mc P_{\mu\nu\alpha\beta}\bar h^{\alpha\beta}$,
\be\label{genLFE}
\Box \bar h_{\mu\nu}-2\doe_\alpha \doe_{(\mu}\bar h_{\nu)}{}^\alpha+\eta_{\mu\nu}\doe_\alpha\doe_\beta \bar h^{\alpha\beta}=-16\pi T_{\mu\nu},
\ee
$h_{\mu\nu}=2\doe_{(\mu}\xi_{\nu)}$ is a solution with $T^{\mu\nu}=0$ for any $\xi_\mu$, and thus $h_{\mu\nu}=(2m/r)\ell_\mu\ell_\nu$ is a solution with (\ref{TSchw}), or is a vacuum solution away from $r=0$.  Thus, given the fact that any metric of the form $g_{\mu\nu}=\eta_{\mu\nu}+h_{\mu\nu}$ with $h_{\mu\nu}=U\ell_{\mu}\ell_{\nu}$ satisfying the vacuum linearized field equation [(\ref{genLFE}) with $T^{\mu\nu}=0$] and $\ell^2=0$ is an exact solution of the full nonlinear vacuum Einstein equation \cite{doi:10.1063/1.523851,Harte:2016vwo}, it follows that the Schwarzschild metric (\ref{KSSchw}) is an exact vacuum solution away from $r=0$.

\subsection{(Twisted) Oblate spheroidal coordinates and the twisted-radial congruence}

An analogous specification of the Kerr metric in terms of a geodesic congruence in flat space, one which is twisted about a ring of radius $a$ in the $x$-$y$ plane centered at the origin, can be given in terms of new coordinates (on flat 3-space and flat spacetime) adapted to its geometry.  From the cylindrical coordinates $(\rho,z,\phi)$ of (\ref{Cartesian})--(\ref{spherical}), we define \emph{oblate spheroidal} (OS) coordinates $(\tilde r,\tilde\theta,\phi)$,
\be
\rho=\sqrt{\tilde r^2+a^2}\sin\tilde\theta,\qquad z=\tilde r\cos\tilde\theta,\qquad\phi=\phi,
\ee
and then \emph{twisted oblate spheroidal} (TOS) coordinates  $(\tilde r,\tilde\theta,\tilde\phi)$,
\be
\tilde r=\tilde r,\qquad\tilde\theta=\tilde\theta,\qquad\tilde\phi=\phi+\tan^{-1}\frac{\tilde r}{a}.
\ee
To summarize all of the coordinate transformations,
\begin{alignat}{3}
x+iy&=\rho e^{i\phi}=r\sin\theta\, e^{i\phi}
\nnm\\\nnm
&=\sqrt{\tilde r^2+a^2}\sin\tilde\theta\, e^{i\phi}=(-i\tilde r+a)\sin\tilde\theta\, 
e^{i\tilde\phi},\phantom{\Big|}
\\\label{coordsumm}
z&=r\cos\theta=\tilde r\cos\tilde\theta.\phantom{\Big|}
\end{alignat}
The components of the Euclidean metric $\delta_{ij}$ are given in the OS coordinates by
\begin{alignat}{3}\label{etaOS}
\delta_{ij}dx^idx^j&=\frac{\Sigma}{\tilde r^2+a^2}d\tilde r^2+\Sigma\, d\tilde\theta^2+\rho^2d\phi^2,
\end{alignat}
where
\be
\Sigma=\tilde r^2+a^2\cos^2\tilde\theta,
\ee
and in the TOS coordinates by
\begin{alignat}{3}
\delta_{ij}dx^idx^j&=d\tilde r^2+\Sigma\, d\tilde \theta^2+\rho^2d\tilde \phi^2-2a\sin^2\tilde \theta\, d\tilde r\,d\tilde\phi
\nnm\\
&=(d\tilde r-a\sin^2\tilde \theta\,d\tilde \phi)^2+\Sigma(d\tilde \theta^2+\sin^2\tilde \theta\,d\tilde \phi^2).\phantom{\bigg|}
\end{alignat}

\begin{figure}
\begin{center}
\includegraphics[scale=.29]{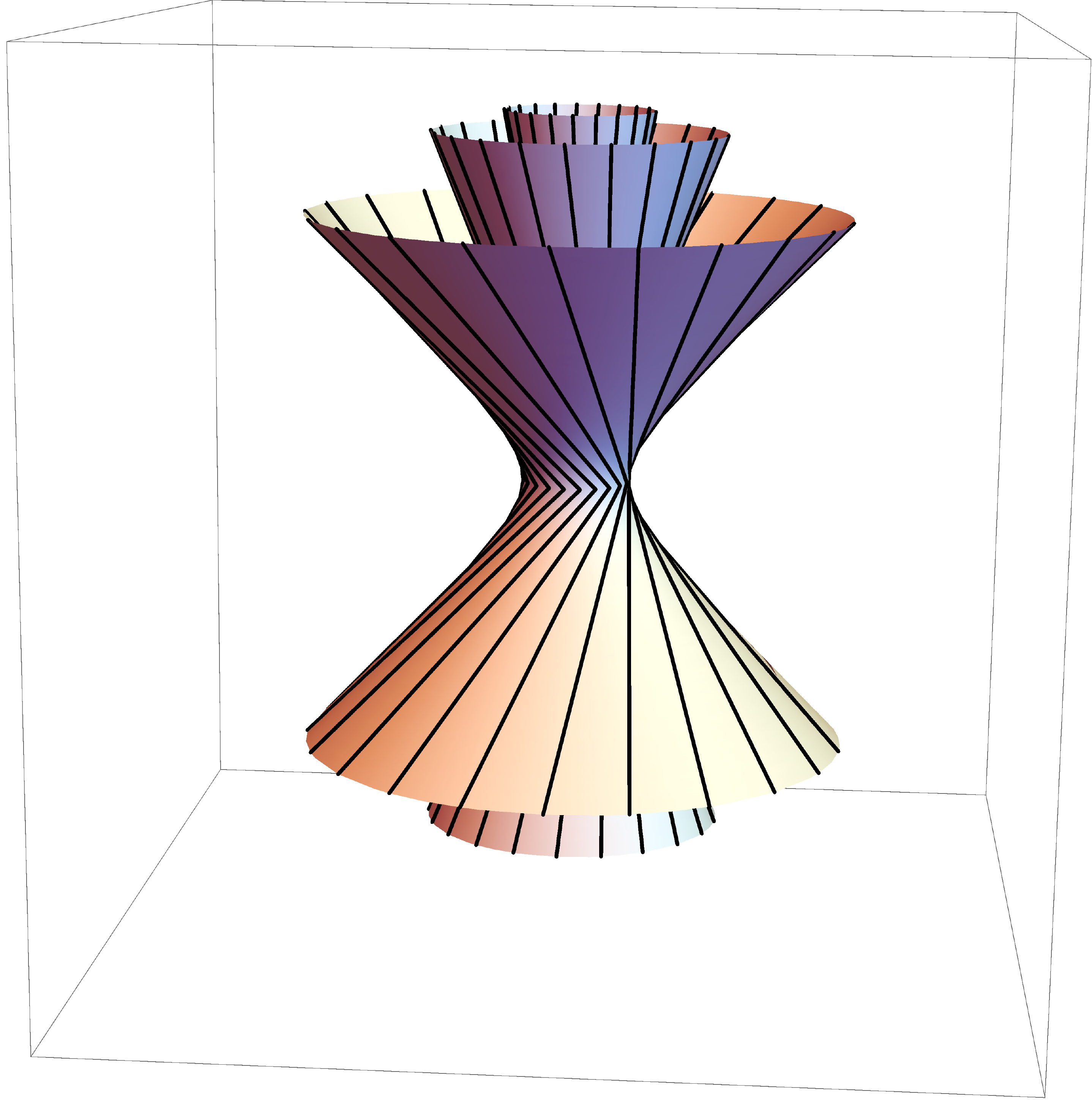} 
\caption{The twisted-radial congruence of straight lines in Euclidean 3-space, given by the lines of constant $\tilde\theta$ and $\tilde\phi$ in the TOS (twisted oblate spheroidal) coordinates $(\tilde r,\tilde\theta,\tilde \phi)$, shown on the surfaces of constant $\tilde\theta$ which are half one-sheeted hyperboloids (instead of the half cones $\theta=\mr{const.}$).  The ring, at $\rho=a$, $z=0$ or $\tilde r=0$, $\tilde\theta=\pi/2$, through which the waists of all the hyperboloids are threaded, is not pictured above, but is the inner boundary of the half hyperboloid $\tilde\theta=\pi/2$ shown below, which is the equatorial plane minus the disk $\rho<a$, $z=0$ or $\tilde r=0$.}
\includegraphics[scale=.393]{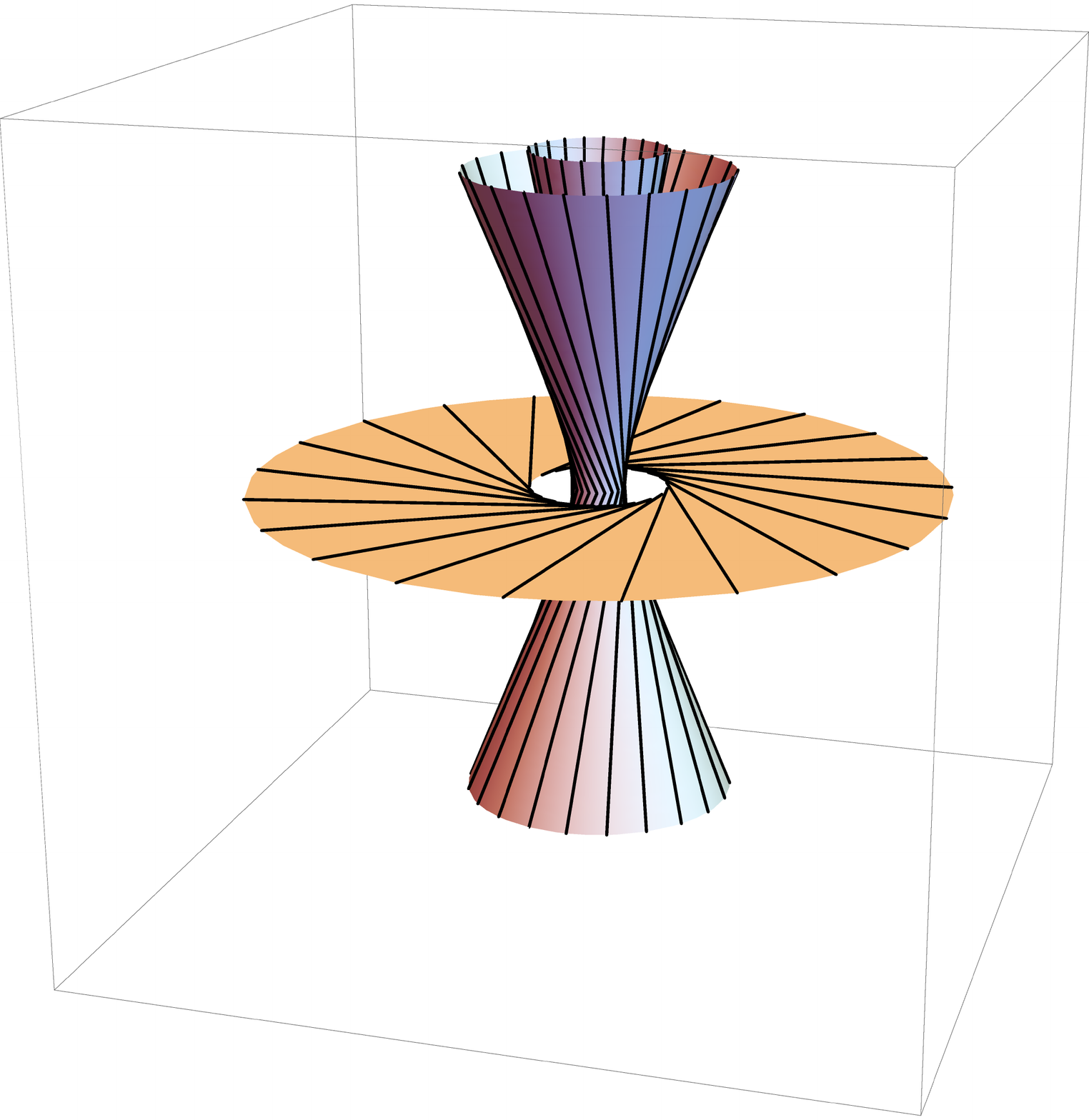}\label{fig:twist}
\end{center}
\end{figure}

The twisted congruence of straight lines in flat 3-space, pictured in Fig.~\ref{fig:twist}, is given by the (straight) curves of constant $\tilde\theta$ and $\tilde\phi$ in the TOS coordinates, and $\tilde r$ is the proper distance along them.  The lines' tangents are given by the \emph{twisted-radial unit vector field}
\begin{alignat}{3}\label{tnOS}
\tilde{\bs n}=\tilde n^i\doe_i&=\left(\frac{\doe}{\doe\tilde r}\right)_{\tilde\theta,\phi}-\frac{a}{\tilde r^2+a^2}\left(\frac{\doe}{\doe\phi}\right)_{\tilde r,\tilde\theta}
\\
&=\left(\frac{\doe}{\doe\tilde r}\right)_{\tilde\theta,\tilde\phi},
\end{alignat}
in OS and TOS coordinates respectively.  The corresponding 1-form $\tilde n_i=\delta_{ij}\tilde n^i$ is given by
\begin{alignat}{3}\label{tnd}
\tilde n_idx^i&=\frac{\Sigma}{\tilde r^2+a^2}d\tilde r-a\sin^2\tilde\theta\,d\phi
\\
&=d\tilde r-a\sin^2\tilde\theta\,d\tilde\phi,\phantom{\Big|}
\end{alignat}
satisfying $\tilde{\bs n}^2=1$ and
\be
\doe_i\tilde n_j=\frac{\tilde r}{\Sigma}(\delta_{ij}-\tilde n_i\tilde n_j)+\frac{a\cos\tilde\theta}{\Sigma}\epsilon_{ijk}\tilde n^k,
\ee
(where $\doe_i$ is the covariant derivative on flat 3-space) demonstrating that it is geodesic, since $\tilde n^j\doe_j\tilde n^i=0$, and shear-free,
since $\doe_{(i}\tilde n_{j)}$ is proportional to the projector orthogonal to $\tilde{\bs n}$, with nonzero twist $\doe_{[i}\tilde n_{j]}$ proportional to $a$.

In the Minkowski spacetime with line element $\eta_{\mu\nu}dx^\mu dx^\nu=-dt^2+\delta_{ij}dx^idx^j$, we have the corresponding null congruence given by the ingoing twisted-radial null vector field
\be\label{tell}
\tilde\ell^\mu\doe_\mu=(u^\mu-\tilde n^\mu)\doe_\mu=\doe_t-\tilde n^i\doe_i,
\ee
satisfying
\begin{alignat}{3}\label{doetell}
\doe_\mu\tilde\ell_\nu=\doe_\mu \tilde n_\nu&=\frac{\tilde r}{\Sigma}(\eta_{\mu\nu}+u_\mu u_\nu-\tilde n_\mu\tilde n_\nu)
\nnm\\
&\quad+\frac{a\cos\tilde\theta}{\Sigma}\epsilon_{\mu\nu\alpha\beta}u^\alpha\tilde n^\beta,
\end{alignat}
with the divergence
\be
\doe_\mu\tilde\ell^\mu=\bs\nabla\cdot\tilde{\bs n}=\frac{2\tilde r}{\Sigma}
\ee
being harmonic, $\Box(\tilde r/\Sigma)=\bs\nabla^2(\tilde r/\Sigma)=0$, away from $\tilde r=0$.
Defining the spacelike (mass-rescaled) covariant spin vector
\be\label{a3vec}
a^\mu\doe_\mu=a^i\doe_i=a\left(\frac{\doe}{\doe z}\right)_{x,y}=\bs a,
\ee
and noting
\mbox{$-a(\tilde r^2+a^2)\sin^2\tilde\theta\,d\phi$} $=$ \mbox{$
-a\rho^2\,d\phi$} $=$ \mbox{$(\bs r\times\bs a)_i dx^i$} \mbox{$=u^\nu \epsilon_{\nu\mu\alpha\beta} r^\alpha a^\beta dx^\mu
$},
with $r_\mu=rn_\mu=\doe_\mu r^2/2$, we can express (\ref{tell}) with (\ref{tnd}) as
\begin{alignat}{3}\label{exell}
\tilde\ell_\mu&=u_\mu-\tilde n_\mu,
\\\nnm
\tilde n_\mu&=\frac{\Sigma}{2\tilde r}\doe_\mu \ln(\tilde r^2+a^2)+u^\nu \epsilon_{\nu\mu\alpha\beta} \frac{r^\alpha a^\beta}{\tilde r^2+a^2}.
\end{alignat}

\subsection{The Kerr metric in oblate-spheroidal-Kerr-Schild coordinates and in Boyer-Lindquist coordinates}

In close analogy to (\ref{gSchw})--(\ref{divSchw}) for Schwarzschild, the (ingoing) Kerr-Schild form of the Kerr metric with mass $m$ and spin $ma$ (determining the radius of the ring $\rho=a$, $z=0$) is generated from the null vector field $\tilde\ell^\mu$ and the scalar field \mbox{$2\tilde r/\Sigma=\doe\cd\tilde\ell$} according to
\be\label{gKerr}
g^\mr{Kerr}_{\mu\nu}=\eta_{\mu\nu}+\frac{2m\tilde r}{\Sigma}\tilde\ell_\mu\tilde\ell_\nu.
\ee
Using (\ref{etaOS}) and (\ref{tnOS}) with the (non-twisted) OS coordinates $(\tilde r,\tilde\theta,\phi)$ on the flat 3-space, the Kerr line element reads
\begin{alignat}{3}
ds^2_\mr{Kerr}&=-dt^2+\frac{\Sigma}{\tilde r^2+a^2}d\tilde r^2+\Sigma\, d\tilde\theta^2+\rho^2d\phi^2
\\\nnm
&\quad+\frac{2m\tilde r}{\Sigma}\left(-dt-\frac{\Sigma}{\tilde r^2+a^2}d\tilde r+a\sin^2\tilde\theta\,d\phi\right)^2,
\end{alignat}
in oblate-spheroidal-Kerr-Schild coordinates $(t,\tilde r,\tilde\theta,\phi)$.  Defining new coordinates $t_\mr{BL}$ and $\phi_\mr{BL}$ via
\begin{alignat}{3}
dt&=dt_\mr{BL}+\frac{2m{\tilde r}}{\Delta}d{\tilde r},
\\
d\phi&=d{\phi}_\mr{BL}+\frac{a}{{\tilde r}^2+a^2}\frac{2m{\tilde r}}{\Delta}d{\tilde r},
\end{alignat}
with
\be
\Delta={\tilde r}^2+a^2-2m{\tilde r},
\ee
yields the standard form
\begin{alignat}{3}
ds^2_\mr{Kerr}&=-\frac{\Delta}{\Sigma}\Big(d{t_\mr{BL}}-a\sin^2{\tilde\theta}\,d{\phi_\mr{BL}}\Big)^2+\frac{\Sigma}{\Delta}d{\tilde r}^2
\\\nnm
&\quad+\Sigma\, d{\tilde\theta}^2+\frac{\sin^2{\tilde\theta}}{\Sigma}\Big(({\tilde r}^2+a^2)\,d{\phi_\mr{BL}}-a \,d{t_\mr{BL}}\Big)^2,
\end{alignat}
in Boyer-Lindquist coordinates $(t_\mr{BL},\tilde r,\tilde\theta,\phi_\mr{BL})$.

\subsection{Harmonic-gauge linearization of Kerr}

It follows from (\ref{doetell}) and (\ref{exell}) that the Kerr metric (\ref{gKerr}) can be decomposed according to
\begin{alignat}{3}\label{KSK}
g^\mr{Kerr}_{\mu\nu}&=\eta_{\mu\nu}+\frac{2m\tilde r}{\Sigma}\tilde\ell_\mu\tilde\ell_\nu
\nnm\\
&=\eta_{\mu\nu}+ h_{\mu\nu}^\mr{Kerr}+2\doe_{(\mu}\xi_{\nu)},\phantom{\Big|}
\end{alignat}
where $h_{\mu\nu}^\mr{Kerr}=\mc P_{\mu\nu\alpha\beta}\bar h^{\alpha\beta}_\mr{Kerr}$ with
\be\label{hharmK}
\bar h^{\mu\nu}_\mr{Kerr}=\frac{4m\tilde r}{\Sigma}\left(u^\mu u^\nu+u^{(\mu}\epsilon^{\nu)}{}_{\rho\alpha\beta}u^\rho\frac{r^\alpha a^\beta}{\tilde r^2+a^2}\right),
\ee
and
\be\label{xiK}
\xi_\mu=-m\ln(\tilde r^2+a^2)u_\mu-m\tilde n_\mu.
\ee
The metric perturbation $h^\mr{Kerr}_{\mu\nu}$ exactly solves the harmonic gauge condition (\ref{mcPgauge}) and the harmonic-gauge linearized vacuum field equation (\ref{mcPFE}) with the source $T^{\mu\nu}$ given by (\ref{Tpad}) with (\ref{hatTKerr}), with $z^\mu(t)=u^\mu t$ for the worldline of the origin.
The same arguments following 
(\ref{genLFE}) apply here to show that the Kerr metric (\ref{KSK}) is an exact solution of the full nonlinear vacuum Einstein equation away from $\tilde r=0$.

The expressions in terms of OS coordinates can be traded for expressions involving complex vectors, and then complex translation operators, as follows.  Note from (\ref{coordsumm}), (\ref{r3vec}), and (\ref{a3vec}) that
\begin{alignat}{3}
\tilde r+ia\cos\tilde\theta&=\sqrt{(\tilde r^2+a^2)\sin^2\tilde\theta+(\tilde r\cos\tilde\theta+ ia)^2}
\nnm\\
&=\sqrt{x^2+y^2+(z+ ia)^2}
\nnm\\
&=\sqrt{\bs r^2-\bs a^2+2i\bs r\cdot\bs a}
\nnm\\
&\equiv|\bs r+ i\bs a|,
\end{alignat}
where we take the branch of the square root which makes the real part positive, $\tilde r>0$.  It follows that
\begin{alignat}{3}\label{COS}
\frac{\tilde r}{\Sigma}=\Re\,\frac{1}{\tilde r+ia\cos\tilde\theta}
&=\Re\,\frac{1}{|\bs r+i\bs a|}
\nnm\\
&=\Re\,\exp(i\bs a\cd\bs\nabla)\frac{1}{r}
\nnm\\
&=\cos(\bs a\cd\bs\nabla)\frac{1}{r}.
\end{alignat}
Similarly, using (\ref{coordsumm}) in the first equality,
\begin{alignat}{3}\label{SIN}
\frac{\tilde r}{\Sigma}\,\frac{\bs r\times\bs a}{\tilde r^2+a^2}
&=\Re\,\frac{z+ia}{2ia\rho^2}\, \frac{\bs r\times\bs a}{|\bs r+i\bs a|}
\nnm\\\nnm
&=\sin(\bs a\cd\bs\nabla)\frac{z\,\bs r\times\bs a}{a\rho^2r}
\\
&=\bs a\times\bs\nabla\frac{\sin(\bs a\cd\bs\nabla)}{\bs a\cd\bs\nabla}\frac{1}{r}
\\\nnm
&=\sinh(\bs a\times\bs\nabla)\frac{1}{r},
\end{alignat}
since \mbox{$\bs a\times\bs \nabla(1/r)$} = \mbox{$\bs r\times\bs a/r^3$} = \mbox{$\bs a\cd\bs\nabla((z/a\rho^2r)\bs r\times\bs a)$}, noting that \mbox{$\bs a\cd\bs\nabla$} = \mbox{$a(\doe/\doe z)_{x,y}$} and \mbox{$\doe_ia_j=0$}.  In the last line, it is understood that products of pairs of the vector operator \mbox{$\bs a\times\bs\nabla$} are contracted to become \mbox{$(\bs a\times\bs\nabla)^2=\bs a^2\bs \nabla^2-(\bs a\cd\bs\nabla)^2$}, where we can drop the first term since $\bs\nabla^2(1/r)=0$ away from $r=0$.  With (\ref{COS})--(\ref{SIN}), \mbox{$(\bs r\times\bs a)_idx_i=u^\nu\epsilon_{\nu\mu\alpha\beta}r^\alpha a^\beta$}, and \mbox{$\bs a\cd\bs\nabla\, f(x^i)=a\cd\doe \,f(x^i)$}, the harmonic-gauge linearized metric perturbation (\ref{hharmK}) can be written as
\begin{alignat}{3}\label{hKfin}
\bar h^{\mu\nu}_\mr{Kerr}&=\bigg(u^\mu u^\nu\cos(a\cd\doe)
\\\nnm
&\quad+u^{(\mu}\epsilon^{\nu)}{}_{\rho\alpha\beta}u^\rho a^\alpha\doe^\beta\frac{\sin(a\cd\doe)}{a\cd\doe}\bigg)\frac{4m}{r},
\end{alignat}
which is as in (\ref{hbarKerr})--(\ref{hexp}).

\section{Generally covariant worldline-skeleton effective action principle}\label{app:gencov}

%As discussed above, it will be sufficient for a treatment of arbitrary-mass-ratio 1PM scattering to consider the motion test bodies in given background spacetimes.  We thus begin by considering an effective worldline action functional $\mc S_\mr{wl}$ which encodes the dynamics of a spinning extended test body in a curved background, specializing here to the case of a body with only spin-induced multipole moments.  The action approach (including the body's full multipole series) dates back to \cite{Bailey1975}, following work on the multipole expansion of the equations of motion culminating in \cite{Dixon:1979}, and it has subsequently been developed and reviewed by several authors, e.g.\ in \cite{Porto:2005ac,Steinhoff:2014,Marsat:2014xea,Levi:Steinhoff:2015:1,Porto:2016pyg}, to which we refer the reader for many further details.

Under the no-hair assumption, that a localized body has only translational and rotational degrees of freedom, it is natural to take these to be an arbitrarily parametrized timelike worldline $x=z(\sigma)$ with tangent $\dot z^\mu=dz^\mu/d\sigma$ and a body-fixed orthonormal frame field, or tetrad, $\Lambda_a{}^\mu(\sigma)$ along the worldline.   The tetrad satisfies $g_{\mu\nu}\Lambda_{a}{}^\mu\Lambda_{ b}{}^\nu=\eta_{a b}$, where $g_{\mu\nu}$ is the spacetime metric (evaluated at $x=z$) and $\eta_{ab}$ is the frame Minkowski metric, and we define from $\Lambda_a{}^\mu$ the angular velocity tensor
\be
\Omega^{\mu\nu}=\eta^{ab}\Lambda_a{}^{\mu}\frac{D\Lambda_{b}{}^{\nu}}{d\sigma},
\ee
%As in \cite{Porto:2005ac}, one considers a worldline action functional $\mc S_\mr{wl}=\int d\sigma\,\mc L(z,\dot z,\Omega)[g]$, the form of which is determined by appropriate symmetries.  The fact that the Lagrangian $\mc L$ (assumed first-order in $\sigma$-derivatives) can depend on $\Lambda_a{}^\mu$ and $D\Lambda_a{}^\mu/d\sigma$ only through $\Omega^{\mu\nu}$ is implied by invariance under (global) internal Lorentz transformations of the body-fixed tetrad, $\Lambda_a{}^\mu\to L_a{}^b\Lambda_b{}^\mu$; effectively, all frame indices appear only on $\Lambda$ and $D\Lambda/d\sigma$ and must all be contracted, and noting $\Lambda_a{}^\mu\Lambda^{a\nu}=g^{\mu\nu}$ and $(D\Lambda_a{}^\mu/d\sigma)(D\Lambda^{a\nu}/d\sigma)=\Omega^{\mu}{}_\rho\Omega^{\nu\rho}$ leaves only the combination (\ref{Omega}).  General covariance (and the assumption that the body interacts only gravitationally with its environment) implies that $\mc L$ can depend on the worldline point $z$ only through the metric and invariant curvature tensors evaluated at $z$.  Finally, the action should be invariant under reparametrizations of the worldline.
satisfying $\Omega^{\mu\nu}=\Omega^{[\mu\nu]}$.
The form of a worldline action functional $\mc S_\mr{wl}$ governing the body's (assumed purely gravitational) dynamics is determined by appropriate symmetries:  general covariance, worldline reparametrization invariance, and invariance under (global) internal Lorentz transformations of the body-fixed tetrad \cite{Porto:2005ac,Steinhoff:2014,Marsat:2014xea,Levi:Steinhoff:2015:1,Porto:2016pyg}.
We can work with the following phase-space action, having undergone Legendre transformations in $\dot z^\mu$ and $\Omega^{\mu\nu}$ \cite{Steinhoff:2014,Porto:2016pyg},
\begin{align}\label{Agen_action}
&\mc S_\mr{wl}[z,p,\Lambda,S,g]=
\\\nnm
&\quad\int d\sigma\bigg[p_\mu\dot z^\mu+\frac{1}{2}S_{\mu\nu}\Omega^{\mu\nu}
-\chi^\mu S_{\mu\nu} p^\nu-\frac{\lambda}{2}(p^2+\mc M^2)\bigg],
\end{align}
where $p_\mu(\sigma)$ is the linear momentum and $S_{\mu\nu}(\sigma)$ is the antisymmetric spin tensor, conjugate to the worldline $z(\sigma)$ and the tetrad $\Lambda_a{}^\mu(\sigma)$ respectively, all of which are to be independently varied along with the two Lagrange multipliers $\chi^\mu(\sigma)$ and $\lambda(\sigma)$.  The two constraints enforced by $\chi^\mu$ and $\lambda$ are respectively the covariant SSC, 
\be\label{AcovSSC}
p_\mu S^{\mu\nu}=0,
\ee
giving a special case of the ``spin gauge constraint'' discussed in \cite{Steinhoff:2015}, and the ``mass-shell constraint'' 
\be\label{AMSC}
p^2=-\mc M^2(p,S,z).
\ee
While all other ingredients of the action are kinematical, the ``dynamical mass function'' $\mc M^2(p,S,z)$ encodes the dynamics.  Symmetries dictate that it should depend on the worldline point $z$ only through the metric and invariant curvature tensors evaluated at $z$, 
%and a  fully general form {\color{red} (see further discussion in Appendix \ref{app:var})} for $\mc M^2$ which is linear in the curvature and its derivatives is
\be\label{AmcMgen}
\mc M^2(p,S,z)=\mc M^2\Big(p_\mu,S_{\mu\nu},g_{\mu\nu}(z),\big\{R_{\mu\nu\rho\sigma;(N)}(z)\big\}_{n=0}^\infty\Big).
\ee
Here, the multi-index $N$ represents the string $\alpha_1\ldots\alpha_n$ of $n$ (ordered) spacetime indices, and $R_{\mu\nu\rho\sigma;(N)}$ is the $n$th symmetrized covariant derivative of the Riemann tensor.  Our sign convention for the Riemann tensor is fixed by the form $2\nabla_{[a}\nabla_{b]}w_c=R_{abc}{}^dw_d$ of the Ricci identity.  We need only consider symmetrized derivatives of the Riemann tensor because any antisymmetric parts can be eliminated by use of the Ricci identity.

Variation of the action (\ref{Agen_action}) with respect to the fields $\{z,p,\Lambda,S,\chi,\lambda\}$ along the worldline (which can be accomplished e.g.\ precisely as in Appendix A of \cite{VKSH}, with similar calculations discussed e.g. in \cite{Bailey1975,Blanchet:2013haa,Steinhoff:2014,Marsat:2014xea}),
 leads to a form of the MPD equations,
\begin{align}\label{AMPDp}
\frac{Dp_\mu}{d\sigma}&=-\frac{1}{2}R_{\mu\nu\alpha\beta}\dot z^\nu S^{\alpha\beta}-\frac{\lambda}{2}\frac{\mc D\mc M^2}{\doe z^\mu},\phantom{\bigg|}
\\\label{AMPDS}
\frac{DS^{\mu\nu}}{d\sigma}&=2p^{[\mu}\dot z^{\nu]}-\lambda
\left(p^{[\mu}\frac{\partial\mathcal M^2}{\partial p_{\nu]}}
+2S^{[\mu}{}_\alpha\frac{\partial\mathcal M^2}{\partial S_{\nu]\alpha}}\right),
\end{align}
along with the covariant SSC (\ref{AcovSSC}) and, for our purposes,
\be\label{Azdot}
\dot z^\mu=\lambda p^\mu+{O}(R),\quad \lambda=\dfrac{p\cd\dot z}{p^2}=\frac{\sqrt{-\dot z^2}}{\mc M}+{O}(R),
\ee
where ${O}(R)$ denotes terms with one or more explicit factors of the curvature and its derivatives.  
In (\ref{AMPDp}), $\mc D/\doe z^\mu$ denotes the ``horizontal'' derivative \cite{Dixon:1979} which covariantly differentiates with respect to $z$ while parallel transporting $p$ and $S$.\footnote{
Given (\ref{AmcMgen}), we have the explicit expression
\be\label{Ahoriz}
\frac{\mc D\mc M^2}{\doe z^\mu}=\sum_{n=0}^\infty\frac{\doe\mc M^2}{\doe R_{\alpha\beta\gamma\delta;(N)}}R_{\alpha\beta\gamma\delta;(N)\mu},
\ee
with no term corresponding to the dependence of $\mc M^2$ on $g_{\mu\nu}(z)$.
}
%Equation (\ref{AMPDS}) can also be written explicitly in terms of the multipole and curvature tensors, as in (\ref{Atorque_A}), via the condition (\ref{Ascalar}) that $\mc M^2$ be a scalar  [without knowledge of the multipoles $J(p,S)$]. 

With $S^{\mu\nu}=\epsilon^{\mu\nu}{}_{\alpha\beta}p^\alpha a^\beta$, as in (\ref{acovS}), (\ref{AMPDS}) and (\ref{Azdot}) yield
\be\label{Asdot}
\frac{Da^\mu}{d\sigma}=\frac{\sqrt{-\dot z^2}}{2\mc M^3}\epsilon^{\mu\nu}{}_{\alpha\beta}p^\alpha a^\beta \frac{\doe\mc M^2}{\doe a^\nu}+\frac{p^\mu}{\mc M^2}a^\nu\frac{Dp_\nu}{d\sigma}.
\ee
This together with (\ref{AMPDp}) and (\ref{Azdot}) become (\ref{MPD1PMcov})--(\ref{puzd}) when specialized to a 1PM background spacetime, with
\begin{alignat}{3}
\mc L_\mr{int}&=\frac{m}{2}\Big(u^\mu-\epsilon^\mu{}_{\rho\alpha\beta}u^\rho a^\alpha\doe^\beta\Big)u^\nu h_{\mu\nu}
\nnm\\
&\quad-\frac{\mc M^2-m^2}{2m}+O(G^2),
\end{alignat}
using $\mc M^2=m^2+O(G)$ with $m$ constant, and neglecting total derivatives in the action.  The monopole and dipole couplings in the first line here arise from expanding the covariant derivatives in (\ref{AMPDp}) and (\ref{Asdot}), with 
\be\label{Gamma1PM}
\Gamma^\mu{}_{\alpha\beta}=\doe_{(\alpha}h_{\beta)}{}^\mu-\frac{1}{2}\doe^\mu h_{\alpha\beta}+{O}(G^2),
\ee
and from the Riemann tensor term in (\ref{AMPDp}), with 
\be\label{R1PM}
R^{\mu\nu}{}_{\alpha\beta}=-2\doe^{[\mu}\doe_{[\alpha}h^{\nu]}{}_{\beta]}+O(G^2).
\ee  The quadrupole and higher-order terms are the ``non-minimal couplings'' given by $\mc M^2-m^2$, containing the Riemann tensor and its derivatives.  The form of $\mc M^2$ appropriate for a no-hair body to all orders in the spin-multipole expansion at linear order in curvature (linear order in the Riemann tensor and its derivatives) is given in Eq.~(4.16) of \cite{Levi:Steinhoff:2015:1}, with $L_\mr{SI}=-(\mc M^2-m^2)/2m+{O}(R^2)$.  With all of the $C$ coefficients in \cite{Levi:Steinhoff:2015:1} set to 1 for a BH, specializing to a 1PM background with (\ref{Gamma1PM})--(\ref{R1PM}), the interaction terms of \cite{Levi:Steinhoff:2015:1}'s action become our Eqs.~(\ref{SLint}) with (\ref{hatTKerr}).

Variation of $\mc S_\mr{wl}$ with respect to the metric leads to the effective stress-energy tensor
\begin{alignat}{3}
T^{\mu\nu}(x)&=\int d\sigma\,\Bigg\{p^{(\mu}\dot z^{\nu)}\delta_4(x,z)-\nabla_\alpha\Big[S^{\alpha(\mu}\dot z^{\nu)}\delta_4(x,z)\Big]
\nnm\\\nnm
&\quad-\frac{\lambda}{2}\sum_{n=0}^\infty\frac{\doe\mc M^2}{\doe R_{\alpha\beta\gamma\delta;(N)}}
\Bigg[\dw^{(\mu\nu)}R_{\alpha\beta\gamma\delta;(N)}\delta_4(x,z)
\\\label{Asetgen}
&\qquad\qquad\qquad+\frac{2}{\sqrt{-g}}\frac{\delta R_{\alpha\beta\gamma\delta;(N)}(z)}{\delta g_{\mu\nu}(x)}\Bigg]\Bigg\}, 
\end{alignat}
where $\delta_4(x,z)=\delta^4(x-z)/\sqrt{-g}$, $\dw^\alpha{}_\beta$ is DeWitt's index-suffling operator \cite{DeWitt:1965jb,Steinhoff:2014} such that 
$\nabla_\mu=\doe_\mu+\Gamma^\beta{}_{\alpha\mu}\dw^\alpha{}_\beta
$,
and the last line contains the functional derivatives of the Riemann tensor derivatives with respect to the metric.\footnote{
The variation with respect to $g_{\mu\nu}$ proceeds mostly as described e.g.\ in \cite{Blanchet:2013haa,Steinhoff:2014,Marsat:2014xea}.  The only necessary ingredient not explicitly given in \cite{Steinhoff:2014} is the extension of the ``scalar condition'' (for $\mc M^2$ here, for the Lagrangian in those references) to include all symmetrized derivatives of the curvature.  This can be derived by temporarily treating $p$ and $S$ as fields
in $\mc M^2(p(x),S(x),g(x),\{R(x)\})$, and then expanding $\mc L_\xi\mc M^2=\xi^\mu\nabla_\mu\mc M^2$, holding for an arbitrary vector field $\xi$, using 
$
\mc L_\xi=\xi^\alpha\nabla_\alpha-(\nabla_\alpha\xi^\beta)\dw^\alpha{}_\beta
$
%\be
%\dw^\alpha{}_\beta T^\mu{}_\nu=\delta^\mu{}_\beta T^\alpha{}_\nu-\delta^\alpha{}_\nu T^\mu{}_\beta,
%\ee
for the Lie derivatives, yielding
\begin{alignat}{3}\label{Ascalar}
0&=2\frac{\doe \mc M^2}{\doe g_{\mu\nu}}+p^\mu\frac{\doe \mc M^2}{\doe p_\nu}+2S^\mu{}_\alpha\frac{\doe \mc M^2}{\doe S_{\nu\alpha}}
\\\nnm
&\quad-\sum_{n=0}^\infty\Big(\mc G^{\mu\nu} R_{\alpha\beta\gamma\delta;(N)}\Big)\frac{\doe \mc M^2}{\doe R_{\alpha\beta\gamma\delta;(N)}}.
\end{alignat}
}  

Importantly, it can be verified that the equations of motion (\ref{AMPDp})--(\ref{AMPDS}) imply the conservation this stress-energy tensor according to $\nabla_\mu T^{\mu\nu}=0$.
This can be accomplished e.g.\ by showing that (\ref{AMPDp})--(\ref{AMPDS}) imply
\be
\int d^4x\sqrt{-g}\,\psi_\nu\nabla_\mu T^{\mu\nu}=0,
\ee
for any vector field $\psi$ with compact support including the body's worldline, using integrations by parts.  A key observation is that
\be
\frac{\delta R_{\alpha\beta\gamma\delta;(N)}(z)}{\delta g_{\mu\nu}(x)}2\nabla_{\mu}\psi_{\nu}(x)
=\delta^4(x-z)\mc L_\psi R_{\alpha\beta\gamma\delta;(N)}(x),
\ee
which follows from the fact that a variation of the metric given by $\delta g_{\mu\nu}=\mc L_\psi g_{\mu\nu}=2\nabla_{(\mu}\psi_{\nu)}$ corresponds simply to a linearized coordinate transformation, thus leaving the curvature tensors invariant (apart from being Lie-dragged).

\py

\section{On the reduced action}\label{app:Sred}

Continuing the discussion at the end of Sec.~\ref{sec:linint}:

The total action (\ref{mcStot}) for a two-BH system takes the form
\begin{alignat}{3}
\mc S_\mr{tot}&=\mc S_\mr{grav}[h_1+h_2]+\mc S_\mr{kin}[\Psi_1]+\mc S_\mr{kin}[\Psi_2]
\\\nnm
&\quad+\mc S_\mr{int}[\Psi_1,h_1+h_2]+\mc S_\mr{int}[\Psi_2,h_1+h_2].\phantom{\Big|}
\end{alignat}
With $\mc S_\mr{grav}$ (\ref{mcSgrav}) being symmetrically quadratic in $h$, $S_\mr{grav}[h]=\mc S_\mr{g}[h,h]$, we have
\be
S_\mr{grav}[h_1+h_2]=\mc S_\mr{g}[h_1,h_1]+\mc S_\mr{g}[h_2,h_2]
+2\mc S_\mr{g}[h_1,h_2],
\ee
and with $\mc S_\mr{int}$ being linear in $h$,
\be
\mc S_\mr{int}[\Psi_\ms{1},h_1+h_2]=\mc S_\mr{int}[\Psi_\ms{1},h_1]+\mc S_\mr{int}[\Psi_1,h_2],
\ee
sim.\ $\Psi_1\to\Psi_2$.  We ``integrate out'' the field by solving for $h_1[\Psi_1]$ and $h_2[\Psi_2]$, and ``renormalize'' by dropping the divergent 1--1 and 2--2 terms, and obtain the reduced action
\begin{alignat}{3}
\mc S^\mr{red}_\mr{tot}&=2\mc S_\mr{g}[h_1,h_2]+\mc S_\mr{kin}[\Psi_1]+\mc S_\mr{kin}[\Psi_2]
\\\nnm
&\quad+\mc S_\mr{int}[\Psi_1,h_2]+\mc S_\mr{int}[\Psi_2,h_1],
\end{alignat}
with $h_\ms{a}=h_\ms{a}[\Psi_\ms{a}]$ understood.  

It turns out to be the case for the 1PM two-BH system that
\be\label{actionsymms}
\mc S_\mr{int}[\Psi_1,h_2]\;\dot{=}\;\mc S_\mr{int}[\Psi_2,h_1]\;\dot{=}\;-2\mc S_\mr{g}[h_1,h_2],
\ee
neglecting $O(G^2)$ and total derivative terms.  Thus,
\be
\mc S^\mr{red}_\mr{tot}\;\dot{=}\;\mc S_\mr{kin}[\Psi_1]+\mc S_\mr{kin}[\Psi_2]
+\mc S_\mr{int}[\Psi_1,h_2],
\ee
which leads to the conclusions in the paragraph following (\ref{Lint12}).

While the property given by the first equality of (\ref{actionsymms}), e.g., is not manifest in the form of the the result for $
\mc S_\mr{int}[\Psi_1,h_2]=\int d\tau_1\,\mc L_\mr{int}
$ from (\ref{Lint12}), it can be verfied as follows.  Returning to (\ref{SintL}) and (\ref{hmunuhatT}), after some manipulation, with (\ref{mcGex}), one finds
\begin{alignat}{3}\label{L12sym}
&\mc S_\mr{int}[\Psi_1,h_2]=\frac{1}{2}\int d^4x\;T_1^{\mu\nu}(x)\;h_{2\mu\nu}(x)
\\\nnm
&=2G\,\mc P_{\mu\nu\alpha\beta}\int d\tau_1\int d\tau_2
\\\nnm
&\qquad\;\;\hat{\mc T}^{\mu\nu}(p_1,a_1,\doe_1)\,\hat{\mc T}^{\alpha\beta}(p_2,a_2,\doe_2)\,\mc G(z_1,z_2),
\end{alignat}
where $\doe_\ms{a}=\doe/\doe z_\ms{a}$.  We should have $\mc G\to\mc G_\mr{ret}$ here, as in (\ref{hmunuhatT}).  However, if both BHs are taken in the zeroth-order states, with their worldlines being Minkowski geodesics---as can always be done in using this interaction integral (and its derivatives) at 1PM order, dropping $O(G^2)$ corrections---then $\mc G_\mr{ret}$ can replaced with $\mc G_\mr{sym}$, as in (\ref{rzx}) where $1/r=\int d\tau\, \mc G_\mr{ret}=\int d\tau\,\mc G_\mr{sym}$.  Then, with $\mc G(z_1,z_2)\to\mc G_\mr{sym}(z_1,z_2)=\mc G_\mr{sym}(z_2,z_1)$ in (\ref{L12sym}), the symmetry under $1\leftrightarrow 2$ is manifest,
\be
\mc S_\mr{int}[\Psi_1,h_2]\,\;\dot{=}\;\,\mc S_\mr{int}[\Psi_2,h_1],
\ee
neglecting $O(G^2)$ and total derivative terms, with $\Psi_\ms{a}=(z_\ms{a},p_\ms{a},a_\ms{a})$.

%\addtocontents{toc}{\endgroup}

%\bibliographystyle{utphys}
%\bibliography{testspin}

%\bibliographystyle{apsrev4-1.bst}
%\bibliography{testspin}

\begin{thebibliography}{104}%
\makeatletter
\providecommand \@ifxundefined [1]{%
 \@ifx{#1\undefined}
}%
\providecommand \@ifnum [1]{%
 \ifnum #1\expandafter \@firstoftwo
 \else \expandafter \@secondoftwo
 \fi
}%
\providecommand \@ifx [1]{%
 \ifx #1\expandafter \@firstoftwo
 \else \expandafter \@secondoftwo
 \fi
}%
\providecommand \natexlab [1]{#1}%
\providecommand \enquote  [1]{``#1''}%
\providecommand \bibnamefont  [1]{#1}%
\providecommand \bibfnamefont [1]{#1}%
\providecommand \citenamefont [1]{#1}%
\providecommand \href@noop [0]{\@secondoftwo}%
\providecommand \href [0]{\begingroup \@sanitize@url \@href}%
\providecommand \@href[1]{\@@startlink{#1}\@@href}%
\providecommand \@@href[1]{\endgroup#1\@@endlink}%
\providecommand \@sanitize@url [0]{\catcode `\\12\catcode `\$12\catcode
  `\&12\catcode `\#12\catcode `\^12\catcode `\_12\catcode `\%12\relax}%
\providecommand \@@startlink[1]{}%
\providecommand \@@endlink[0]{}%
\providecommand \url  [0]{\begingroup\@sanitize@url \@url }%
\providecommand \@url [1]{\endgroup\@href {#1}{\urlprefix }}%
\providecommand \urlprefix  [0]{URL }%
\providecommand \Eprint [0]{\href }%
\providecommand \doibase [0]{http://dx.doi.org/}%
\providecommand \selectlanguage [0]{\@gobble}%
\providecommand \bibinfo  [0]{\@secondoftwo}%
\providecommand \bibfield  [0]{\@secondoftwo}%
\providecommand \translation [1]{[#1]}%
\providecommand \BibitemOpen [0]{}%
\providecommand \bibitemStop [0]{}%
\providecommand \bibitemNoStop [0]{.\EOS\space}%
\providecommand \EOS [0]{\spacefactor3000\relax}%
\providecommand \BibitemShut  [1]{\csname bibitem#1\endcsname}%
\let\auto@bib@innerbib\@empty
%</preamble>
\bibitem [{\citenamefont {Abbott}\ \emph
  {et~al.}(2016{\natexlab{a}})\citenamefont {Abbott} \emph
  {et~al.}}]{Abbott:2016blz}%
  \BibitemOpen
  \bibfield  {author} {\bibinfo {author} {\bibfnamefont {B.~P.}\ \bibnamefont
  {Abbott}} \emph {et~al.} (\bibinfo {collaboration} {Virgo, LIGO
  Scientific}),\ }\href {\doibase 10.1103/PhysRevLett.116.061102} {\bibfield
  {journal} {\bibinfo  {journal} {Phys. Rev. Lett.}\ }\textbf {\bibinfo
  {volume} {116}},\ \bibinfo {pages} {061102} (\bibinfo {year}
  {2016}{\natexlab{a}})},\ \Eprint {http://arxiv.org/abs/1602.03837}
  {arXiv:1602.03837 [gr-qc]} \BibitemShut {NoStop}%
%%CITATION = ARXIV:1602.03837;%%
\bibitem [{\citenamefont {Abbott}\ \emph
  {et~al.}(2016{\natexlab{b}})\citenamefont {Abbott} \emph
  {et~al.}}]{Abbott:2016nmj}%
  \BibitemOpen
  \bibfield  {author} {\bibinfo {author} {\bibfnamefont {B.~P.}\ \bibnamefont
  {Abbott}} \emph {et~al.} (\bibinfo {collaboration} {Virgo, LIGO
  Scientific}),\ }\href {\doibase 10.1103/PhysRevLett.116.241103} {\bibfield
  {journal} {\bibinfo  {journal} {Phys. Rev. Lett.}\ }\textbf {\bibinfo
  {volume} {116}},\ \bibinfo {pages} {241103} (\bibinfo {year}
  {2016}{\natexlab{b}})},\ \Eprint {http://arxiv.org/abs/1606.04855}
  {arXiv:1606.04855 [gr-qc]} \BibitemShut {NoStop}%
%%CITATION = ARXIV:1606.04855;%%
\bibitem [{\citenamefont {Abbott}\ \emph
  {et~al.}(2017{\natexlab{a}})\citenamefont {Abbott} \emph
  {et~al.}}]{Abbott:2017vtc}%
  \BibitemOpen
  \bibfield  {author} {\bibinfo {author} {\bibfnamefont {B.~P.}\ \bibnamefont
  {Abbott}} \emph {et~al.} (\bibinfo {collaboration} {VIRGO, LIGO
  Scientific}),\ }\href {\doibase 10.1103/PhysRevLett.118.221101} {\bibfield
  {journal} {\bibinfo  {journal} {Phys. Rev. Lett.}\ }\textbf {\bibinfo
  {volume} {118}},\ \bibinfo {pages} {221101} (\bibinfo {year}
  {2017}{\natexlab{a}})},\ \Eprint {http://arxiv.org/abs/1706.01812}
  {arXiv:1706.01812 [gr-qc]} \BibitemShut {NoStop}%
%%CITATION = ARXIV:1706.01812;%%
\bibitem [{\citenamefont {Abbott}\ \emph
  {et~al.}(2017{\natexlab{b}})\citenamefont {Abbott} \emph
  {et~al.}}]{Abbott:2017oio}%
  \BibitemOpen
  \bibfield  {author} {\bibinfo {author} {\bibfnamefont {B.~P.}\ \bibnamefont
  {Abbott}} \emph {et~al.} (\bibinfo {collaboration} {Virgo, LIGO
  Scientific}),\ }\href@noop {} {\bibfield  {journal} {\bibinfo  {journal}
  {Submitted to: Phys. Rev. Lett.}\ } (\bibinfo {year} {2017}{\natexlab{b}})},\
  \Eprint {http://arxiv.org/abs/1709.09660} {arXiv:1709.09660 [gr-qc]}
  \BibitemShut {NoStop}%
%%CITATION = ARXIV:1709.09660;%%
\bibitem [{\citenamefont {{Poisson}}\ \emph {et~al.}(2011)\citenamefont
  {{Poisson}}, \citenamefont {{Pound}},\ and\ \citenamefont
  {{Vega}}}]{Poisson:Pound:Vega:2011}%
  \BibitemOpen
  \bibfield  {author} {\bibinfo {author} {\bibfnamefont {E.}~\bibnamefont
  {{Poisson}}}, \bibinfo {author} {\bibfnamefont {A.}~\bibnamefont {{Pound}}},
  \ and\ \bibinfo {author} {\bibfnamefont {I.}~\bibnamefont {{Vega}}},\ }\href
  {\doibase 10.12942/lrr-2011-7} {\bibfield  {journal} {\bibinfo  {journal}
  {Living Reviews in Relativity}\ }\textbf {\bibinfo {volume} {14}},\ \bibinfo
  {pages} {7} (\bibinfo {year} {2011})},\ \Eprint
  {http://arxiv.org/abs/1102.0529} {arXiv:1102.0529 [gr-qc]} \BibitemShut
  {NoStop}%
\bibitem [{\citenamefont {Harte}(2015)}]{Harte:review}%
  \BibitemOpen
  \bibfield  {author} {\bibinfo {author} {\bibfnamefont {A.~I.}\ \bibnamefont
  {Harte}},\ }\bibfield  {booktitle} {\emph {\bibinfo {booktitle}
  {{Proceedings, 524th WE-Heraeus-Seminar: Equations of Motion in Relativistic
  Gravity (EOM 2013)}}},\ }\href {\doibase 10.1007/978-3-319-18335-0_12}
  {\bibfield  {journal} {\bibinfo  {journal} {Fund. Theor. Phys.}\ }\textbf
  {\bibinfo {volume} {179}},\ \bibinfo {pages} {327} (\bibinfo {year}
  {2015})},\ \Eprint {http://arxiv.org/abs/1405.5077} {arXiv:1405.5077 [gr-qc]}
  \BibitemShut {NoStop}%
%%CITATION = ARXIV:1405.5077;%%
\bibitem [{\citenamefont {Barack}(2014)}]{Barack:2014}%
  \BibitemOpen
  \bibfield  {author} {\bibinfo {author} {\bibfnamefont {L.}~\bibnamefont
  {Barack}},\ }in\ \href {\doibase 10.1007/978-3-319-06349-2_6} {\emph
  {\bibinfo {booktitle} {General Relativity, Cosmology and Astrophysics}}},\
  \bibinfo {series} {Fundamental Theories of Physics}, Vol.\ \bibinfo {volume}
  {177}\ (\bibinfo  {publisher} {Springer International Publishing},\ \bibinfo
  {year} {2014})\ pp.\ \bibinfo {pages} {147--168}\BibitemShut {NoStop}%
\bibitem [{\citenamefont {Pound}(2015)}]{Pound:review}%
  \BibitemOpen
  \bibfield  {author} {\bibinfo {author} {\bibfnamefont {A.}~\bibnamefont
  {Pound}},\ }\bibfield  {booktitle} {\emph {\bibinfo {booktitle}
  {{Proceedings, 524th WE-Heraeus-Seminar: Equations of Motion in Relativistic
  Gravity (EOM 2013): Bad Honnef, Germany, February 17-23, 2013}}},\ }\href
  {\doibase 10.1007/978-3-319-18335-0_13} {\bibfield  {journal} {\bibinfo
  {journal} {Fund. Theor. Phys.}\ }\textbf {\bibinfo {volume} {179}},\ \bibinfo
  {pages} {399} (\bibinfo {year} {2015})},\ \Eprint
  {http://arxiv.org/abs/1506.06245} {arXiv:1506.06245 [gr-qc]} \BibitemShut
  {NoStop}%
%%CITATION = ARXIV:1506.06245;%%
\bibitem [{\citenamefont {Futamase}\ and\ \citenamefont
  {Itoh}(2007)}]{Futamase:2007zz}%
  \BibitemOpen
  \bibfield  {author} {\bibinfo {author} {\bibfnamefont {T.}~\bibnamefont
  {Futamase}}\ and\ \bibinfo {author} {\bibfnamefont {Y.}~\bibnamefont
  {Itoh}},\ }\href@noop {} {\bibfield  {journal} {\bibinfo  {journal} {Living
  Rev. Rel.}\ }\textbf {\bibinfo {volume} {10}},\ \bibinfo {pages} {2}
  (\bibinfo {year} {2007})}\BibitemShut {NoStop}%
%%CITATION = 00222,10,2;%%
\bibitem [{\citenamefont {Schaefer}(2011)}]{Schafer:2009dq}%
  \BibitemOpen
  \bibfield  {author} {\bibinfo {author} {\bibfnamefont {G.}~\bibnamefont
  {Schaefer}},\ }\bibfield  {booktitle} {\emph {\bibinfo {booktitle} {{Mass and
  motion in general relativity. Proceedings, School on Mass, Orleans, France,
  June 23-25, 2008}}},\ }\href@noop {} {\bibfield  {journal} {\bibinfo
  {journal} {Fundam. Theor. Phys.}\ }\textbf {\bibinfo {volume} {162}},\
  \bibinfo {pages} {167} (\bibinfo {year} {2011})},\ \bibinfo {note}
  {[,167(2009)]},\ \Eprint {http://arxiv.org/abs/0910.2857} {arXiv:0910.2857
  [gr-qc]} \BibitemShut {NoStop}%
%%CITATION = ARXIV:0910.2857;%%
\bibitem [{\citenamefont {Blanchet}(2014)}]{Blanchet:2013haa}%
  \BibitemOpen
  \bibfield  {author} {\bibinfo {author} {\bibfnamefont {L.}~\bibnamefont
  {Blanchet}},\ }\href {\doibase 10.12942/lrr-2014-2} {\bibfield  {journal}
  {\bibinfo  {journal} {Living Rev. Rel.}\ }\textbf {\bibinfo {volume} {17}},\
  \bibinfo {pages} {2} (\bibinfo {year} {2014})},\ \Eprint
  {http://arxiv.org/abs/1310.1528} {arXiv:1310.1528 [gr-qc]} \BibitemShut
  {NoStop}%
%%CITATION = ARXIV:1310.1528;%%
\bibitem [{\citenamefont {Poisson}\ and\ \citenamefont
  {Will}(2014)}]{poisson2014gravity}%
  \BibitemOpen
  \bibfield  {author} {\bibinfo {author} {\bibfnamefont {E.}~\bibnamefont
  {Poisson}}\ and\ \bibinfo {author} {\bibfnamefont {C.}~\bibnamefont {Will}},\
  }\href {https://books.google.de/books?id=PZ5cAwAAQBAJ} {\emph {\bibinfo
  {title} {Gravity: Newtonian, Post-Newtonian, Relativistic}}}\ (\bibinfo
  {publisher} {Cambridge University Press},\ \bibinfo {year}
  {2014})\BibitemShut {NoStop}%
\bibitem [{\citenamefont {Rothstein}(2014)}]{Rothstein:2014sra}%
  \BibitemOpen
  \bibfield  {author} {\bibinfo {author} {\bibfnamefont {I.~Z.}\ \bibnamefont
  {Rothstein}},\ }\href {\doibase 10.1007/s10714-014-1726-y} {\bibfield
  {journal} {\bibinfo  {journal} {Gen. Rel. Grav.}\ }\textbf {\bibinfo {volume}
  {46}},\ \bibinfo {pages} {1726} (\bibinfo {year} {2014})}\BibitemShut
  {NoStop}%
%%CITATION = GRGVA,46,1726;%%
\bibitem [{\citenamefont {Porto}(2016)}]{Porto:2016pyg}%
  \BibitemOpen
  \bibfield  {author} {\bibinfo {author} {\bibfnamefont {R.~A.}\ \bibnamefont
  {Porto}},\ }\href {\doibase 10.1016/j.physrep.2016.04.003} {\bibfield
  {journal} {\bibinfo  {journal} {Phys. Rept.}\ }\textbf {\bibinfo {volume}
  {633}},\ \bibinfo {pages} {1} (\bibinfo {year} {2016})},\ \Eprint
  {http://arxiv.org/abs/1601.04914} {arXiv:1601.04914 [hep-th]} \BibitemShut
  {NoStop}%
%%CITATION = ARXIV:1601.04914;%%
\bibitem [{\citenamefont {{Bertotti}}(1956)}]{Bertotti:1956}%
  \BibitemOpen
  \bibfield  {author} {\bibinfo {author} {\bibfnamefont {B.}~\bibnamefont
  {{Bertotti}}},\ }\href {\doibase 10.1007/BF02746175} {\bibfield  {journal}
  {\bibinfo  {journal} {Il Nuovo Cimento}\ }\textbf {\bibinfo {volume} {4}},\
  \bibinfo {pages} {898} (\bibinfo {year} {1956})}\BibitemShut {NoStop}%
\bibitem [{\citenamefont {{Bertotti}}\ and\ \citenamefont
  {{Plebanski}}(1960)}]{Bertotti:1960}%
  \BibitemOpen
  \bibfield  {author} {\bibinfo {author} {\bibfnamefont {B.}~\bibnamefont
  {{Bertotti}}}\ and\ \bibinfo {author} {\bibfnamefont {J.}~\bibnamefont
  {{Plebanski}}},\ }\href {\doibase 10.1016/0003-4916(60)90132-9} {\bibfield
  {journal} {\bibinfo  {journal} {Annals of Physics}\ }\textbf {\bibinfo
  {volume} {11}},\ \bibinfo {pages} {169} (\bibinfo {year} {1960})}\BibitemShut
  {NoStop}%
\bibitem [{\citenamefont {{Rosenblum}}(1978)}]{Rosenblum:1978}%
  \BibitemOpen
  \bibfield  {author} {\bibinfo {author} {\bibfnamefont {A.}~\bibnamefont
  {{Rosenblum}}},\ }\href {\doibase 10.1103/PhysRevLett.41.1140.2} {\bibfield
  {journal} {\bibinfo  {journal} {Physical Review Letters}\ }\textbf {\bibinfo
  {volume} {41}},\ \bibinfo {pages} {1140} (\bibinfo {year}
  {1978})}\BibitemShut {NoStop}%
\bibitem [{\citenamefont {{Bel}}\ \emph {et~al.}(1981)\citenamefont {{Bel}},
  \citenamefont {{Deruelle}}, \citenamefont {{Damour}}, \citenamefont
  {{Ibanez}},\ and\ \citenamefont {{Martin}}}]{Bel:1981}%
  \BibitemOpen
  \bibfield  {author} {\bibinfo {author} {\bibfnamefont {L.}~\bibnamefont
  {{Bel}}}, \bibinfo {author} {\bibfnamefont {N.}~\bibnamefont {{Deruelle}}},
  \bibinfo {author} {\bibfnamefont {T.}~\bibnamefont {{Damour}}}, \bibinfo
  {author} {\bibfnamefont {J.}~\bibnamefont {{Ibanez}}}, \ and\ \bibinfo
  {author} {\bibfnamefont {J.}~\bibnamefont {{Martin}}},\ }\href {\doibase
  10.1007/BF00756073} {\bibfield  {journal} {\bibinfo  {journal} {General
  Relativity and Gravitation}\ }\textbf {\bibinfo {volume} {13}},\ \bibinfo
  {pages} {963} (\bibinfo {year} {1981})}\BibitemShut {NoStop}%
\bibitem [{\citenamefont {{Damour}}\ and\ \citenamefont
  {{Deruelle}}(1981)}]{Damour:1981PhLA}%
  \BibitemOpen
  \bibfield  {author} {\bibinfo {author} {\bibfnamefont {T.}~\bibnamefont
  {{Damour}}}\ and\ \bibinfo {author} {\bibfnamefont {N.}~\bibnamefont
  {{Deruelle}}},\ }\href {\doibase 10.1016/0375-9601(81)90567-3} {\bibfield
  {journal} {\bibinfo  {journal} {Physics Letters A}\ }\textbf {\bibinfo
  {volume} {87}},\ \bibinfo {pages} {81} (\bibinfo {year} {1981})}\BibitemShut
  {NoStop}%
\bibitem [{\citenamefont {{Portilla}}(1979)}]{Portilla:1979}%
  \BibitemOpen
  \bibfield  {author} {\bibinfo {author} {\bibfnamefont {M.}~\bibnamefont
  {{Portilla}}},\ }\href {\doibase 10.1088/0305-4470/12/7/025} {\bibfield
  {journal} {\bibinfo  {journal} {Journal of Physics A Mathematical General}\
  }\textbf {\bibinfo {volume} {12}},\ \bibinfo {pages} {1075} (\bibinfo {year}
  {1979})}\BibitemShut {NoStop}%
\bibitem [{\citenamefont {Portilla}(1980)}]{Portilla:1980}%
  \BibitemOpen
  \bibfield  {author} {\bibinfo {author} {\bibfnamefont {M.}~\bibnamefont
  {Portilla}},\ }\href {http://stacks.iop.org/0305-4470/13/i=12/a=017}
  {\bibfield  {journal} {\bibinfo  {journal} {Journal of Physics A:
  Mathematical and General}\ }\textbf {\bibinfo {volume} {13}},\ \bibinfo
  {pages} {3677} (\bibinfo {year} {1980})}\BibitemShut {NoStop}%
\bibitem [{\citenamefont {Westpfahl}\ and\ \citenamefont
  {Goller}(1979)}]{Westpfahl1979}%
  \BibitemOpen
  \bibfield  {author} {\bibinfo {author} {\bibfnamefont {K.}~\bibnamefont
  {Westpfahl}}\ and\ \bibinfo {author} {\bibfnamefont {M.}~\bibnamefont
  {Goller}},\ }\href {\doibase 10.1007/BF02817047} {\bibfield  {journal}
  {\bibinfo  {journal} {Lettere al Nuovo Cimento (1971-1985)}\ }\textbf
  {\bibinfo {volume} {26}},\ \bibinfo {pages} {573} (\bibinfo {year}
  {1979})}\BibitemShut {NoStop}%
\bibitem [{\citenamefont {{Westpfahl}}(1985)}]{Westpfahl:1985}%
  \BibitemOpen
  \bibfield  {author} {\bibinfo {author} {\bibfnamefont {K.}~\bibnamefont
  {{Westpfahl}}},\ }\href {\doibase 10.1002/prop.2190330802} {\bibfield
  {journal} {\bibinfo  {journal} {Fortschritte der Physik}\ }\textbf {\bibinfo
  {volume} {33}},\ \bibinfo {pages} {417} (\bibinfo {year} {1985})}\BibitemShut
  {NoStop}%
\bibitem [{\citenamefont {{Westpfahl}}\ \emph {et~al.}(1987)\citenamefont
  {{Westpfahl}}, \citenamefont {{Mohles}},\ and\ \citenamefont
  {{Simonis}}}]{Wespfahl:1987}%
  \BibitemOpen
  \bibfield  {author} {\bibinfo {author} {\bibfnamefont {K.}~\bibnamefont
  {{Westpfahl}}}, \bibinfo {author} {\bibfnamefont {R.}~\bibnamefont
  {{Mohles}}}, \ and\ \bibinfo {author} {\bibfnamefont {H.}~\bibnamefont
  {{Simonis}}},\ }\href {\doibase 10.1088/0264-9381/4/5/006} {\bibfield
  {journal} {\bibinfo  {journal} {Classical and Quantum Gravity}\ }\textbf
  {\bibinfo {volume} {4}},\ \bibinfo {pages} {L185} (\bibinfo {year}
  {1987})}\BibitemShut {NoStop}%
\bibitem [{\citenamefont {{Ledvinka}}\ \emph {et~al.}(2008)\citenamefont
  {{Ledvinka}}, \citenamefont {{Sch{\"a}fer}},\ and\ \citenamefont {{Bi{\v
  c}{\'a}k}}}]{Ledvinka:2008}%
  \BibitemOpen
  \bibfield  {author} {\bibinfo {author} {\bibfnamefont {T.}~\bibnamefont
  {{Ledvinka}}}, \bibinfo {author} {\bibfnamefont {G.}~\bibnamefont
  {{Sch{\"a}fer}}}, \ and\ \bibinfo {author} {\bibfnamefont {J.}~\bibnamefont
  {{Bi{\v c}{\'a}k}}},\ }\href {\doibase 10.1103/PhysRevLett.100.251101}
  {\bibfield  {journal} {\bibinfo  {journal} {Physical Review Letters}\
  }\textbf {\bibinfo {volume} {100}},\ \bibinfo {eid} {251101} (\bibinfo {year}
  {2008})},\ \Eprint {http://arxiv.org/abs/0807.0214} {arXiv:0807.0214 [gr-qc]}
  \BibitemShut {NoStop}%
\bibitem [{\citenamefont {Foffa}(2014)}]{Foffa:2013gja}%
  \BibitemOpen
  \bibfield  {author} {\bibinfo {author} {\bibfnamefont {S.}~\bibnamefont
  {Foffa}},\ }\href {\doibase 10.1103/PhysRevD.89.024019} {\bibfield  {journal}
  {\bibinfo  {journal} {Phys. Rev.}\ }\textbf {\bibinfo {volume} {D89}},\
  \bibinfo {pages} {024019} (\bibinfo {year} {2014})},\ \Eprint
  {http://arxiv.org/abs/1309.3956} {arXiv:1309.3956 [gr-qc]} \BibitemShut
  {NoStop}%
%%CITATION = ARXIV:1309.3956;%%
\bibitem [{\citenamefont {{Damour}}(2016)}]{Damour:2016s}%
  \BibitemOpen
  \bibfield  {author} {\bibinfo {author} {\bibfnamefont {T.}~\bibnamefont
  {{Damour}}},\ }\href {\doibase 10.1103/PhysRevD.94.104015} {\bibfield
  {journal} {\bibinfo  {journal} {\prd}\ }\textbf {\bibinfo {volume} {94}},\
  \bibinfo {eid} {104015} (\bibinfo {year} {2016})},\ \Eprint
  {http://arxiv.org/abs/1609.00354} {arXiv:1609.00354 [gr-qc]} \BibitemShut
  {NoStop}%
\bibitem [{\citenamefont {Damour}(2010)}]{Damour:2009sm}%
  \BibitemOpen
  \bibfield  {author} {\bibinfo {author} {\bibfnamefont {T.}~\bibnamefont
  {Damour}},\ }\href {\doibase 10.1103/PhysRevD.81.024017} {\bibfield
  {journal} {\bibinfo  {journal} {Phys. Rev.}\ }\textbf {\bibinfo {volume}
  {D81}},\ \bibinfo {pages} {024017} (\bibinfo {year} {2010})},\ \Eprint
  {http://arxiv.org/abs/0910.5533} {arXiv:0910.5533 [gr-qc]} \BibitemShut
  {NoStop}%
%%CITATION = ARXIV:0910.5533;%%
\bibitem [{\citenamefont {Bini}\ and\ \citenamefont
  {Damour}(2012)}]{Bini:2012ji}%
  \BibitemOpen
  \bibfield  {author} {\bibinfo {author} {\bibfnamefont {D.}~\bibnamefont
  {Bini}}\ and\ \bibinfo {author} {\bibfnamefont {T.}~\bibnamefont {Damour}},\
  }\href {\doibase 10.1103/PhysRevD.86.124012} {\bibfield  {journal} {\bibinfo
  {journal} {Phys. Rev.}\ }\textbf {\bibinfo {volume} {D86}},\ \bibinfo {pages}
  {124012} (\bibinfo {year} {2012})},\ \Eprint {http://arxiv.org/abs/1210.2834}
  {arXiv:1210.2834 [gr-qc]} \BibitemShut {NoStop}%
%%CITATION = ARXIV:1210.2834;%%
\bibitem [{\citenamefont {Damour}\ \emph
  {et~al.}(2014{\natexlab{a}})\citenamefont {Damour}, \citenamefont
  {Guercilena}, \citenamefont {Hinder}, \citenamefont {Hopper}, \citenamefont
  {Nagar},\ and\ \citenamefont {Rezzolla}}]{Damour:2014afa}%
  \BibitemOpen
  \bibfield  {author} {\bibinfo {author} {\bibfnamefont {T.}~\bibnamefont
  {Damour}}, \bibinfo {author} {\bibfnamefont {F.}~\bibnamefont {Guercilena}},
  \bibinfo {author} {\bibfnamefont {I.}~\bibnamefont {Hinder}}, \bibinfo
  {author} {\bibfnamefont {S.}~\bibnamefont {Hopper}}, \bibinfo {author}
  {\bibfnamefont {A.}~\bibnamefont {Nagar}}, \ and\ \bibinfo {author}
  {\bibfnamefont {L.}~\bibnamefont {Rezzolla}},\ }\href {\doibase
  10.1103/PhysRevD.89.081503} {\bibfield  {journal} {\bibinfo  {journal} {Phys.
  Rev.}\ }\textbf {\bibinfo {volume} {D89}},\ \bibinfo {pages} {081503}
  (\bibinfo {year} {2014}{\natexlab{a}})},\ \Eprint
  {http://arxiv.org/abs/1402.7307} {arXiv:1402.7307 [gr-qc]} \BibitemShut
  {NoStop}%
%%CITATION = ARXIV:1402.7307;%%
\bibitem [{\citenamefont {Bini}\ and\ \citenamefont
  {Damour}(2017{\natexlab{a}})}]{Bini:2017wfr}%
  \BibitemOpen
  \bibfield  {author} {\bibinfo {author} {\bibfnamefont {D.}~\bibnamefont
  {Bini}}\ and\ \bibinfo {author} {\bibfnamefont {T.}~\bibnamefont {Damour}},\
  }\href@noop {} {\  (\bibinfo {year} {2017}{\natexlab{a}})},\ \Eprint
  {http://arxiv.org/abs/1706.06877} {arXiv:1706.06877 [gr-qc]} \BibitemShut
  {NoStop}%
%%CITATION = ARXIV:1706.06877;%%
\bibitem [{\citenamefont {Hansen}(1974)}]{Hansen:1974zz}%
  \BibitemOpen
  \bibfield  {author} {\bibinfo {author} {\bibfnamefont {R.~O.}\ \bibnamefont
  {Hansen}},\ }\href {\doibase 10.1063/1.1666501} {\bibfield  {journal}
  {\bibinfo  {journal} {J. Math. Phys.}\ }\textbf {\bibinfo {volume} {15}},\
  \bibinfo {pages} {46} (\bibinfo {year} {1974})}\BibitemShut {NoStop}%
%%CITATION = JMAPA,15,46;%%
\bibitem [{\citenamefont {Buonanno}\ and\ \citenamefont
  {Damour}(1999)}]{Buonanno99}%
  \BibitemOpen
  \bibfield  {author} {\bibinfo {author} {\bibfnamefont {A.}~\bibnamefont
  {Buonanno}}\ and\ \bibinfo {author} {\bibfnamefont {T.}~\bibnamefont
  {Damour}},\ }\href {\doibase 10.1103/PhysRevD.59.084006} {\bibfield
  {journal} {\bibinfo  {journal} {\prd}\ }\textbf {\bibinfo {volume} {59}},\
  \bibinfo {pages} {084006} (\bibinfo {year} {1999})},\ \Eprint
  {http://arxiv.org/abs/gr-qc/9811091} {gr-qc/9811091} \BibitemShut {NoStop}%
%%CITATION = GR-QC/9811091;%%
\bibitem [{\citenamefont {Buonanno}\ and\ \citenamefont
  {Damour}(2000)}]{Buonanno00}%
  \BibitemOpen
  \bibfield  {author} {\bibinfo {author} {\bibfnamefont {A.}~\bibnamefont
  {Buonanno}}\ and\ \bibinfo {author} {\bibfnamefont {T.}~\bibnamefont
  {Damour}},\ }\href {\doibase 10.1103/PhysRevD.62.064015} {\bibfield
  {journal} {\bibinfo  {journal} {\prd}\ }\textbf {\bibinfo {volume} {62}},\
  \bibinfo {pages} {064015} (\bibinfo {year} {2000})},\ \Eprint
  {http://arxiv.org/abs/gr-qc/0001013} {gr-qc/0001013} \BibitemShut {NoStop}%
%%CITATION = GR-QC/0001013;%%
\bibitem [{\citenamefont {{Damour}}(2001)}]{Damour1SEOB}%
  \BibitemOpen
  \bibfield  {author} {\bibinfo {author} {\bibfnamefont {T.}~\bibnamefont
  {{Damour}}},\ }\href {\doibase 10.1103/PhysRevD.64.124013} {\bibfield
  {journal} {\bibinfo  {journal} {\prd}\ }\textbf {\bibinfo {volume} {64}},\
  \bibinfo {pages} {124013} (\bibinfo {year} {2001})},\ \Eprint
  {http://arxiv.org/abs/gr-qc/0103018} {gr-qc/0103018} \BibitemShut {NoStop}%
\bibitem [{\citenamefont {{Damour}}\ \emph {et~al.}(2008)\citenamefont
  {{Damour}}, \citenamefont {{Jaranowski}},\ and\ \citenamefont
  {{Sch{\"a}fer}}}]{DJS}%
  \BibitemOpen
  \bibfield  {author} {\bibinfo {author} {\bibfnamefont {T.}~\bibnamefont
  {{Damour}}}, \bibinfo {author} {\bibfnamefont {P.}~\bibnamefont
  {{Jaranowski}}}, \ and\ \bibinfo {author} {\bibfnamefont {G.}~\bibnamefont
  {{Sch{\"a}fer}}},\ }\href {\doibase 10.1103/PhysRevD.78.024009} {\bibfield
  {journal} {\bibinfo  {journal} {\prd}\ }\textbf {\bibinfo {volume} {78}},\
  \bibinfo {eid} {024009} (\bibinfo {year} {2008})},\ \Eprint
  {http://arxiv.org/abs/0803.0915} {arXiv:0803.0915 [gr-qc]} \BibitemShut
  {NoStop}%
\bibitem [{\citenamefont {Barausse}\ and\ \citenamefont
  {Buonanno}(2010)}]{Barausse:2009xi}%
  \BibitemOpen
  \bibfield  {author} {\bibinfo {author} {\bibfnamefont {E.}~\bibnamefont
  {Barausse}}\ and\ \bibinfo {author} {\bibfnamefont {A.}~\bibnamefont
  {Buonanno}},\ }\href {\doibase 10.1103/PhysRevD.81.084024} {\bibfield
  {journal} {\bibinfo  {journal} {\prd}\ }\textbf {\bibinfo {volume} {81}},\
  \bibinfo {pages} {084024} (\bibinfo {year} {2010})},\ \Eprint
  {http://arxiv.org/abs/gr-qc/0912.3517} {gr-qc/0912.3517} \BibitemShut
  {NoStop}%
%%CITATION = ARXIV:0912.3517;%%
\bibitem [{\citenamefont {Damour}(2014)}]{Damour:2012mv}%
  \BibitemOpen
  \bibfield  {author} {\bibinfo {author} {\bibfnamefont {T.}~\bibnamefont
  {Damour}},\ }\bibfield  {booktitle} {\emph {\bibinfo {booktitle}
  {{Proceedings, Relativity and Gravitation: Perspectives 100 years after
  Einstein's stay in Prague: Prague, Czech Republic, June 25-29, 2012}}},\
  }\href {\doibase 10.1007/978-3-319-06349-2_5} {\bibfield  {journal} {\bibinfo
   {journal} {Fundam. Theor. Phys.}\ }\textbf {\bibinfo {volume} {177}},\
  \bibinfo {pages} {111} (\bibinfo {year} {2014})},\ \Eprint
  {http://arxiv.org/abs/1212.3169} {arXiv:1212.3169 [gr-qc]} \BibitemShut
  {NoStop}%
%%CITATION = ARXIV:1212.3169;%%
\bibitem [{\citenamefont {Pan}\ \emph {et~al.}(2014)\citenamefont {Pan},
  \citenamefont {Buonanno}, \citenamefont {Taracchini}, \citenamefont {Kidder},
  \citenamefont {Mrou{\'e}}, \citenamefont {Pfeiffer}, \citenamefont {Scheel},\
  and\ \citenamefont {Szil{\'a}gyi}}]{pan2014}%
  \BibitemOpen
  \bibfield  {author} {\bibinfo {author} {\bibfnamefont {Y.}~\bibnamefont
  {Pan}}, \bibinfo {author} {\bibfnamefont {A.}~\bibnamefont {Buonanno}},
  \bibinfo {author} {\bibfnamefont {A.}~\bibnamefont {Taracchini}}, \bibinfo
  {author} {\bibfnamefont {L.~E.}\ \bibnamefont {Kidder}}, \bibinfo {author}
  {\bibfnamefont {A.~H.}\ \bibnamefont {Mrou{\'e}}}, \bibinfo {author}
  {\bibfnamefont {H.~P.}\ \bibnamefont {Pfeiffer}}, \bibinfo {author}
  {\bibfnamefont {M.~A.}\ \bibnamefont {Scheel}}, \ and\ \bibinfo {author}
  {\bibfnamefont {B.}~\bibnamefont {Szil{\'a}gyi}},\ }\href@noop {} {\bibfield
  {journal} {\bibinfo  {journal} {\prd}\ }\textbf {\bibinfo {volume} {89}},\
  \bibinfo {pages} {084006} (\bibinfo {year} {2014})}\BibitemShut {NoStop}%
\bibitem [{\citenamefont {{Damour}}\ and\ \citenamefont
  {{Nagar}}(2016)}]{2016LNP...905..273D}%
  \BibitemOpen
  \bibfield  {author} {\bibinfo {author} {\bibfnamefont {T.}~\bibnamefont
  {{Damour}}}\ and\ \bibinfo {author} {\bibfnamefont {A.}~\bibnamefont
  {{Nagar}}},\ }in\ \href {\doibase 10.1007/978-3-319-19416-5_7} {\emph
  {\bibinfo {booktitle} {Lecture Notes in Physics, Berlin Springer Verlag}}},\
  \bibinfo {series} {Lecture Notes in Physics, Berlin Springer Verlag}, Vol.\
  \bibinfo {volume} {905},\ \bibinfo {editor} {edited by\ \bibinfo {editor}
  {\bibfnamefont {F.}~\bibnamefont {{Haardt}}}, \bibinfo {editor}
  {\bibfnamefont {V.}~\bibnamefont {{Gorini}}}, \bibinfo {editor}
  {\bibfnamefont {U.}~\bibnamefont {{Moschella}}}, \bibinfo {editor}
  {\bibfnamefont {A.}~\bibnamefont {{Treves}}}, \ and\ \bibinfo {editor}
  {\bibfnamefont {M.}~\bibnamefont {{Colpi}}}}\ (\bibinfo {year} {2016})\ p.\
  \bibinfo {pages} {273}\BibitemShut {NoStop}%
\bibitem [{\citenamefont {{Balmelli}}\ and\ \citenamefont
  {{Damour}}(2015)}]{Balmelli:Damour:NLOSS}%
  \BibitemOpen
  \bibfield  {author} {\bibinfo {author} {\bibfnamefont {S.}~\bibnamefont
  {{Balmelli}}}\ and\ \bibinfo {author} {\bibfnamefont {T.}~\bibnamefont
  {{Damour}}},\ }\href {\doibase 10.1103/PhysRevD.92.124022} {\bibfield
  {journal} {\bibinfo  {journal} {\prd}\ }\textbf {\bibinfo {volume} {92}},\
  \bibinfo {eid} {124022} (\bibinfo {year} {2015})},\ \Eprint
  {http://arxiv.org/abs/1509.08135} {arXiv:1509.08135 [gr-qc]} \BibitemShut
  {NoStop}%
\bibitem [{\citenamefont {Abbott}\ \emph
  {et~al.}(2016{\natexlab{c}})\citenamefont {Abbott} \emph
  {et~al.}}]{TheLIGOScientific:2016wfe}%
  \BibitemOpen
  \bibfield  {author} {\bibinfo {author} {\bibfnamefont {B.~P.}\ \bibnamefont
  {Abbott}} \emph {et~al.} (\bibinfo {collaboration} {Virgo, LIGO
  Scientific}),\ }\href {\doibase 10.1103/PhysRevLett.116.241102} {\bibfield
  {journal} {\bibinfo  {journal} {Phys. Rev. Lett.}\ }\textbf {\bibinfo
  {volume} {116}},\ \bibinfo {pages} {241102} (\bibinfo {year}
  {2016}{\natexlab{c}})},\ \Eprint {http://arxiv.org/abs/1602.03840}
  {arXiv:1602.03840 [gr-qc]} \BibitemShut {NoStop}%
%%CITATION = ARXIV:1602.03840;%%
\bibitem [{\citenamefont {Abbott}\ \emph
  {et~al.}(2016{\natexlab{d}})\citenamefont {Abbott} \emph
  {et~al.}}]{Abbott:2016izl}%
  \BibitemOpen
  \bibfield  {author} {\bibinfo {author} {\bibfnamefont {B.~P.}\ \bibnamefont
  {Abbott}} \emph {et~al.} (\bibinfo {collaboration} {Virgo, LIGO
  Scientific}),\ }\href {\doibase 10.1103/PhysRevX.6.041014} {\bibfield
  {journal} {\bibinfo  {journal} {Phys. Rev.}\ }\textbf {\bibinfo {volume}
  {X6}},\ \bibinfo {pages} {041014} (\bibinfo {year} {2016}{\natexlab{d}})},\
  \Eprint {http://arxiv.org/abs/1606.01210} {arXiv:1606.01210 [gr-qc]}
  \BibitemShut {NoStop}%
%%CITATION = ARXIV:1606.01210;%%
\bibitem [{\citenamefont {Abbott}\ \emph
  {et~al.}(2016{\natexlab{e}})\citenamefont {Abbott} \emph
  {et~al.}}]{TheLIGOScientific:2016src}%
  \BibitemOpen
  \bibfield  {author} {\bibinfo {author} {\bibfnamefont {B.~P.}\ \bibnamefont
  {Abbott}} \emph {et~al.} (\bibinfo {collaboration} {Virgo, LIGO
  Scientific}),\ }\href {\doibase 10.1103/PhysRevLett.116.221101} {\bibfield
  {journal} {\bibinfo  {journal} {Phys. Rev. Lett.}\ }\textbf {\bibinfo
  {volume} {116}},\ \bibinfo {pages} {221101} (\bibinfo {year}
  {2016}{\natexlab{e}})},\ \Eprint {http://arxiv.org/abs/1602.03841}
  {arXiv:1602.03841 [gr-qc]} \BibitemShut {NoStop}%
%%CITATION = ARXIV:1602.03841;%%
\bibitem [{\citenamefont {Mathisson}(1937)}]{Mathisson:1937}%
  \BibitemOpen
  \bibfield  {author} {\bibinfo {author} {\bibfnamefont {M.}~\bibnamefont
  {Mathisson}},\ }\href@noop {} {\bibfield  {journal} {\bibinfo  {journal}
  {Acta Physica Polonica}\ }\textbf {\bibinfo {volume} {6}},\ \bibinfo {pages}
  {163} (\bibinfo {year} {1937})}\BibitemShut {NoStop}%
\bibitem [{\citenamefont {Mathisson}(2010)}]{Mathisson:2010}%
  \BibitemOpen
  \bibfield  {author} {\bibinfo {author} {\bibfnamefont {M.}~\bibnamefont
  {Mathisson}},\ }\href {\doibase 10.1007/s10714-010-0939-y} {\bibfield
  {journal} {\bibinfo  {journal} {Gen. Relativ. Gravit.}\ }\textbf {\bibinfo
  {volume} {42}},\ \bibinfo {pages} {1011} (\bibinfo {year}
  {2010})}\BibitemShut {NoStop}%
\bibitem [{\citenamefont {Papapetrou}(1951)}]{Papapetrou:1951pa}%
  \BibitemOpen
  \bibfield  {author} {\bibinfo {author} {\bibfnamefont {A.}~\bibnamefont
  {Papapetrou}},\ }\href@noop {} {\bibfield  {journal} {\bibinfo  {journal}
  {Proc.\ R.\ Soc.\ London\ A}\ }\textbf {\bibinfo {volume} {209}},\ \bibinfo
  {pages} {248} (\bibinfo {year} {1951})}\BibitemShut {NoStop}%
\bibitem [{\citenamefont {Dixon}(1979)}]{Dixon:1979}%
  \BibitemOpen
  \bibfield  {author} {\bibinfo {author} {\bibfnamefont {W.~G.}\ \bibnamefont
  {Dixon}},\ }in\ \href@noop {} {\emph {\bibinfo {booktitle} {Proceedings of
  the International School of Physics Enrico Fermi LXVII}}},\ \bibinfo {editor}
  {edited by\ \bibinfo {editor} {\bibfnamefont {J.}~\bibnamefont {Ehlers}}}\
  (\bibinfo  {publisher} {North Holland},\ \bibinfo {address} {Amsterdam},\
  \bibinfo {year} {1979})\ pp.\ \bibinfo {pages} {156--219}\BibitemShut
  {NoStop}%
\bibitem [{\citenamefont {Dixon}(2015)}]{Dixon:2015vxa}%
  \BibitemOpen
  \bibfield  {author} {\bibinfo {author} {\bibfnamefont {W.~G.}\ \bibnamefont
  {Dixon}},\ }\bibfield  {booktitle} {\emph {\bibinfo {booktitle}
  {{Proceedings, 524th WE-Heraeus-Seminar: Equations of Motion in Relativistic
  Gravity (EOM 2013)}}},\ }\href {\doibase 10.1007/978-3-319-18335-0_1}
  {\bibfield  {journal} {\bibinfo  {journal} {Fund. Theor. Phys.}\ }\textbf
  {\bibinfo {volume} {179}},\ \bibinfo {pages} {1} (\bibinfo {year}
  {2015})}\BibitemShut {NoStop}%
%%CITATION = INSPIRE-1381011;%%
\bibitem [{\citenamefont {Vines}\ and\ \citenamefont
  {Steinhoff}(2016)}]{Vines:2016qwa}%
  \BibitemOpen
  \bibfield  {author} {\bibinfo {author} {\bibfnamefont {J.}~\bibnamefont
  {Vines}}\ and\ \bibinfo {author} {\bibfnamefont {J.}~\bibnamefont
  {Steinhoff}},\ }\href@noop {} {\  (\bibinfo {year} {2016})},\ \Eprint
  {http://arxiv.org/abs/1606.08832} {arXiv:1606.08832 [gr-qc]} \BibitemShut
  {NoStop}%
%%CITATION = ARXIV:1606.08832;%%
\bibitem [{\citenamefont {Hanson}\ and\ \citenamefont
  {Regge}(1974)}]{Hanson:Regge:1974}%
  \BibitemOpen
  \bibfield  {author} {\bibinfo {author} {\bibfnamefont {A.}~\bibnamefont
  {Hanson}}\ and\ \bibinfo {author} {\bibfnamefont {T.}~\bibnamefont {Regge}},\
  }\href {\doibase http://dx.doi.org/10.1016/0003-4916(74)90046-3} {\bibfield
  {journal} {\bibinfo  {journal} {Annals of Physics}\ }\textbf {\bibinfo
  {volume} {87}},\ \bibinfo {pages} {498 } (\bibinfo {year}
  {1974})}\BibitemShut {NoStop}%
\bibitem [{\citenamefont {Bailey}\ and\ \citenamefont
  {Israel}(1975)}]{Bailey1975}%
  \BibitemOpen
  \bibfield  {author} {\bibinfo {author} {\bibfnamefont {I.}~\bibnamefont
  {Bailey}}\ and\ \bibinfo {author} {\bibfnamefont {W.}~\bibnamefont
  {Israel}},\ }\href {\doibase 10.1007/BF01609434} {\bibfield  {journal}
  {\bibinfo  {journal} {Communications in Mathematical Physics}\ }\textbf
  {\bibinfo {volume} {42}},\ \bibinfo {pages} {65} (\bibinfo {year}
  {1975})}\BibitemShut {NoStop}%
\bibitem [{\citenamefont {Porto}(2006)}]{Porto:2005ac}%
  \BibitemOpen
  \bibfield  {author} {\bibinfo {author} {\bibfnamefont {R.~A.}\ \bibnamefont
  {Porto}},\ }\href {\doibase 10.1103/PhysRevD.73.104031} {\bibfield  {journal}
  {\bibinfo  {journal} {Phys. Rev.}\ }\textbf {\bibinfo {volume} {D73}},\
  \bibinfo {pages} {104031} (\bibinfo {year} {2006})},\ \Eprint
  {http://arxiv.org/abs/gr-qc/0511061} {arXiv:gr-qc/0511061 [gr-qc]}
  \BibitemShut {NoStop}%
%%CITATION = GR-QC/0511061;%%
\bibitem [{\citenamefont {Steinhoff}(2015)}]{Steinhoff:2014}%
  \BibitemOpen
  \bibfield  {author} {\bibinfo {author} {\bibfnamefont {J.}~\bibnamefont
  {Steinhoff}},\ }\bibfield  {booktitle} {\emph {\bibinfo {booktitle}
  {{Proceedings, 524th WE-Heraeus-Seminar: Equations of Motion in Relativistic
  Gravity (EOM 2013)}}},\ }\href {\doibase 10.1007/978-3-319-18335-0_19}
  {\bibfield  {journal} {\bibinfo  {journal} {Fund. Theor. Phys.}\ }\textbf
  {\bibinfo {volume} {179}},\ \bibinfo {pages} {615} (\bibinfo {year}
  {2015})},\ \Eprint {http://arxiv.org/abs/1412.3251} {arXiv:1412.3251 [gr-qc]}
  \BibitemShut {NoStop}%
%%CITATION = ARXIV:1412.3251;%%
\bibitem [{\citenamefont {Marsat}(2015)}]{Marsat:2014xea}%
  \BibitemOpen
  \bibfield  {author} {\bibinfo {author} {\bibfnamefont {S.}~\bibnamefont
  {Marsat}},\ }\href {\doibase 10.1088/0264-9381/32/8/085008} {\bibfield
  {journal} {\bibinfo  {journal} {Class. Quant. Grav.}\ }\textbf {\bibinfo
  {volume} {32}},\ \bibinfo {pages} {085008} (\bibinfo {year} {2015})},\
  \Eprint {http://arxiv.org/abs/1411.4118} {arXiv:1411.4118 [gr-qc]}
  \BibitemShut {NoStop}%
%%CITATION = ARXIV:1411.4118;%%
\bibitem [{\citenamefont {Levi}\ and\ \citenamefont
  {Steinhoff}(2015{\natexlab{a}})}]{Levi:Steinhoff:2015:1}%
  \BibitemOpen
  \bibfield  {author} {\bibinfo {author} {\bibfnamefont {M.}~\bibnamefont
  {Levi}}\ and\ \bibinfo {author} {\bibfnamefont {J.}~\bibnamefont
  {Steinhoff}},\ }\href {\doibase 10.1007/JHEP09(2015)219} {\bibfield
  {journal} {\bibinfo  {journal} {JHEP}\ }\textbf {\bibinfo {volume} {09}},\
  \bibinfo {pages} {219} (\bibinfo {year} {2015}{\natexlab{a}})},\ \Eprint
  {http://arxiv.org/abs/1501.04956} {arXiv:1501.04956 [gr-qc]} \BibitemShut
  {NoStop}%
%%CITATION = ARXIV:1501.04956;%%
\bibitem [{\citenamefont {Holstein}\ and\ \citenamefont
  {Ross}(2008)}]{Holstein:2008sx}%
  \BibitemOpen
  \bibfield  {author} {\bibinfo {author} {\bibfnamefont {B.~R.}\ \bibnamefont
  {Holstein}}\ and\ \bibinfo {author} {\bibfnamefont {A.}~\bibnamefont
  {Ross}},\ }\href@noop {} {\  (\bibinfo {year} {2008})},\ \Eprint
  {http://arxiv.org/abs/0802.0716} {arXiv:0802.0716 [hep-ph]} \BibitemShut
  {NoStop}%
%%CITATION = ARXIV:0802.0716;%%
\bibitem [{\citenamefont {Vaidya}(2015)}]{Vaidya:2014kza}%
  \BibitemOpen
  \bibfield  {author} {\bibinfo {author} {\bibfnamefont {V.}~\bibnamefont
  {Vaidya}},\ }\href {\doibase 10.1103/PhysRevD.91.024017} {\bibfield
  {journal} {\bibinfo  {journal} {Phys. Rev.}\ }\textbf {\bibinfo {volume}
  {D91}},\ \bibinfo {pages} {024017} (\bibinfo {year} {2015})},\ \Eprint
  {http://arxiv.org/abs/1410.5348} {arXiv:1410.5348 [hep-th]} \BibitemShut
  {NoStop}%
%%CITATION = ARXIV:1410.5348;%%
\bibitem [{\citenamefont {Guevara}(2017)}]{Guevara:2017csg}%
  \BibitemOpen
  \bibfield  {author} {\bibinfo {author} {\bibfnamefont {A.}~\bibnamefont
  {Guevara}},\ }\href@noop {} {\  (\bibinfo {year} {2017})},\ \Eprint
  {http://arxiv.org/abs/1706.02314} {arXiv:1706.02314 [hep-th]} \BibitemShut
  {NoStop}%
%%CITATION = ARXIV:1706.02314;%%
\bibitem [{\citenamefont {Kawai}\ \emph {et~al.}(1986)\citenamefont {Kawai},
  \citenamefont {Lewellen},\ and\ \citenamefont {Tye}}]{Kawai:1985xq}%
  \BibitemOpen
  \bibfield  {author} {\bibinfo {author} {\bibfnamefont {H.}~\bibnamefont
  {Kawai}}, \bibinfo {author} {\bibfnamefont {D.~C.}\ \bibnamefont {Lewellen}},
  \ and\ \bibinfo {author} {\bibfnamefont {S.~H.~H.}\ \bibnamefont {Tye}},\
  }\href {\doibase 10.1016/0550-3213(86)90362-7} {\bibfield  {journal}
  {\bibinfo  {journal} {Nucl. Phys.}\ }\textbf {\bibinfo {volume} {B269}},\
  \bibinfo {pages} {1} (\bibinfo {year} {1986})}\BibitemShut {NoStop}%
%%CITATION = NUPHA,B269,1;%%
\bibitem [{\citenamefont {Bern}\ \emph {et~al.}(2008)\citenamefont {Bern},
  \citenamefont {Carrasco},\ and\ \citenamefont {Johansson}}]{Bern:2008qj}%
  \BibitemOpen
  \bibfield  {author} {\bibinfo {author} {\bibfnamefont {Z.}~\bibnamefont
  {Bern}}, \bibinfo {author} {\bibfnamefont {J.~J.~M.}\ \bibnamefont
  {Carrasco}}, \ and\ \bibinfo {author} {\bibfnamefont {H.}~\bibnamefont
  {Johansson}},\ }\href {\doibase 10.1103/PhysRevD.78.085011} {\bibfield
  {journal} {\bibinfo  {journal} {Phys. Rev.}\ }\textbf {\bibinfo {volume}
  {D78}},\ \bibinfo {pages} {085011} (\bibinfo {year} {2008})},\ \Eprint
  {http://arxiv.org/abs/0805.3993} {arXiv:0805.3993 [hep-ph]} \BibitemShut
  {NoStop}%
%%CITATION = ARXIV:0805.3993;%%
\bibitem [{\citenamefont {Monteiro}\ \emph {et~al.}(2014)\citenamefont
  {Monteiro}, \citenamefont {O'Connell},\ and\ \citenamefont
  {White}}]{Monteiro:2014cda}%
  \BibitemOpen
  \bibfield  {author} {\bibinfo {author} {\bibfnamefont {R.}~\bibnamefont
  {Monteiro}}, \bibinfo {author} {\bibfnamefont {D.}~\bibnamefont {O'Connell}},
  \ and\ \bibinfo {author} {\bibfnamefont {C.~D.}\ \bibnamefont {White}},\
  }\href {\doibase 10.1007/JHEP12(2014)056} {\bibfield  {journal} {\bibinfo
  {journal} {JHEP}\ }\textbf {\bibinfo {volume} {12}},\ \bibinfo {pages} {056}
  (\bibinfo {year} {2014})},\ \Eprint {http://arxiv.org/abs/1410.0239}
  {arXiv:1410.0239 [hep-th]} \BibitemShut {NoStop}%
%%CITATION = ARXIV:1410.0239;%%
\bibitem [{\citenamefont {Monteiro}\ \emph {et~al.}(2015)\citenamefont
  {Monteiro}, \citenamefont {O'Connell},\ and\ \citenamefont
  {White}}]{Monteiro:2015bna}%
  \BibitemOpen
  \bibfield  {author} {\bibinfo {author} {\bibfnamefont {R.}~\bibnamefont
  {Monteiro}}, \bibinfo {author} {\bibfnamefont {D.}~\bibnamefont {O'Connell}},
  \ and\ \bibinfo {author} {\bibfnamefont {C.~D.}\ \bibnamefont {White}},\
  }\bibfield  {booktitle} {\emph {\bibinfo {booktitle} {{Proceedings, 7th Black
  Holes Workshop 2014: Aveiro, Portugal, December 18-19, 2014}}},\ }\href
  {\doibase 10.1142/S0218271815420080} {\bibfield  {journal} {\bibinfo
  {journal} {Int. J. Mod. Phys.}\ }\textbf {\bibinfo {volume} {D24}},\ \bibinfo
  {pages} {1542008} (\bibinfo {year} {2015})}\BibitemShut {NoStop}%
%%CITATION = IMPAE,D24,1542008;%%
\bibitem [{\citenamefont {Luna}\ \emph {et~al.}(2017)\citenamefont {Luna},
  \citenamefont {Monteiro}, \citenamefont {Nicholson}, \citenamefont {Ochirov},
  \citenamefont {O'Connell}, \citenamefont {Westerberg},\ and\ \citenamefont
  {White}}]{Luna:2016hge}%
  \BibitemOpen
  \bibfield  {author} {\bibinfo {author} {\bibfnamefont {A.}~\bibnamefont
  {Luna}}, \bibinfo {author} {\bibfnamefont {R.}~\bibnamefont {Monteiro}},
  \bibinfo {author} {\bibfnamefont {I.}~\bibnamefont {Nicholson}}, \bibinfo
  {author} {\bibfnamefont {A.}~\bibnamefont {Ochirov}}, \bibinfo {author}
  {\bibfnamefont {D.}~\bibnamefont {O'Connell}}, \bibinfo {author}
  {\bibfnamefont {N.}~\bibnamefont {Westerberg}}, \ and\ \bibinfo {author}
  {\bibfnamefont {C.~D.}\ \bibnamefont {White}},\ }\href {\doibase
  10.1007/JHEP04(2017)069} {\bibfield  {journal} {\bibinfo  {journal} {JHEP}\
  }\textbf {\bibinfo {volume} {04}},\ \bibinfo {pages} {069} (\bibinfo {year}
  {2017})},\ \Eprint {http://arxiv.org/abs/1611.07508} {arXiv:1611.07508
  [hep-th]} \BibitemShut {NoStop}%
%%CITATION = ARXIV:1611.07508;%%
\bibitem [{\citenamefont {Goldberger}\ and\ \citenamefont
  {Ridgway}(2017)}]{Goldberger:2016iau}%
  \BibitemOpen
  \bibfield  {author} {\bibinfo {author} {\bibfnamefont {W.~D.}\ \bibnamefont
  {Goldberger}}\ and\ \bibinfo {author} {\bibfnamefont {A.~K.}\ \bibnamefont
  {Ridgway}},\ }\href {\doibase 10.1103/PhysRevD.95.125010} {\bibfield
  {journal} {\bibinfo  {journal} {Phys. Rev.}\ }\textbf {\bibinfo {volume}
  {D95}},\ \bibinfo {pages} {125010} (\bibinfo {year} {2017})},\ \Eprint
  {http://arxiv.org/abs/1611.03493} {arXiv:1611.03493 [hep-th]} \BibitemShut
  {NoStop}%
%%CITATION = ARXIV:1611.03493;%%
\bibitem [{\citenamefont {Goldberger}\ \emph {et~al.}(2017)\citenamefont
  {Goldberger}, \citenamefont {Prabhu},\ and\ \citenamefont
  {Thompson}}]{Goldberger:2017frp}%
  \BibitemOpen
  \bibfield  {author} {\bibinfo {author} {\bibfnamefont {W.~D.}\ \bibnamefont
  {Goldberger}}, \bibinfo {author} {\bibfnamefont {S.~G.}\ \bibnamefont
  {Prabhu}}, \ and\ \bibinfo {author} {\bibfnamefont {J.~O.}\ \bibnamefont
  {Thompson}},\ }\href {\doibase 10.1103/PhysRevD.96.065009} {\bibfield
  {journal} {\bibinfo  {journal} {Phys. Rev.}\ }\textbf {\bibinfo {volume}
  {D96}},\ \bibinfo {pages} {065009} (\bibinfo {year} {2017})},\ \Eprint
  {http://arxiv.org/abs/1705.09263} {arXiv:1705.09263 [hep-th]} \BibitemShut
  {NoStop}%
%%CITATION = ARXIV:1705.09263;%%
\bibitem [{\citenamefont {Cheung}\ and\ \citenamefont
  {Remmen}(2017)}]{Cheung:2016say}%
  \BibitemOpen
  \bibfield  {author} {\bibinfo {author} {\bibfnamefont {C.}~\bibnamefont
  {Cheung}}\ and\ \bibinfo {author} {\bibfnamefont {G.~N.}\ \bibnamefont
  {Remmen}},\ }\href {\doibase 10.1007/JHEP01(2017)104} {\bibfield  {journal}
  {\bibinfo  {journal} {JHEP}\ }\textbf {\bibinfo {volume} {01}},\ \bibinfo
  {pages} {104} (\bibinfo {year} {2017})},\ \Eprint
  {http://arxiv.org/abs/1612.03927} {arXiv:1612.03927 [hep-th]} \BibitemShut
  {NoStop}%
%%CITATION = ARXIV:1612.03927;%%
\bibitem [{\citenamefont {Flanagan}\ \emph {et~al.}(2016)\citenamefont
  {Flanagan}, \citenamefont {Nichols}, \citenamefont {Stein},\ and\
  \citenamefont {Vines}}]{FNSV}%
  \BibitemOpen
  \bibfield  {author} {\bibinfo {author} {\bibfnamefont {{\'E}.~{\'E}.}\
  \bibnamefont {Flanagan}}, \bibinfo {author} {\bibfnamefont {D.~A.}\
  \bibnamefont {Nichols}}, \bibinfo {author} {\bibfnamefont {L.~C.}\
  \bibnamefont {Stein}}, \ and\ \bibinfo {author} {\bibfnamefont
  {J.}~\bibnamefont {Vines}},\ }\href {\doibase 10.1103/PhysRevD.93.104007}
  {\bibfield  {journal} {\bibinfo  {journal} {Phys. Rev.}\ }\textbf {\bibinfo
  {volume} {D93}},\ \bibinfo {pages} {104007} (\bibinfo {year} {2016})},\
  \Eprint {http://arxiv.org/abs/1602.01847} {arXiv:1602.01847 [gr-qc]}
  \BibitemShut {NoStop}%
%%CITATION = ARXIV:1602.01847;%%
\bibitem [{\citenamefont {Harte}\ and\ \citenamefont
  {Vines}(2016)}]{Harte:2016vwo}%
  \BibitemOpen
  \bibfield  {author} {\bibinfo {author} {\bibfnamefont {A.~I.}\ \bibnamefont
  {Harte}}\ and\ \bibinfo {author} {\bibfnamefont {J.}~\bibnamefont {Vines}},\
  }\href {\doibase 10.1103/PhysRevD.94.084009} {\bibfield  {journal} {\bibinfo
  {journal} {Phys. Rev.}\ }\textbf {\bibinfo {volume} {D94}},\ \bibinfo {pages}
  {084009} (\bibinfo {year} {2016})},\ \Eprint
  {http://arxiv.org/abs/1608.04359} {arXiv:1608.04359 [gr-qc]} \BibitemShut
  {NoStop}%
%%CITATION = ARXIV:1608.04359;%%
\bibitem [{\citenamefont {Harte}(2017)}]{Harte:2017gpd}%
  \BibitemOpen
  \bibfield  {author} {\bibinfo {author} {\bibfnamefont {A.~I.}\ \bibnamefont
  {Harte}},\ }\href {\doibase 10.1103/PhysRevLett.118.141101} {\bibfield
  {journal} {\bibinfo  {journal} {Phys. Rev. Lett.}\ }\textbf {\bibinfo
  {volume} {118}},\ \bibinfo {pages} {141101} (\bibinfo {year} {2017})},\
  \Eprint {http://arxiv.org/abs/1701.05257} {arXiv:1701.05257 [gr-qc]}
  \BibitemShut {NoStop}%
%%CITATION = ARXIV:1701.05257;%%
\bibitem [{\citenamefont {Bini}\ and\ \citenamefont
  {Damour}(2017{\natexlab{b}})}]{Bini:2017xzy}%
  \BibitemOpen
  \bibfield  {author} {\bibinfo {author} {\bibfnamefont {D.}~\bibnamefont
  {Bini}}\ and\ \bibinfo {author} {\bibfnamefont {T.}~\bibnamefont {Damour}},\
  }\href@noop {} {\  (\bibinfo {year} {2017}{\natexlab{b}})},\ \Eprint
  {http://arxiv.org/abs/1709.00590} {arXiv:1709.00590 [gr-qc]} \BibitemShut
  {NoStop}%
%%CITATION = ARXIV:1709.00590;%%
\bibitem [{\citenamefont {Damour}\ \emph
  {et~al.}(2014{\natexlab{b}})\citenamefont {Damour}, \citenamefont
  {Jaranowski},\ and\ \citenamefont {Sch{\"a}fer}}]{Damour:2014jta}%
  \BibitemOpen
  \bibfield  {author} {\bibinfo {author} {\bibfnamefont {T.}~\bibnamefont
  {Damour}}, \bibinfo {author} {\bibfnamefont {P.}~\bibnamefont {Jaranowski}},
  \ and\ \bibinfo {author} {\bibfnamefont {G.}~\bibnamefont {Sch{\"a}fer}},\
  }\href {\doibase 10.1103/PhysRevD.89.064058} {\bibfield  {journal} {\bibinfo
  {journal} {Phys. Rev.}\ }\textbf {\bibinfo {volume} {D89}},\ \bibinfo {pages}
  {064058} (\bibinfo {year} {2014}{\natexlab{b}})},\ \Eprint
  {http://arxiv.org/abs/1401.4548} {arXiv:1401.4548 [gr-qc]} \BibitemShut
  {NoStop}%
%%CITATION = ARXIV:1401.4548;%%
\bibitem [{\citenamefont {Damour}\ \emph {et~al.}(2015)\citenamefont {Damour},
  \citenamefont {Jaranowski},\ and\ \citenamefont
  {Sch{\"a}fer}}]{Damour:2015isa}%
  \BibitemOpen
  \bibfield  {author} {\bibinfo {author} {\bibfnamefont {T.}~\bibnamefont
  {Damour}}, \bibinfo {author} {\bibfnamefont {P.}~\bibnamefont {Jaranowski}},
  \ and\ \bibinfo {author} {\bibfnamefont {G.}~\bibnamefont {Sch{\"a}fer}},\
  }\href {\doibase 10.1103/PhysRevD.91.084024} {\bibfield  {journal} {\bibinfo
  {journal} {Phys. Rev.}\ }\textbf {\bibinfo {volume} {D91}},\ \bibinfo {pages}
  {084024} (\bibinfo {year} {2015})},\ \Eprint
  {http://arxiv.org/abs/1502.07245} {arXiv:1502.07245 [gr-qc]} \BibitemShut
  {NoStop}%
%%CITATION = ARXIV:1502.07245;%%
\bibitem [{\citenamefont {Bernard}\ \emph {et~al.}(2016)\citenamefont
  {Bernard}, \citenamefont {Blanchet}, \citenamefont {BohŽ}, \citenamefont
  {Faye},\ and\ \citenamefont {Marsat}}]{Bernard:2015njp}%
  \BibitemOpen
  \bibfield  {author} {\bibinfo {author} {\bibfnamefont {L.}~\bibnamefont
  {Bernard}}, \bibinfo {author} {\bibfnamefont {L.}~\bibnamefont {Blanchet}},
  \bibinfo {author} {\bibfnamefont {A.}~\bibnamefont {BohŽ}}, \bibinfo {author}
  {\bibfnamefont {G.}~\bibnamefont {Faye}}, \ and\ \bibinfo {author}
  {\bibfnamefont {S.}~\bibnamefont {Marsat}},\ }\href {\doibase
  10.1103/PhysRevD.93.084037} {\bibfield  {journal} {\bibinfo  {journal} {Phys.
  Rev.}\ }\textbf {\bibinfo {volume} {D93}},\ \bibinfo {pages} {084037}
  (\bibinfo {year} {2016})},\ \Eprint {http://arxiv.org/abs/1512.02876}
  {arXiv:1512.02876 [gr-qc]} \BibitemShut {NoStop}%
%%CITATION = ARXIV:1512.02876;%%
\bibitem [{\citenamefont {Damour}\ \emph {et~al.}(2016)\citenamefont {Damour},
  \citenamefont {Jaranowski},\ and\ \citenamefont
  {Sch{\"a}fer}}]{Damour:2016abl}%
  \BibitemOpen
  \bibfield  {author} {\bibinfo {author} {\bibfnamefont {T.}~\bibnamefont
  {Damour}}, \bibinfo {author} {\bibfnamefont {P.}~\bibnamefont {Jaranowski}},
  \ and\ \bibinfo {author} {\bibfnamefont {G.}~\bibnamefont {Sch{\"a}fer}},\
  }\href {\doibase 10.1103/PhysRevD.93.084014} {\bibfield  {journal} {\bibinfo
  {journal} {Phys. Rev.}\ }\textbf {\bibinfo {volume} {D93}},\ \bibinfo {pages}
  {084014} (\bibinfo {year} {2016})},\ \Eprint
  {http://arxiv.org/abs/1601.01283} {arXiv:1601.01283 [gr-qc]} \BibitemShut
  {NoStop}%
%%CITATION = ARXIV:1601.01283;%%
\bibitem [{\citenamefont {Hartung}\ and\ \citenamefont
  {Steinhoff}(2011)}]{Hartung:2011te}%
  \BibitemOpen
  \bibfield  {author} {\bibinfo {author} {\bibfnamefont {J.}~\bibnamefont
  {Hartung}}\ and\ \bibinfo {author} {\bibfnamefont {J.}~\bibnamefont
  {Steinhoff}},\ }\href {\doibase 10.1002/andp.201100094} {\bibfield  {journal}
  {\bibinfo  {journal} {Annalen Phys.}\ }\textbf {\bibinfo {volume} {523}},\
  \bibinfo {pages} {783} (\bibinfo {year} {2011})},\ \Eprint
  {http://arxiv.org/abs/1104.3079} {arXiv:1104.3079 [gr-qc]} \BibitemShut
  {NoStop}%
%%CITATION = ARXIV:1104.3079;%%
\bibitem [{\citenamefont {Marchand}\ \emph {et~al.}(2017)\citenamefont
  {Marchand}, \citenamefont {Bernard}, \citenamefont {Blanchet},\ and\
  \citenamefont {Faye}}]{Marchand:2017pir}%
  \BibitemOpen
  \bibfield  {author} {\bibinfo {author} {\bibfnamefont {T.}~\bibnamefont
  {Marchand}}, \bibinfo {author} {\bibfnamefont {L.}~\bibnamefont {Bernard}},
  \bibinfo {author} {\bibfnamefont {L.}~\bibnamefont {Blanchet}}, \ and\
  \bibinfo {author} {\bibfnamefont {G.}~\bibnamefont {Faye}},\ }\href@noop {}
  {\  (\bibinfo {year} {2017})},\ \Eprint {http://arxiv.org/abs/1707.09289}
  {arXiv:1707.09289 [gr-qc]} \BibitemShut {NoStop}%
%%CITATION = ARXIV:1707.09289;%%
\bibitem [{\citenamefont {Hartung}\ \emph {et~al.}(2013)\citenamefont
  {Hartung}, \citenamefont {Steinhoff},\ and\ \citenamefont
  {Sch{\"a}fer}}]{Hartung:Steinhoff:Schafer:2012}%
  \BibitemOpen
  \bibfield  {author} {\bibinfo {author} {\bibfnamefont {J.}~\bibnamefont
  {Hartung}}, \bibinfo {author} {\bibfnamefont {J.}~\bibnamefont {Steinhoff}},
  \ and\ \bibinfo {author} {\bibfnamefont {G.}~\bibnamefont {Sch{\"a}fer}},\
  }\href {\doibase 10.1002/andp.201200271} {\bibfield  {journal} {\bibinfo
  {journal} {Ann. Phys. (Berlin)}\ }\textbf {\bibinfo {volume} {525}},\
  \bibinfo {pages} {359} (\bibinfo {year} {2013})},\ \Eprint
  {http://arxiv.org/abs/1302.6723} {arXiv:1302.6723 [gr-qc]} \BibitemShut
  {NoStop}%
\bibitem [{\citenamefont {Marsat}\ \emph {et~al.}(2013)\citenamefont {Marsat},
  \citenamefont {Bohe}, \citenamefont {Faye},\ and\ \citenamefont
  {Blanchet}}]{Marsat:2012fn}%
  \BibitemOpen
  \bibfield  {author} {\bibinfo {author} {\bibfnamefont {S.}~\bibnamefont
  {Marsat}}, \bibinfo {author} {\bibfnamefont {A.}~\bibnamefont {Bohe}},
  \bibinfo {author} {\bibfnamefont {G.}~\bibnamefont {Faye}}, \ and\ \bibinfo
  {author} {\bibfnamefont {L.}~\bibnamefont {Blanchet}},\ }\href {\doibase
  10.1088/0264-9381/30/5/055007} {\bibfield  {journal} {\bibinfo  {journal}
  {Class. Quant. Grav.}\ }\textbf {\bibinfo {volume} {30}},\ \bibinfo {pages}
  {055007} (\bibinfo {year} {2013})},\ \Eprint {http://arxiv.org/abs/1210.4143}
  {arXiv:1210.4143 [gr-qc]} \BibitemShut {NoStop}%
%%CITATION = ARXIV:1210.4143;%%
\bibitem [{\citenamefont {Bohe}\ \emph {et~al.}(2013)\citenamefont {Bohe},
  \citenamefont {Marsat}, \citenamefont {Faye},\ and\ \citenamefont
  {Blanchet}}]{Bohe:2012mr}%
  \BibitemOpen
  \bibfield  {author} {\bibinfo {author} {\bibfnamefont {A.}~\bibnamefont
  {Bohe}}, \bibinfo {author} {\bibfnamefont {S.}~\bibnamefont {Marsat}},
  \bibinfo {author} {\bibfnamefont {G.}~\bibnamefont {Faye}}, \ and\ \bibinfo
  {author} {\bibfnamefont {L.}~\bibnamefont {Blanchet}},\ }\href {\doibase
  10.1088/0264-9381/30/7/075017} {\bibfield  {journal} {\bibinfo  {journal}
  {Class. Quant. Grav.}\ }\textbf {\bibinfo {volume} {30}},\ \bibinfo {pages}
  {075017} (\bibinfo {year} {2013})},\ \Eprint {http://arxiv.org/abs/1212.5520}
  {arXiv:1212.5520} \BibitemShut {NoStop}%
%%CITATION = ARXIV:1212.5520;%%
\bibitem [{\citenamefont {Levi}\ and\ \citenamefont
  {Steinhoff}(2016{\natexlab{a}})}]{Levi:Steinhoff:2015:2}%
  \BibitemOpen
  \bibfield  {author} {\bibinfo {author} {\bibfnamefont {M.}~\bibnamefont
  {Levi}}\ and\ \bibinfo {author} {\bibfnamefont {J.}~\bibnamefont
  {Steinhoff}},\ }\href {\doibase 10.1088/1475-7516/2016/01/011} {\bibfield
  {journal} {\bibinfo  {journal} {JCAP}\ }\textbf {\bibinfo {volume} {1601}},\
  \bibinfo {pages} {011} (\bibinfo {year} {2016}{\natexlab{a}})},\ \Eprint
  {http://arxiv.org/abs/1506.05056} {arXiv:1506.05056 [gr-qc]} \BibitemShut
  {NoStop}%
%%CITATION = ARXIV:1506.05056;%%
\bibitem [{\citenamefont {Levi}\ and\ \citenamefont
  {Steinhoff}(2016{\natexlab{b}})}]{Levi:Steinhoff:2015:3}%
  \BibitemOpen
  \bibfield  {author} {\bibinfo {author} {\bibfnamefont {M.}~\bibnamefont
  {Levi}}\ and\ \bibinfo {author} {\bibfnamefont {J.}~\bibnamefont
  {Steinhoff}},\ }\href {\doibase 10.1088/1475-7516/2016/01/008} {\bibfield
  {journal} {\bibinfo  {journal} {JCAP}\ }\textbf {\bibinfo {volume} {1601}},\
  \bibinfo {pages} {008} (\bibinfo {year} {2016}{\natexlab{b}})},\ \Eprint
  {http://arxiv.org/abs/1506.05794} {arXiv:1506.05794 [gr-qc]} \BibitemShut
  {NoStop}%
%%CITATION = ARXIV:1506.05794;%%
\bibitem [{\citenamefont {Hergt}\ and\ \citenamefont
  {Sch{\"a}fer}(2008{\natexlab{a}})}]{Hergt:2007ha}%
  \BibitemOpen
  \bibfield  {author} {\bibinfo {author} {\bibfnamefont {S.}~\bibnamefont
  {Hergt}}\ and\ \bibinfo {author} {\bibfnamefont {G.}~\bibnamefont
  {Sch{\"a}fer}},\ }\href {\doibase 10.1103/PhysRevD.77.104001} {\bibfield
  {journal} {\bibinfo  {journal} {Phys. Rev.}\ }\textbf {\bibinfo {volume}
  {D77}},\ \bibinfo {pages} {104001} (\bibinfo {year} {2008}{\natexlab{a}})},\
  \Eprint {http://arxiv.org/abs/0712.1515} {arXiv:0712.1515 [gr-qc]}
  \BibitemShut {NoStop}%
%%CITATION = ARXIV:0712.1515;%%
\bibitem [{\citenamefont {Hergt}\ and\ \citenamefont
  {Sch{\"a}fer}(2008{\natexlab{b}})}]{Hergt:2008jn}%
  \BibitemOpen
  \bibfield  {author} {\bibinfo {author} {\bibfnamefont {S.}~\bibnamefont
  {Hergt}}\ and\ \bibinfo {author} {\bibfnamefont {G.}~\bibnamefont
  {Sch{\"a}fer}},\ }\href {\doibase 10.1103/PhysRevD.78.124004} {\bibfield
  {journal} {\bibinfo  {journal} {Phys. Rev.}\ }\textbf {\bibinfo {volume}
  {D78}},\ \bibinfo {pages} {124004} (\bibinfo {year} {2008}{\natexlab{b}})},\
  \Eprint {http://arxiv.org/abs/0809.2208} {arXiv:0809.2208 [gr-qc]}
  \BibitemShut {NoStop}%
%%CITATION = ARXIV:0809.2208;%%
\bibitem [{\citenamefont {Levi}\ and\ \citenamefont
  {Steinhoff}(2015{\natexlab{b}})}]{Levi:Steinhoff:2014:2}%
  \BibitemOpen
  \bibfield  {author} {\bibinfo {author} {\bibfnamefont {M.}~\bibnamefont
  {Levi}}\ and\ \bibinfo {author} {\bibfnamefont {J.}~\bibnamefont
  {Steinhoff}},\ }\href {\doibase 10.1007/JHEP06(2015)059} {\bibfield
  {journal} {\bibinfo  {journal} {JHEP}\ }\textbf {\bibinfo {volume} {06}},\
  \bibinfo {pages} {059} (\bibinfo {year} {2015}{\natexlab{b}})},\ \Eprint
  {http://arxiv.org/abs/1410.2601} {arXiv:1410.2601 [gr-qc]} \BibitemShut
  {NoStop}%
%%CITATION = ARXIV:1410.2601;%%
\bibitem [{\citenamefont {Tulczyjew}(1959)}]{Tulczyjew:1959}%
  \BibitemOpen
  \bibfield  {author} {\bibinfo {author} {\bibfnamefont {W.~M.}\ \bibnamefont
  {Tulczyjew}},\ }\href@noop {} {\bibfield  {journal} {\bibinfo  {journal}
  {Acta Phys. Pol.}\ }\textbf {\bibinfo {volume} {18}},\ \bibinfo {pages} {393}
  (\bibinfo {year} {1959})}\BibitemShut {NoStop}%
\bibitem [{\citenamefont {Barker}\ and\ \citenamefont
  {O'Connell}(1975)}]{Barker:1975ae}%
  \BibitemOpen
  \bibfield  {author} {\bibinfo {author} {\bibfnamefont {B.~M.}\ \bibnamefont
  {Barker}}\ and\ \bibinfo {author} {\bibfnamefont {R.~F.}\ \bibnamefont
  {O'Connell}},\ }\href {\doibase 10.1103/PhysRevD.12.329} {\bibfield
  {journal} {\bibinfo  {journal} {Phys. Rev.}\ }\textbf {\bibinfo {volume}
  {D12}},\ \bibinfo {pages} {329} (\bibinfo {year} {1975})}\BibitemShut
  {NoStop}%
%%CITATION = PHRVA,D12,329;%%
\bibitem [{\citenamefont {Barker}\ and\ \citenamefont
  {O'Connell}(1979)}]{Barker1979}%
  \BibitemOpen
  \bibfield  {author} {\bibinfo {author} {\bibfnamefont {B.~M.}\ \bibnamefont
  {Barker}}\ and\ \bibinfo {author} {\bibfnamefont {R.~F.}\ \bibnamefont
  {O'Connell}},\ }\href {\doibase 10.1007/BF00756587} {\bibfield  {journal}
  {\bibinfo  {journal} {General Relativity and Gravitation}\ }\textbf {\bibinfo
  {volume} {11}},\ \bibinfo {pages} {149} (\bibinfo {year} {1979})}\BibitemShut
  {NoStop}%
\bibitem [{\citenamefont {Kyrian}\ and\ \citenamefont
  {Semerak}(2007)}]{Kyrian:2007zz}%
  \BibitemOpen
  \bibfield  {author} {\bibinfo {author} {\bibfnamefont {K.}~\bibnamefont
  {Kyrian}}\ and\ \bibinfo {author} {\bibfnamefont {O.}~\bibnamefont
  {Semerak}},\ }\href {\doibase 10.1111/j.1365-2966.2007.12502.x} {\bibfield
  {journal} {\bibinfo  {journal} {Mon. Not. Roy. Astron. Soc.}\ }\textbf
  {\bibinfo {volume} {382}},\ \bibinfo {pages} {1922} (\bibinfo {year}
  {2007})}\BibitemShut {NoStop}%
%%CITATION = MNRAA,382,1922;%%
\bibitem [{\citenamefont {Taylor}\ and\ \citenamefont
  {Poisson}(2008)}]{Taylor:2008xy}%
  \BibitemOpen
  \bibfield  {author} {\bibinfo {author} {\bibfnamefont {S.}~\bibnamefont
  {Taylor}}\ and\ \bibinfo {author} {\bibfnamefont {E.}~\bibnamefont
  {Poisson}},\ }\href {\doibase 10.1103/PhysRevD.78.084016} {\bibfield
  {journal} {\bibinfo  {journal} {Phys. Rev.}\ }\textbf {\bibinfo {volume}
  {D78}},\ \bibinfo {pages} {084016} (\bibinfo {year} {2008})},\ \Eprint
  {http://arxiv.org/abs/0806.3052} {arXiv:0806.3052 [gr-qc]} \BibitemShut
  {NoStop}%
%%CITATION = ARXIV:0806.3052;%%
\bibitem [{\citenamefont {Damour}\ and\ \citenamefont
  {Nagar}(2009)}]{Damour:2009vw}%
  \BibitemOpen
  \bibfield  {author} {\bibinfo {author} {\bibfnamefont {T.}~\bibnamefont
  {Damour}}\ and\ \bibinfo {author} {\bibfnamefont {A.}~\bibnamefont {Nagar}},\
  }\href {\doibase 10.1103/PhysRevD.80.084035} {\bibfield  {journal} {\bibinfo
  {journal} {Phys. Rev.}\ }\textbf {\bibinfo {volume} {D80}},\ \bibinfo {pages}
  {084035} (\bibinfo {year} {2009})},\ \Eprint {http://arxiv.org/abs/0906.0096}
  {arXiv:0906.0096 [gr-qc]} \BibitemShut {NoStop}%
%%CITATION = ARXIV:0906.0096;%%
\bibitem [{\citenamefont {Kol}\ and\ \citenamefont
  {Smolkin}(2012)}]{Kol:2011vg}%
  \BibitemOpen
  \bibfield  {author} {\bibinfo {author} {\bibfnamefont {B.}~\bibnamefont
  {Kol}}\ and\ \bibinfo {author} {\bibfnamefont {M.}~\bibnamefont {Smolkin}},\
  }\href {\doibase 10.1007/JHEP02(2012)010} {\bibfield  {journal} {\bibinfo
  {journal} {JHEP}\ }\textbf {\bibinfo {volume} {02}},\ \bibinfo {pages} {010}
  (\bibinfo {year} {2012})},\ \Eprint {http://arxiv.org/abs/1110.3764}
  {arXiv:1110.3764 [hep-th]} \BibitemShut {NoStop}%
%%CITATION = ARXIV:1110.3764;%%
\bibitem [{\citenamefont {Pani}\ \emph {et~al.}(2015)\citenamefont {Pani},
  \citenamefont {Gualtieri}, \citenamefont {Maselli},\ and\ \citenamefont
  {Ferrari}}]{Pani:2015hfa}%
  \BibitemOpen
  \bibfield  {author} {\bibinfo {author} {\bibfnamefont {P.}~\bibnamefont
  {Pani}}, \bibinfo {author} {\bibfnamefont {L.}~\bibnamefont {Gualtieri}},
  \bibinfo {author} {\bibfnamefont {A.}~\bibnamefont {Maselli}}, \ and\
  \bibinfo {author} {\bibfnamefont {V.}~\bibnamefont {Ferrari}},\ }\href
  {\doibase 10.1103/PhysRevD.92.024010} {\bibfield  {journal} {\bibinfo
  {journal} {Phys. Rev.}\ }\textbf {\bibinfo {volume} {D92}},\ \bibinfo {pages}
  {024010} (\bibinfo {year} {2015})},\ \Eprint
  {http://arxiv.org/abs/1503.07365} {arXiv:1503.07365 [gr-qc]} \BibitemShut
  {NoStop}%
%%CITATION = ARXIV:1503.07365;%%
\bibitem [{\citenamefont {Bini}\ \emph {et~al.}(2012)\citenamefont {Bini},
  \citenamefont {Damour},\ and\ \citenamefont {Faye}}]{Bini:2012gu}%
  \BibitemOpen
  \bibfield  {author} {\bibinfo {author} {\bibfnamefont {D.}~\bibnamefont
  {Bini}}, \bibinfo {author} {\bibfnamefont {T.}~\bibnamefont {Damour}}, \ and\
  \bibinfo {author} {\bibfnamefont {G.}~\bibnamefont {Faye}},\ }\href {\doibase
  10.1103/PhysRevD.85.124034} {\bibfield  {journal} {\bibinfo  {journal} {Phys.
  Rev.}\ }\textbf {\bibinfo {volume} {D85}},\ \bibinfo {pages} {124034}
  (\bibinfo {year} {2012})},\ \Eprint {http://arxiv.org/abs/1202.3565}
  {arXiv:1202.3565 [gr-qc]} \BibitemShut {NoStop}%
%%CITATION = ARXIV:1202.3565;%%
\bibitem [{\citenamefont {Brezin}\ \emph {et~al.}(1970)\citenamefont {Brezin},
  \citenamefont {Itzykson},\ and\ \citenamefont {Zinn-Justin}}]{Brezin:1970zr}%
  \BibitemOpen
  \bibfield  {author} {\bibinfo {author} {\bibfnamefont {E.}~\bibnamefont
  {Brezin}}, \bibinfo {author} {\bibfnamefont {C.}~\bibnamefont {Itzykson}}, \
  and\ \bibinfo {author} {\bibfnamefont {J.}~\bibnamefont {Zinn-Justin}},\
  }\href {\doibase 10.1103/PhysRevD.1.2349} {\bibfield  {journal} {\bibinfo
  {journal} {Phys. Rev.}\ }\textbf {\bibinfo {volume} {D1}},\ \bibinfo {pages}
  {2349} (\bibinfo {year} {1970})}\BibitemShut {NoStop}%
%%CITATION = PHRVA,D1,2349;%%
\bibitem [{\citenamefont {Barausse}\ \emph {et~al.}(2009)\citenamefont
  {Barausse}, \citenamefont {Racine},\ and\ \citenamefont
  {Buonanno}}]{Barausse:Racine:Buonanno:2009}%
  \BibitemOpen
  \bibfield  {author} {\bibinfo {author} {\bibfnamefont {E.}~\bibnamefont
  {Barausse}}, \bibinfo {author} {\bibfnamefont {{\'E}.}~\bibnamefont
  {Racine}}, \ and\ \bibinfo {author} {\bibfnamefont {A.}~\bibnamefont
  {Buonanno}},\ }\href {\doibase 10.1103/PhysRevD.80.104025} {\bibfield
  {journal} {\bibinfo  {journal} {Phys. Rev. D}\ }\textbf {\bibinfo {volume}
  {80}},\ \bibinfo {pages} {104025} (\bibinfo {year} {2009})},\ \Eprint
  {http://arxiv.org/abs/0907.4745} {arXiv:0907.4745 [gr-qc]} \BibitemShut
  {NoStop}%
%%CITATION = 0907.4745;%%
\bibitem [{\citenamefont {Vines}\ \emph {et~al.}(2016)\citenamefont {Vines},
  \citenamefont {Kunst}, \citenamefont {Steinhoff},\ and\ \citenamefont
  {Hinderer}}]{VKSH}%
  \BibitemOpen
  \bibfield  {author} {\bibinfo {author} {\bibfnamefont {J.}~\bibnamefont
  {Vines}}, \bibinfo {author} {\bibfnamefont {D.}~\bibnamefont {Kunst}},
  \bibinfo {author} {\bibfnamefont {J.}~\bibnamefont {Steinhoff}}, \ and\
  \bibinfo {author} {\bibfnamefont {T.}~\bibnamefont {Hinderer}},\ }\href
  {\doibase 10.1103/PhysRevD.93.103008} {\bibfield  {journal} {\bibinfo
  {journal} {Phys. Rev.}\ }\textbf {\bibinfo {volume} {D93}},\ \bibinfo {pages}
  {103008} (\bibinfo {year} {2016})},\ \Eprint
  {http://arxiv.org/abs/1601.07529} {arXiv:1601.07529 [gr-qc]} \BibitemShut
  {NoStop}%
%%CITATION = ARXIV:1601.07529;%%
\bibitem [{\citenamefont {Pryce}(1935)}]{Pryce:1935}%
  \BibitemOpen
  \bibfield  {author} {\bibinfo {author} {\bibfnamefont {M.~H.~L.}\
  \bibnamefont {Pryce}},\ }\href {\doibase doi:10.1098/rspa.1935.0094}
  {\bibfield  {journal} {\bibinfo  {journal} {Proc. R. Soc. A}\ }\textbf
  {\bibinfo {volume} {150}},\ \bibinfo {pages} {166} (\bibinfo {year}
  {1935})}\BibitemShut {NoStop}%
\bibitem [{\citenamefont {Pryce}(1948)}]{Pryce:1948}%
  \BibitemOpen
  \bibfield  {author} {\bibinfo {author} {\bibfnamefont {M.~H.~L.}\
  \bibnamefont {Pryce}},\ }\href {\doibase 10.1098/rspa.1948.0103} {\bibfield
  {journal} {\bibinfo  {journal} {Proc. R. Soc. A}\ }\textbf {\bibinfo {volume}
  {195}},\ \bibinfo {pages} {62} (\bibinfo {year} {1948})}\BibitemShut
  {NoStop}%
%%CITATION = PRSLA,A195,62;%%
\bibitem [{\citenamefont {Newton}\ and\ \citenamefont
  {Wigner}(1949)}]{Newton:Wigner:1949}%
  \BibitemOpen
  \bibfield  {author} {\bibinfo {author} {\bibfnamefont {T.~D.}\ \bibnamefont
  {Newton}}\ and\ \bibinfo {author} {\bibfnamefont {E.~P.}\ \bibnamefont
  {Wigner}},\ }\href {\doibase 10.1103/RevModPhys.21.400} {\bibfield  {journal}
  {\bibinfo  {journal} {Rev. Mod. Phys.}\ }\textbf {\bibinfo {volume} {21}},\
  \bibinfo {pages} {400} (\bibinfo {year} {1949})}\BibitemShut {NoStop}%
%%CITATION = RMPHA,21,400;%%
\bibitem [{\citenamefont {Kerr}\ and\ \citenamefont {Schild}(2009)}]{Kerr2009}%
  \BibitemOpen
  \bibfield  {author} {\bibinfo {author} {\bibfnamefont {R.~P.}\ \bibnamefont
  {Kerr}}\ and\ \bibinfo {author} {\bibfnamefont {A.}~\bibnamefont {Schild}},\
  }\href {\doibase 10.1007/s10714-009-0857-z} {\bibfield  {journal} {\bibinfo
  {journal} {General Relativity and Gravitation}\ }\textbf {\bibinfo {volume}
  {41}},\ \bibinfo {pages} {2485} (\bibinfo {year} {2009})}\BibitemShut
  {NoStop}%
\bibitem [{\citenamefont {Xanthopoulos}(1978)}]{doi:10.1063/1.523851}%
  \BibitemOpen
  \bibfield  {author} {\bibinfo {author} {\bibfnamefont {B.~C.}\ \bibnamefont
  {Xanthopoulos}},\ }\href {\doibase 10.1063/1.523851} {\bibfield  {journal}
  {\bibinfo  {journal} {Journal of Mathematical Physics}\ }\textbf {\bibinfo
  {volume} {19}},\ \bibinfo {pages} {1607} (\bibinfo {year} {1978})},\ \Eprint
  {http://arxiv.org/abs/http://dx.doi.org/10.1063/1.523851}
  {http://dx.doi.org/10.1063/1.523851} \BibitemShut {NoStop}%
\bibitem [{\citenamefont {{Steinhoff}}(2015)}]{Steinhoff:2015}%
  \BibitemOpen
  \bibfield  {author} {\bibinfo {author} {\bibfnamefont {J.}~\bibnamefont
  {{Steinhoff}}},\ }\href@noop {} {\bibfield  {journal} {\bibinfo  {journal}
  {ArXiv e-prints}\ } (\bibinfo {year} {2015})},\ \Eprint
  {http://arxiv.org/abs/1501.04951} {arXiv:1501.04951 [gr-qc]} \BibitemShut
  {NoStop}%
\bibitem [{\citenamefont {DeWitt}(1964)}]{DeWitt:1965jb}%
  \BibitemOpen
  \bibfield  {author} {\bibinfo {author} {\bibfnamefont {B.~S.}\ \bibnamefont
  {DeWitt}},\ }\bibfield  {booktitle} {\emph {\bibinfo {booktitle}
  {{RelativitŽ, Groupes et Topologie: Proceedings, Ecole d'ŽtŽ de Physique
  ThŽorique, Session XIII, Les Houches, France, Jul 1 - Aug 24, 1963}}},\
  }\href@noop {} {\bibfield  {journal} {\bibinfo  {journal} {Conf. Proc.}\
  }\textbf {\bibinfo {volume} {C630701}},\ \bibinfo {pages} {585} (\bibinfo
  {year} {1964})},\ \bibinfo {note} {[Les Houches Lect.
  Notes13,585(1964)]}\BibitemShut {NoStop}%
%%CITATION = CONFP,C630701,585;%%
\end{thebibliography}


\end{document}